\documentclass[12pt]{article}
\usepackage{epsf,epsfig,here,amssymb}
\begin{document}
\newcommand{\Kstarr}{\mathrm{K^{*\pm}}}
\newcommand{\Kstarm}{\mathrm{K^{*-}}}
\newcommand{\Kstarmn}{\mathrm{K^{*-}}}
\newcommand{\Km}{\mathrm{K^{-}}}
\newcommand{\Dsl}{{\mathrm{ D_s}}\ell}
\newcommand{\bt}{\begin{tabular}}
\newcommand{\et}{\end{tabular}}
\newcommand{\Mecqt}{{$\mathrm{MeV/c^2}$}}
\newcommand{\Mecqm}{{\mathrm{MeV/c^2}}}
\newcommand{\Gecqt}{{$\mathrm{GeV/c^2}$}}
\newcommand{\Gecqm}{{\mthrm{GeV/c^2}}}
\newcommand{\DB}{\Delta B}
\newcommand{\as}{\alpha_{ s}}
\newcommand{\ep}{\varepsilon}
\newcommand{\rp}{\tau({\rm B^+})/\tau({\rm B^0_d})}
\newcommand{\rs}{\tau({\rm B^0_s})/\tau({\rm B^0_d})}
\newcommand{\rl}{\tau({\rm \Lambda_b})/\tau({\rm B^0_d})}
\newcommand{\dgs}{\Delta \Gamma_{\Bs}}
\newcommand{\dgbs}{\Delta \Gamma_{\rm \Bs}/\Gamma_{\rm \Bs}}
\newcommand{\tbs}{\tau_{\Bs}}
\newcommand{\tbd}{\tau_{\Bd}}
\newcommand{\Gs}{\Gamma_{{\rm B^0_s}}}
\newcommand{\Gd}{\Gamma_{{\rm B^0_d}}}
\newcommand{\rhobar}{\overline {\rho}}
\newcommand{\etabar}{\overline{\eta}}
\newcommand{\epsilonk}{\varepsilon_K}
\newcommand{\vubovcb}{\left | \frac{V_{ub}}{V_{cb}} \right |}
\newcommand{\vubsvcb}{\left | V_{ub}/V_{cb}  \right |}
\newcommand{\vtdovts}{\left | \frac{V_{td}}{V_{ts}} \right |}
\newcommand{\epsp}{\frac{\varepsilon^{'}}{\varepsilon}}
\newcommand{\dmd}{\Delta m_d}
\newcommand{\dms}{\Delta m_s}
\newcommand{\Do}{{\rm D}^0}
\newcommand{\pimora}{\pi^{\ast}}
\newcommand{\Bstar}{{\rm B}^{\ast}}
\newcommand{\Bstarstar}{{\rm B}^{\ast \ast}}
\newcommand{\Dstar}{{\rm D}^{\ast}}
\newcommand{\Dstars}{{\rm D}^{\ast +}_s}
\newcommand{\Dstaro}{{\rm D}^{\ast 0}}
\newcommand{\Dstarp}{{\rm D}^{\ast +}}
\newcommand{\Dstarstar}{{\rm D}^{\ast \ast}}
\newcommand{\pistar}{\pi^{\ast}}
\newcommand{\pisstar}{\pi^{\ast \ast}}
\newcommand{\bptre}{\rm b^{+}_{3}}
\newcommand{\Vcb}{\left | {\rm V}_{cb} \right |}
\newcommand{\Vub}{\left | {\rm V}_{ub} \right |}
\newcommand{\Vcs}{\left | {\rm V}_{cs} \right |}
\newcommand{\Vtd}{\left | {\rm V}_{td} \right |}
\newcommand{\Vts}{\left | {\rm V}_{ts} \right |}
\newcommand{\bp}{\rm b^{+}_{1}}
\newcommand{\bo}{\rm b^0}
\newcommand{\bos}{\rm b^0_s}
\newcommand{\bss}{\rm b^s_s}
\newcommand{\qq}{\rm q \overline{q}}
\newcommand{\cc}{\rm c \overline{c}}
\newcommand{\BsDmX}{{B_{s}^{0}} \rightarrow D \mu X}
\newcommand{\BsDsm}{{B_{s}^{0}} \rightarrow D_{s} \mu X}
\newcommand{\BsDsX}{{B_{s}^{0}} \rightarrow D_{s} X}
\newcommand{\BDsX}{B \rightarrow D_{s} X}
\newcommand{\BDomX}{B \rightarrow D^{0} \mu X}
\newcommand{\BDpmX}{B \rightarrow D^{+} \mu X}
\newcommand{\Dsfmn}{D_{s} \rightarrow \phi \mu \nu}
\newcommand{\Dsfipi}{D_{s} \rightarrow \phi \pi}
\newcommand{\DsfX}{D_{s} \rightarrow \phi X}
\newcommand{\DpfX}{D^{+} \rightarrow \phi X}
\newcommand{\DofX}{D^{0} \rightarrow \phi X}
\newcommand{\DfX}{D \rightarrow \phi X}
\newcommand{\DsD}{B \rightarrow D_{s} D}
\newcommand{\DsmX}{D_{s} \rightarrow \mu X}
\newcommand{\DmX}{D \rightarrow \mu X}
\newcommand{\Zbb}{Z^{0} \rightarrow \rm b \overline{b}}
\newcommand{\Zcc}{Z^{0} \rightarrow \rm c \overline{c}}
\newcommand{\Rbb}{\frac{\Gamma_{Z^0 \rightarrow \rm b \overline{b}}}
{\Gamma_{Z^0 \rightarrow Hadrons}}}
\newcommand{\Rcc}{\frac{\Gamma_{Z^0 \rightarrow \rm c \overline{c}}}
{\Gamma_{Z^0 \rightarrow Hadrons}}}
\newcommand{\bb}{b \overline{b}}
\newcommand{\str}{\rm s \overline{s}}
\newcommand{\Bs}{\rm{B^0_s}}
\newcommand{\Bsb}{\overline{\rm{B}^0_s}}
\newcommand{\Bp}{\rm{B^{+}}}
\newcommand{\Bm}{\rm{B^{-}}}
\newcommand{\Bo}{\rm{B^{0}}}
\newcommand{\Bob}{\overline{\rm{B}^{0}}}
\newcommand{\Bd}{\rm{B^{0}_{d}}}
\newcommand{\Bdb}{\overline{\rm{B^{0}_{d}}}}
\newcommand{\Lb}{\Lambda^0_b}
\newcommand{\Lbb}{\overline{\Lambda^0_b}}
\newcommand{\Kstar}{\rm{K^{\star 0}}}
\newcommand{\phim}{\rm{\phi}}
\newcommand{\Ds}{\rm{D}_s}
\newcommand{\Dsp}{\rm{D}_s^+}
\newcommand{\Dsm}{\rm{D}_s^-}
\newcommand{\Dp}{\rm{D}^+}
\newcommand{\Dn}{\rm{D}^0}
\newcommand{\Dsb}{\overline{\rm{D}_s}}
\newcommand{\Dm}{\rm{D}^-}
\newcommand{\Dnb}{\overline{\rm{D}^0}}
\newcommand{\Lc}{\Lambda_c}
\newcommand{\Lcb}{\overline{\Lambda_c}}
\newcommand{\Dstarm}{\rm{D}^{\ast -}}
\newcommand{\Dsstarp}{\rm{D}_s^{\ast +}}
\newcommand{\Pb}{P_{b-baryon}}
\newcommand{\KKpi}{\rm{ K K \pi }}
\newcommand{\GeV}{\rm{GeV}}
\newcommand{\MeV}{\rm{MeV}}
\newcommand{\nb}{\rm{nb}}
\newcommand{\Zzero}{{\rm Z}^0}
\newcommand{\MZ}{\rm{M_Z}}
\newcommand{\MW}{\rm{M_W}}
\newcommand{\GF}{\rm{G_F}}
\newcommand{\Gm}{\rm{G_{\mu}}}
\newcommand{\MH}{\rm{M_H}}
\newcommand{\MT}{\rm{m_{top}}}
\newcommand{\GZ}{\Gamma_{\rm Z}}
\newcommand{\Afb}{\rm{A_{FB}}}
\newcommand{\Afbs}{\rm{A_{FB}^{s}}}
\newcommand{\sigmaf}{\sigma_{\rm{F}}}
\newcommand{\sigmab}{\sigma_{\rm{B}}}
\newcommand{\NF}{\rm{N_{F}}}
\newcommand{\NB}{\rm{N_{B}}}
\newcommand{\Nnu}{\rm{N_{\nu}}}
\newcommand{\RZ}{\rm{R_Z}}
\newcommand{\rhob}{\rho_{eff}}
\newcommand{\Gammanz}{\rm{\Gamma_{Z}^{new}}}
\newcommand{\Gammani}{\rm{\Gamma_{inv}^{new}}}
\newcommand{\Gammasz}{\rm{\Gamma_{Z}^{SM}}}
\newcommand{\Gammasi}{\rm{\Gamma_{inv}^{SM}}}
\newcommand{\Gammaxz}{\rm{\Gamma_{Z}^{exp}}}
\newcommand{\Gammaxi}{\rm{\Gamma_{inv}^{exp}}}
\newcommand{\rhoZ}{\rho_{\rm Z}}
\newcommand{\thw}{\theta_{\rm W}}
\newcommand{\swsq}{\sin^2\!\thw}
\newcommand{\swsqmsb}{\sin^2\!\theta_{\rm W}^{\overline{\rm MS}}}
\newcommand{\swsqbar}{\sin^2\!\overline{\theta}_{\rm W}}
\newcommand{\cwsqbar}{\cos^2\!\overline{\theta}_{\rm W}}
\newcommand{\swsqb}{\sin^2\!\theta^{eff}_{\rm W}}
\newcommand{\ee}{{e^+e^-}}
\newcommand{\eeX}{{e^+e^-X}}
\newcommand{\gaga}{{\gamma\gamma}}
\newcommand{\mumu}{\ifmmode {\mu^+\mu^-} \else ${\mu^+\mu^-} $ \fi}
\newcommand{\eeg}{{e^+e^-\gamma}}
\newcommand{\mumug}{{\mu^+\mu^-\gamma}}
\newcommand{\tautau}{{\tau^+\tau^-}}
\newcommand{\qqb}{{q\overline{q}}}
\newcommand{\eegg}{e^+e^-\rightarrow \gamma\gamma}
\newcommand{\eeggg}{e^+e^-\rightarrow \gamma\gamma\gamma}
\newcommand{\eeee}{e^+e^-\rightarrow e^+e^-}
\newcommand{\eeeeee}{e^+e^-\rightarrow e^+e^-e^+e^-}
\newcommand{\eeeeg}{e^+e^-\rightarrow e^+e^-(\gamma)}
\newcommand{\eeeegg}{e^+e^-\rightarrow e^+e^-\gamma\gamma}
\newcommand{\eeeg}{e^+e^-\rightarrow (e^+)e^-\gamma}
\newcommand{\eemumu}{e^+e^-\rightarrow \mu^+\mu^-}
\newcommand{\eetautau}{e^+e^-\rightarrow \tau^+\tau^-}
\newcommand{\eehad}{e^+e^-\rightarrow {\rm hadrons}}
\newcommand{\eettg}{e^+e^-\rightarrow \tau^+\tau^-\gamma}
\newcommand{\eell}{e^+e^-\rightarrow l^+l^-}
\newcommand{\Ztopig}{{\rm Z}^0\rightarrow \pi^0\gamma}
\newcommand{\Ztogg}{{\rm Z}^0\rightarrow \gamma\gamma}
\newcommand{\Ztoee}{{\rm Z}^0\rightarrow e^+e^-}
\newcommand{\Ztoggg}{{\rm Z}^0\rightarrow \gamma\gamma\gamma}
\newcommand{\Ztomumu}{{\rm Z}^0\rightarrow \mu^+\mu^-}
\newcommand{\Ztotautau}{{\rm Z}^0\rightarrow \tau^+\tau^-}
\newcommand{\Ztoll}{{\rm Z}^0\rightarrow l^+l^-}
\newcommand{\Ztocc}{{\rm Z^0\rightarrow c \overline c}}
\newcommand{\Lamp}{\Lambda_{+}}
\newcommand{\Lamm}{\Lambda_{-}}
\newcommand{\Pt}{\rm P_{t}}
\newcommand{\Gee}{\Gamma_{ee}}
\newcommand{\Gpig}{\Gamma_{\pi^0\gamma}}
\newcommand{\Ggg}{\Gamma_{\gamma\gamma}}
\newcommand{\Gggg}{\Gamma_{\gamma\gamma\gamma}}
\newcommand{\Gmumu}{\Gamma_{\mu\mu}}
\newcommand{\Gtautau}{\Gamma_{\tau\tau}}
\newcommand{\Ginv}{\Gamma_{\rm inv}}
\newcommand{\Ghad}{\Gamma_{\rm had}}
\newcommand{\Gnu}{\Gamma_{\nu}}
\newcommand{\GnuSM}{\Gamma_{\nu}^{\rm SM}}
\newcommand{\Gll}{\Gamma_{l^+l^-}}
\newcommand{\Gff}{\Gamma_{f\overline{f}}}
\newcommand{\Gtot}{\Gamma_{\rm tot}}
\newcommand{\Rb}{\mbox{R}_b}
\newcommand{\Rc}{\mbox{R}_c}
\newcommand{\al}{a_l}
\newcommand{\vl}{v_l}
\newcommand{\af}{a_f}
\newcommand{\vf}{v_f}
\newcommand{\ael}{a_e}
\newcommand{\ve}{v_e}
\newcommand{\amu}{a_\mu}
\newcommand{\vmu}{v_\mu}
\newcommand{\atau}{a_\tau}
\newcommand{\vtau}{v_\tau}
\newcommand{\ahatl}{\hat{a}_l}
\newcommand{\vhatl}{\hat{v}_l}
\newcommand{\ahate}{\hat{a}_e}
\newcommand{\vhate}{\hat{v}_e}
\newcommand{\ahatmu}{\hat{a}_\mu}
\newcommand{\vhatmu}{\hat{v}_\mu}
\newcommand{\ahattau}{\hat{a}_\tau}
\newcommand{\vhattau}{\hat{v}_\tau}
\newcommand{\vtildel}{\tilde{\rm v}_l}
\newcommand{\avsq}{\ahatl^2\vhatl^2}
\newcommand{\Ahatl}{\hat{A}_l}
\newcommand{\Vhatl}{\hat{V}_l}
\newcommand{\Afer}{A_f}
\newcommand{\Ael}{A_e}
\newcommand{\Aferb}{\overline{A_f}}
\newcommand{\Aelb}{\overline{A_e}}
\newcommand{\AVsq}{\Ahatl^2\Vhatl^2}
\newcommand{\Iwk}{I_{3l}}
\newcommand{\Qch}{|Q_{l}|}
\newcommand{\roots}{\sqrt{s}}
\newcommand{\pT}{p_{\rm T}}
\newcommand{\mt}{m_t}
\newcommand{\Rechi}{{\rm Re} \left\{ \chi (s) \right\}}
\newcommand{\up}{^}
\newcommand{\abscosthe}{|cos\theta|}
\newcommand{\sint}{\mbox{$\sin\theta$}}
\newcommand{\cost}{\mbox{$\cos\theta$}}
\newcommand{\mcost}{|\cos\theta|}
\newcommand{\epair}{\mbox{$e^{+}e^{-}$}}
\newcommand{\mupair}{\mbox{$\mu^{+}\mu^{-}$}}
\newcommand{\taupair}{\mbox{$\tau^{+}\tau^{-}$}}
\newcommand{\gamgam}{\mbox{$e^{+}e^{-}\rightarrow e^{+}e^{-}\mu^{+}\mu^{-}$}}
\newcommand{\fullskip}{\vskip 16cm}
\newcommand{\halfskip}{\vskip  8cm}
\newcommand{\quarskip}{\vskip  6cm}
\newcommand{\abitskip}{\vskip 0.5cm}
\newcommand{\ba}{\begin{array}}
\newcommand{\ea}{\end{array}}
\newcommand{\bc}{\begin{center}}
\newcommand{\ec}{\end{center}}
\newcommand{\be}{\begin{eqnarray}}
\newcommand{\eeq}{\end{eqnarray}}
\newcommand{\bes}{\begin{eqnarray*}}
\newcommand{\ees}{\end{eqnarray*}}
\newcommand{\Kz}{\ifmmode {\rm K^0_s} \else ${\rm K^0_s} $ \fi}
\newcommand{\Zz}{\ifmmode {\rm Z^0} \else ${\rm Z^0 } $ \fi}
\newcommand{\qqbar}{\ifmmode {\rm q\overline{q}} \else ${\rm q\overline{q}} $ \fi}
\newcommand{\ccbar}{\ifmmode {\rm c\overline{c}} \else ${\rm c\overline{c}} $ \fi}
\newcommand{\bbbar}{\ifmmode {\rm b\overline{b}} \else ${\rm b\overline{b}} $ \fi}
\newcommand{\xxbar}{\ifmmode {\rm x\overline{x}} \else ${\rm x\overline{x}} $ \fi}
\newcommand{\rphi}{\ifmmode {\rm R\phi} \else ${\rm R\phi} $ \fi}
\renewcommand\topfraction{1.}
\newcommand{\BK}{B_K}
\newcommand{\nubar}{\overline{\nu_{\ell}}}
\newcommand{\snb}{\sin{2\beta}}
\newcommand{\sna}{\sin{2\alpha}}
\newcommand{\vcb}{\left | {\rm V}_{cb} \right |}
\newcommand{\vub}{\left | {\rm V}_{ub} \right |}
\newcommand{\vus}{\left | V_{us} \right |}
\newcommand{\vud}{\left | V_{ud} \right |}
\newcommand{\vtd}{\left | {V}_{td} \right |}
\newcommand{\vts}{\left | { V}_{ts} \right |}
\newcommand{\fbdsqbd}{f_{B_d} \sqrt{\hat B_{B_d}}}
\newcommand{\fbssqbs}{f_{B_s} \sqrt{\hat B_{B_s}}}
\newcommand{\snbg}{\sin(2\beta + \gamma)}

\renewcommand{\arraystretch}{1.2}
\def\utfit{{\bf{U}}\kern-.24em{\bf{T}}\kern-.21em{\it{fit}}\@}

\pagestyle{empty}
\pagenumbering{arabic}
\vskip  1.5 cm
\begin{center}
{\LARGE {\bf The 2004 \utfit\ Collaboration 
Report}}
\vskip .3 cm 
{\LARGE {\bf on the Status of the Unitarity Triangle}}
\vskip .3 cm 
{\LARGE {\bf in the Standard Model}}
\vskip .1 cm 
\end{center}

\begin{figure}[htb!]
\begin{center}
{\includegraphics[width=2.5cm]{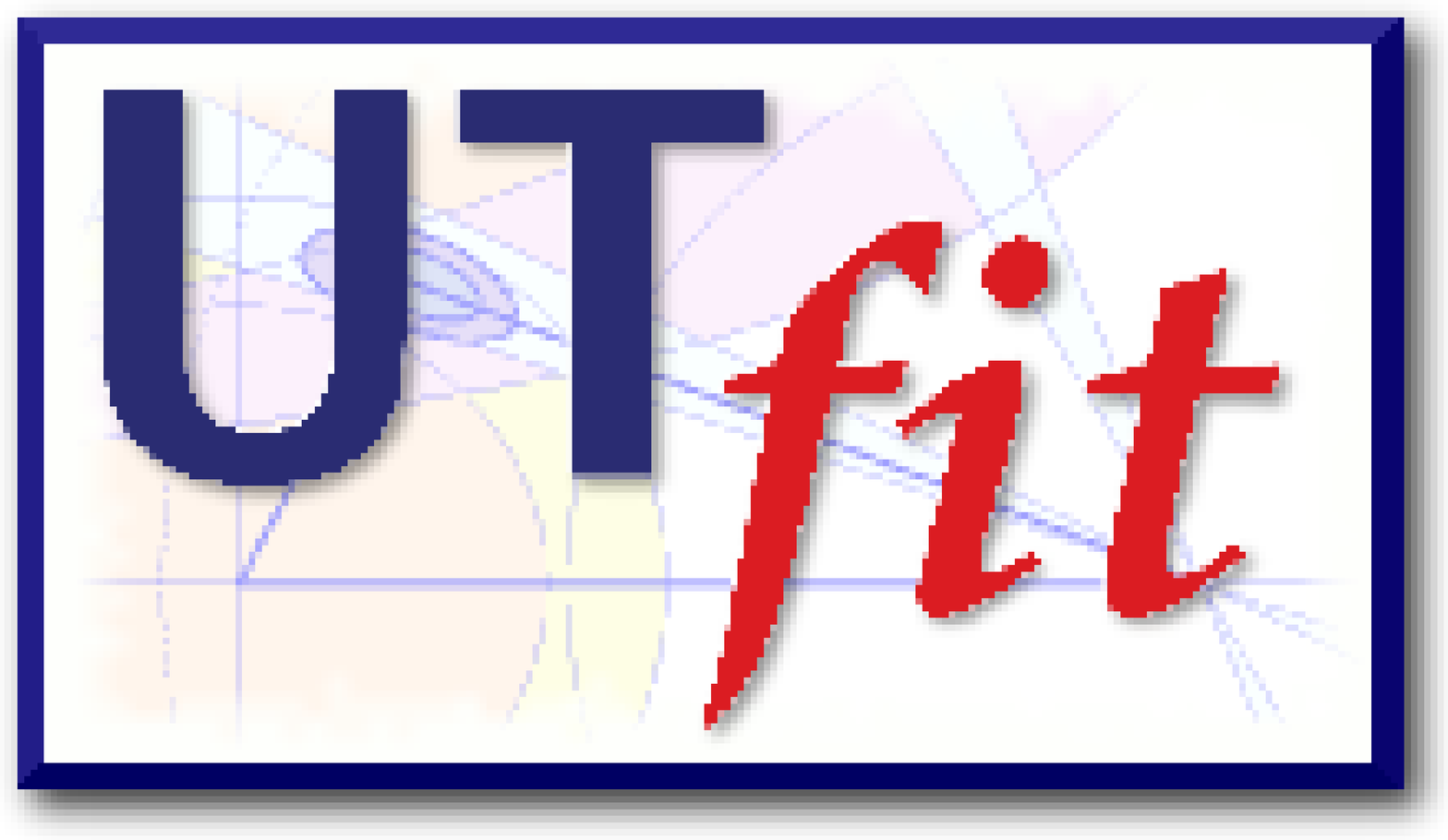}}
\end{center}
\end{figure}

\vspace*{-0.5cm}
\begin{center}
{\Large{\utfit}}{\large ~Collaboration :} \\ 
\end{center}
\begin{center}
{\bf \large M.~Bona$^{(a)}$, M.~Ciuchini$^{(b)}$, E.~Franco$^{(c)}$,
V.~Lubicz$^{(b)}$, } \\
{\bf \large  G. Martinelli$^{(c)}$, F. Parodi$^{(d)}$, M. Pierini$^{(e)}$, P.
Roudeau$^{(e)}$, }\\
{\bf \large C. Schiavi$^{(d)}$, L.~Silvestrini$^{(c)}$, and A. Stocchi$^{(e)}$}
\end{center}

\vspace*{0.3cm}
\begin{center}
\noindent
{\small
\noindent
{\bf $^{(a)}$ Dip. di Fisica, Universit{\`a} di Torino
 and INFN,  Sezione di Torino}\\  
\hspace*{0.5cm}{Via P. Giuria 1, I-10125  Torino, Italy}\\
{\bf $^{(b)}$   Dip. di Fisica, Universit{\`a} di Roma Tre
 and INFN,  Sezione di Roma III}\\  
\hspace*{0.5cm}{Via della Vasca Navale 84, I-00146 Roma, Italy}\\
\noindent
{\bf $^{(c)}$ Dip. di Fisica, Universit\`a di Roma ``La Sapienza'' and INFN, Sezione di Roma}\\
\hspace*{0.5cm}{Piazzale A. Moro 2, 00185 Roma, Italy}\\
\noindent
{\bf $^{(d)}$ Dip. di Fisica, Universit\`a di Genova and INFN, Sezione di Genova}\\
\hspace*{0.5cm}{Via Dodecaneso 33, 16146 Genova, Italy}\\
\noindent
{\bf $^{(e)}$ Laboratoire de l'Acc\'el\'erateur Lin\'eaire}\\
\hspace*{0.5cm}{IN2P3-CNRS et Universit\'e de Paris-Sud, BP 34, 
F-91898 Orsay Cedex}}\\
\noindent
\end{center}

\vspace*{0.5cm}
\begin{abstract}
Using the latest determinations of several theoretical and experimental 
parameters, we update the Unitarity Triangle analysis in the Standard Model.
The basic experimental constraints come from the measurements of $\vubsvcb$,
$\dmd$, the lower limit on $\dms$, $\epsilonk$, and the measurement of the phase
of the $B_d$--$\overline{B}_d$ mixing amplitude through the time-dependent CP asymmetry
in $B^0\to J/\psi K^0$ decays.  
In addition, we consider the direct determination of 
$\alpha$, $\gamma$, $2\beta$+$\gamma$ and $\cos2\beta$ from 
the measurements of new CP-violating quantities, recently 
performed at the $B$ factories.
We also discuss the opportunities offered by improving the precision 
of the various physical quantities entering in the 
determination of the Unitarity Triangle parameters.
The results and the plots presented in this paper can also be found at the
URL {\tt http://www.utfit.org}, where they are continuously updated with the
newest experimental and theoretical results.
\end{abstract}

\newpage
\pagestyle{plain}

\section{Introduction}
\label{sec:intro}

The Standard Model (SM) of electroweak and strong interactions provides 
an excellent description of all observed phenomena in particle physics 
up to the energies presently explored.

The LEP/SLD era of precision electroweak physics has ended, leaving us a 
beautiful legacy of measurements that strongly constrain alternative 
mechanisms of electroweak symmetry breaking \cite{Barbieri:2004qk}, 
pointing strongly to a light Higgs boson. On the other hand, 
$B$ factories have thoroughly developed the study of precision 
$B$ physics, which, after the great achievement of the $A_{CP}(B^0
\to J/\psi K^0)$ measurement, is entering its mature age with many 
analyses aimed at measuring the angles of the Unitarity Triangle (UT) in 
different processes.

This remarkable experimental progress has been paralleled by many 
theoretical novelties: on the one hand, the consolidation of the
Heavy Quark Expansion and Lattice QCD (LQCD) results on heavy mesons;
on the other, the constant refinement of ideas and techniques to extract
information on the UT angles from $B$ decay rates and
CP asymmetries.

Waiting for LHC to uncover the mysteries connected to electroweak 
symmetry breaking and the hierarchy problem, we have the chance to 
indirectly probe scales up to a few TeV through a precision study of 
flavour physics. The aim of the {\utfit} Collaboration is twofold: i) the 
``classic" issue of assessing our present knowledge of flavour physics 
in the SM, combining all available experimental and theoretical 
information; ii) the future-oriented task of constraining new physics
models and their parameters on the ground of flavour and CP-violating
phenomena.

Up to now, the standard UT analysis~\cite{ref:noi, ref:noi2, ref:loro}
relies on the following measurements:
$\left | V_{ub} \right |/\left | V_{cb} \right|$,
$\Delta {m_d}$, the limit on $\Delta {m_s}$, and the measurements
of CP-violating quantities in the kaon ($\epsilonk$) and in the $B$
($\snb$) sectors.  Inputs to this analysis consist of a large body of
both experimental measurements and theoretically determined
parameters, where LQCD calculations play a central r\^ole.  A
careful choice (and a continuous update) of the values of these
parameters is a prerequisite in this study.  The values and errors
attributed to these parameters in the present study are summarized in
Table~\ref{tab:inputs} (Section~\ref{sec:inputs}).
The results of the analysis and the determination of the UT parameters
are presented and discussed in Section~\ref{sec:results} which is an
update of similar analyses performed in~\cite{ref:noi, ref:noi2} to
which the reader can refer for more details.

New CP-violating quantities have been recently measured by the
$B$ factories, allowing for the determination of several combinations of
UT angles. The measurements of $\alpha$ (using $\pi \pi$, $\rho \rho$ 
and $\pi \rho$ modes), $\gamma$ (using $D^{(*)}K^{(*)}$ modes),
$2\beta + \gamma$ (using $D^{(*)}\pi (\rho)$ modes), and $\cos 2\beta$ from
$B^0\to J/\psi K^*$ are now available.
These measurements and their effect on the UT fit are discussed in
Section~\ref{sec:newinputs}.

Finally in Section~\ref{sec:pull} we discuss the perspectives opened
by improving the precision in the measurements of various physical quantities
entering the UT analysis. In particular, we investigate to which extent future
and improved determinations of the experimental constraints, such as
$\sin 2\beta$, $\dms$, $\alpha$ and $\gamma$, could allow us to 
invalidate the SM, thus signalling the presence of new physics effects.

\section{Inputs used for the ``standard'' analysis}
\label{sec:inputs}

The values and errors of the relevant quantities used in this paper for the standard
analysis of the CKM parameters (corresponding to the constraints from
$\left | V_{ub} \right |/\left | V_{cb} \right |$, $\Delta {m_d}$,
$\Delta {m_s}/\Delta {m_d}$, $\epsilonk$ and $\snb$) are summarized in
Table~\ref{tab:inputs}.

The novelties here are the final LEP/SLD likelihood for $\Delta m_s$,
the use of $|V_{ub}|$ measurements from inclusive semileptonic decays
at the $B$ factories~\cite{ref:hfag}, the updated value of $\snb$ and a new
treatment of the non-perturbative QCD parameters as explained in the
following Section~\ref{sec:usefbs}.
In addition, we use updated values of the top mass $m_t^{\rm pole}=178.0\pm 4.3$
GeV~\cite{mtop} and of the CKM parameter $\lambda$. The latter comes from
the average of the following values~\cite{lambda}
\begin{eqnarray}
\lambda(V_{us}~{\rm from }~K_{l3})&=&0.2250\pm 0.0021 \nonumber \\
\lambda(V_{ud}+~{\rm unitarity})&=&0.2265\pm 0.0020.
\end{eqnarray}
Finally, we now calculate QCD corrections to $\Delta B=2$ and $\Delta S=2$ processes, fully taking into account their correlations with the input parameters and, by varying the matching scale, the uncertainty introduced by the residual scale dependence.

\begin{table*}[htbp!]
{\footnotesize
\begin{center}
\begin{tabular}{@{}llll}
\hline\hline
\\
         Parameter                          &  Value                            
     & Gaussian ($\sigma$)      &   Uniform             \\
                                            &                                   
     &                          & (half-width)          \\ \hline\hline
         $\lambda$                          &  0.2258                           
     &  0.0014                  &    -                  \\ \hline
$\left |V_{cb} \right |$(excl.)             & $ 41.4 \times 10^{-3}$            
     & $2.1 \times 10^{-3}$     & -                     \\
$\left |V_{cb} \right |$(incl.)             & $ 41.6 \times 10^{-3}$            
     & $0.7 \times 10^{-3}$     & $0.6 \times 10^{-3}$  \\ 
$\left |V_{ub} \right |$(excl.)             & $ 33.0  \times 10^{-4}$           
     & $2.4 \times 10^{-4}$     & $4.6 \times 10^{-4}$  \\  
$\left |V_{ub} \right |$(incl.)        & $ 47.0  \times 10^{-4}$           
     & $4.4 \times 10^{-4}$     &        -              \\ \hline
$\Delta m_d$                                & $0.502~\mbox{ps}^{-1}$            
     & $0.006~\mbox{ps}^{-1}$   &        -              \\
$\Delta m_s$                                & $>$ 14.5 ps$^{-1}$ at 95\% C.L.   
     & \multicolumn{2}{c}{sensitivity 18.3 ps$^{-1}$}   \\ \hline
$\fbssqbs$                                  & $276$ MeV                        
     & $38$ MeV                 &          -            \\
$\xi=\frac{\fbssqbs}{\fbdsqbd}$             & 1.24                              
     & 0.04                     &  0.06            \\\hline
$\hat B_K$                                  & 0.86                              
     & 0.06                     &     0.14              \\
$\epsilonk$                                 & $2.280 \times 10^{-3}$            
     & $0.013 \times 10^{-3}$   &          -            \\
$f_K$                                       & 0.159 GeV                         
     & \multicolumn{2}{c}{fixed}                        \\
$\Delta m_K$                                & 0.5301 $\times 10^{-2}
~\mbox{ps}^{-1}$ & \multicolumn{2}{c}{fixed}                        \\ \hline
$\snb$                                      &  0.726              
     &  0.037                   &          -            \\ \hline
$\overline m_t$                                       & $168.5$ GeV
     & $4.1$ GeV          &          -            \\
$\overline m_b$                                       & 4.21 GeV
     & 0.08 GeV           &          -            \\
$\overline m_c$                                       & 1.3 GeV
     & 0.1 GeV            &          -            \\
$\alpha_s(M_Z)$                                  & 0.119                             
     & 0.003                    &          -            \\
$G_F $                                      & 1.16639 $\times 10^{-5} \GeV^{-2}$
     & \multicolumn{2}{c}{fixed}                        \\
$ m_{W}$                                    & 80.425 GeV
     & \multicolumn{2}{c}{fixed}                        \\
$ m_{B^0_d}$                                & 5.279 GeV
     & \multicolumn{2}{c}{fixed}                        \\
$ m_{B^0_s}$                                & 5.375 GeV
     & \multicolumn{2}{c}{fixed}                        \\
$ m_K^0$                                   & 0.497648 GeV
     & \multicolumn{2}{c}{fixed}                        \\ \hline\hline
\end{tabular} 
\end{center}
}
\caption {\it {Values of the relevant quantities used in the UT fit.
The Gaussian and the flat contributions to the
uncertainty are given in the third and fourth columns 
respectively (for details on the statistical treatment see~\cite{ref:noi}). 
}}
\label{tab:inputs} 
\end{table*}

\subsection{Use of $\xi$, $f_{B_s}\sqrt{\hat B_{B_s}}$ and $f_{B_d}\sqrt{ \hat
B_{B_d}}$ in 
$\Delta m_s$ and $\Delta m_d$ constraints}
\label{sec:usefbs}

One of the important differences with respect to previous studies is
in the use of the information from non-perturbative QCD parameters
entering the expressions of $\Delta m_s$ and $\Delta m_d$.
The $B_s$--${\overline B}_s$ mass difference is
proportional to the square of the element $|V_{ts}|$. Up to
Cabibbo-suppressed corrections, $|V_{ts}|$ is independent of $\rhobar$ and
$\etabar$. As a consequence, the measurement of $\Delta m_s$ would
provide a strong constraint
on the non-perturbative QCD parameter $f_{B_s}^2 \hat B_{B_s}$.

We propose a new and more appropriate way of treating
the constraints coming from the measurements of $\Delta m_s$ and
$\Delta m_d$. In previous analyses, these constraints were implemented
using the following equations
\begin{eqnarray}
\label{eq:old}
 \Delta m_d &\propto& [(1-\rhobar)^2+\etabar^2]  f_{B_d}^2 \hat B_{B_d}         
          \\ \nonumber
 \Delta m_s &\propto& f_{B_s}^2 \hat B_{B_s}  =  f_{B_d}^2 \hat B_{B_d} \times
\xi^2 
\end{eqnarray}
where $\xi=f_{B_s}\sqrt{\hat B_{B_s}}/f_{B_d}\sqrt{ \hat B_{B_d}}$. In
this case, the input quantities are $f_{B_d}\sqrt{ \hat B_{B_d}}$ and
$\xi$.  The constraints from $\Delta m_s$ and the knowledge of $\xi$
are used to improve the knowledge on $f_{B_d} \sqrt{\hat B_{B_d}}$
which thus makes the constraint on $\Delta m_d$ more effective.

The hadronic parameter that is better determined from lattice calculations,
however, is $f_{B_s}^2 \hat B_{B_s}$, whereas $\xi$ and $f_{B_d}^2 \hat B_{B_d}$
are affected by larger uncertainties coming from the chiral extrapolations.
These uncertainties are strongly correlated. For this reason, a better approach
consists in replacing Eq.~(\ref{eq:old}) with
\begin{eqnarray}
 \Delta m_d &\propto& [(1-\rhobar)^2+\etabar^2] \frac{f_{B_s}^2 \hat
B_{B_s}}{\xi^2} \\ \nonumber
 \Delta m_s &\propto&  f_{B_s}^2 \hat B_{B_s}.
\label{eq:new}
\end{eqnarray}

At present, this new parameterization does not have a large effect on the final 
results. It allows, however, to take into account more accurately the uncertainty from 
the chiral extrapolation in lattice calculations of $f_{B_d}$. Note that in
this way, in order to obtain a more effective constraint on $\Delta m_d$, the
error on $\xi$ should also be improved.

\section{Determination of the Unitarity Triangle parameters: the standard analysis}
\label{sec:results}

\begin{figure}[htbp!]
\begin{center}
{\includegraphics[height=7.cm]{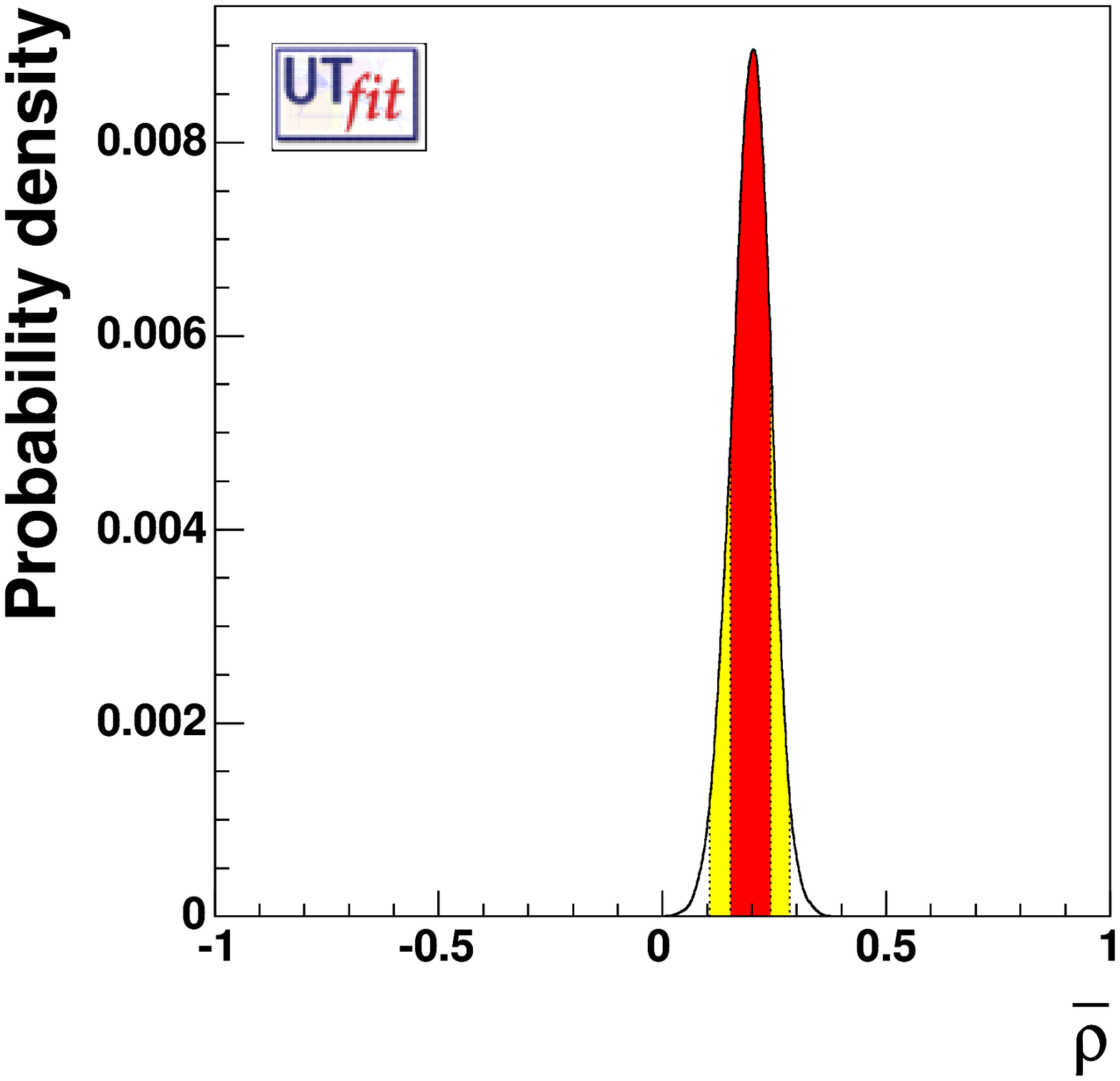}}
{\includegraphics[height=7.cm]{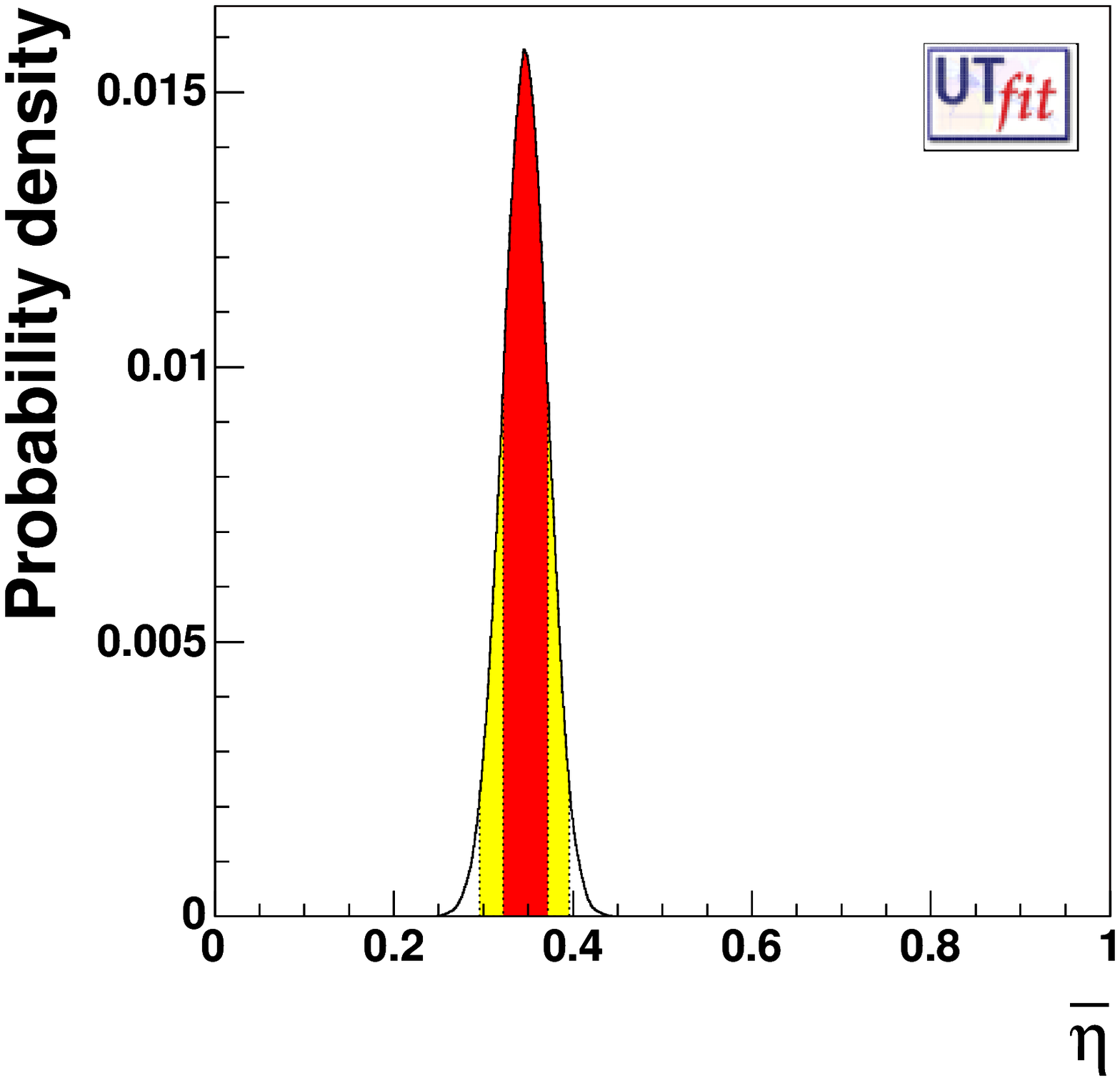}}\\
{\includegraphics[height=7.cm]{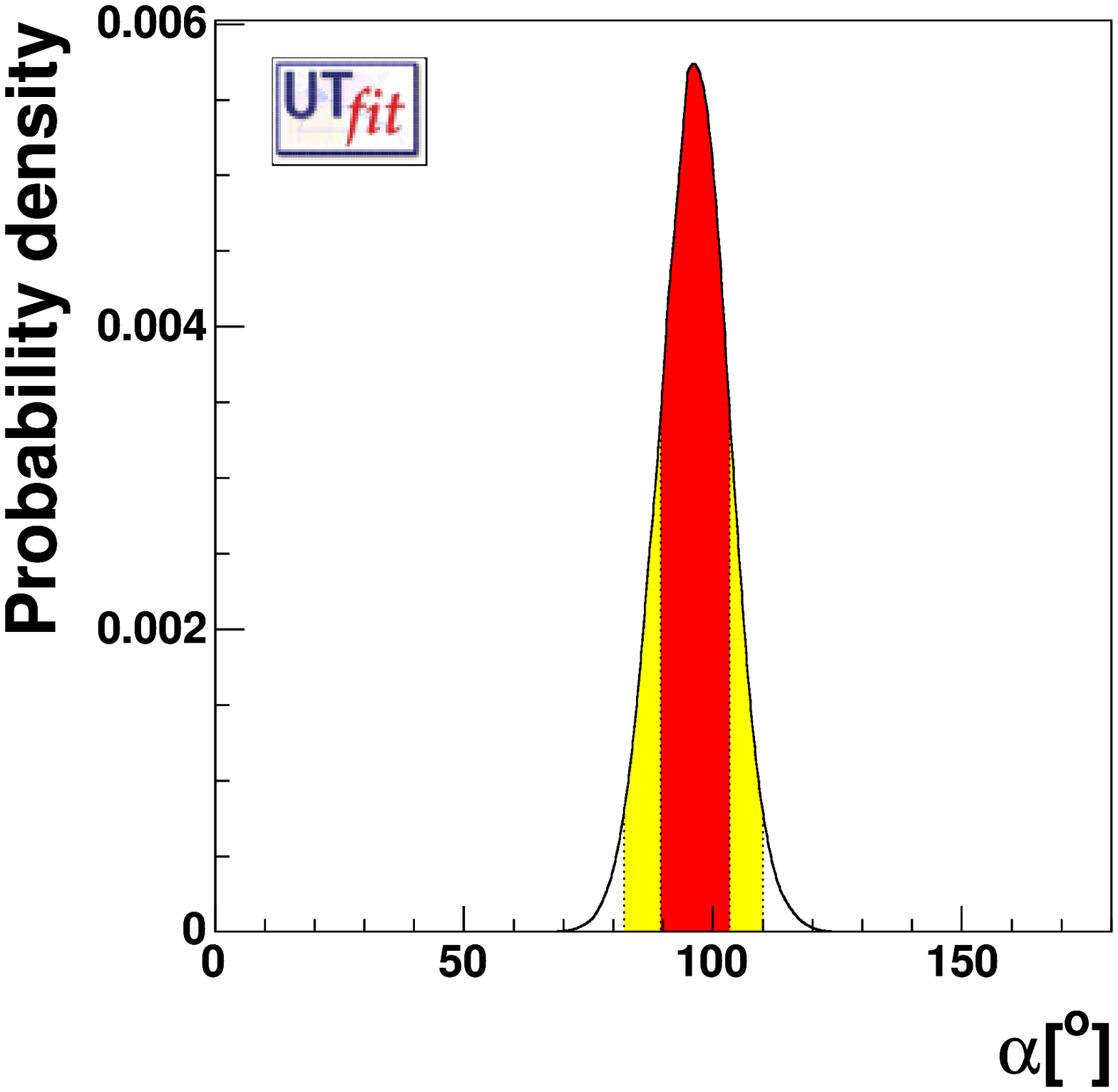}}
{\includegraphics[height=7.cm]{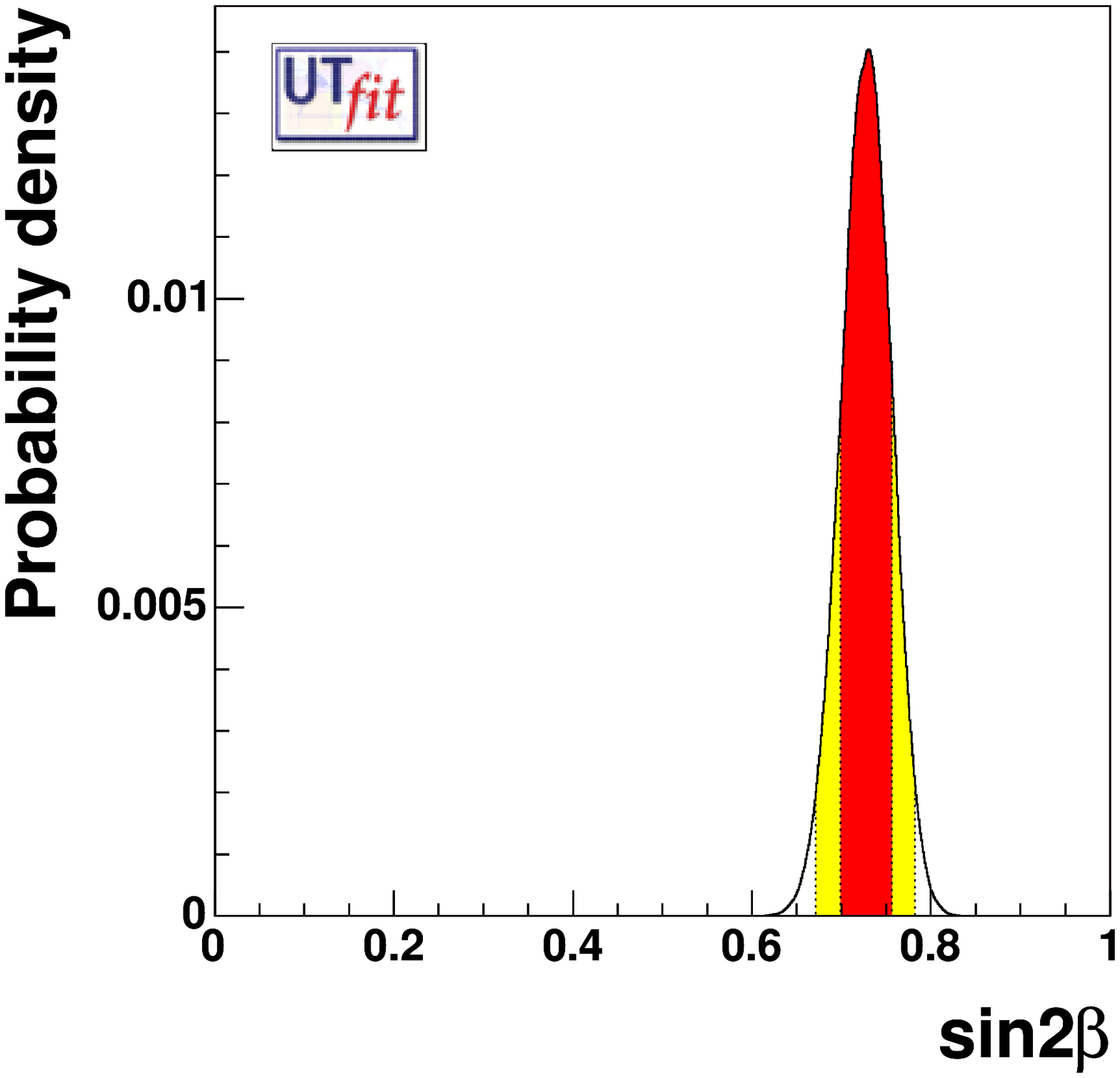}} \\
{\includegraphics[height=7.cm]{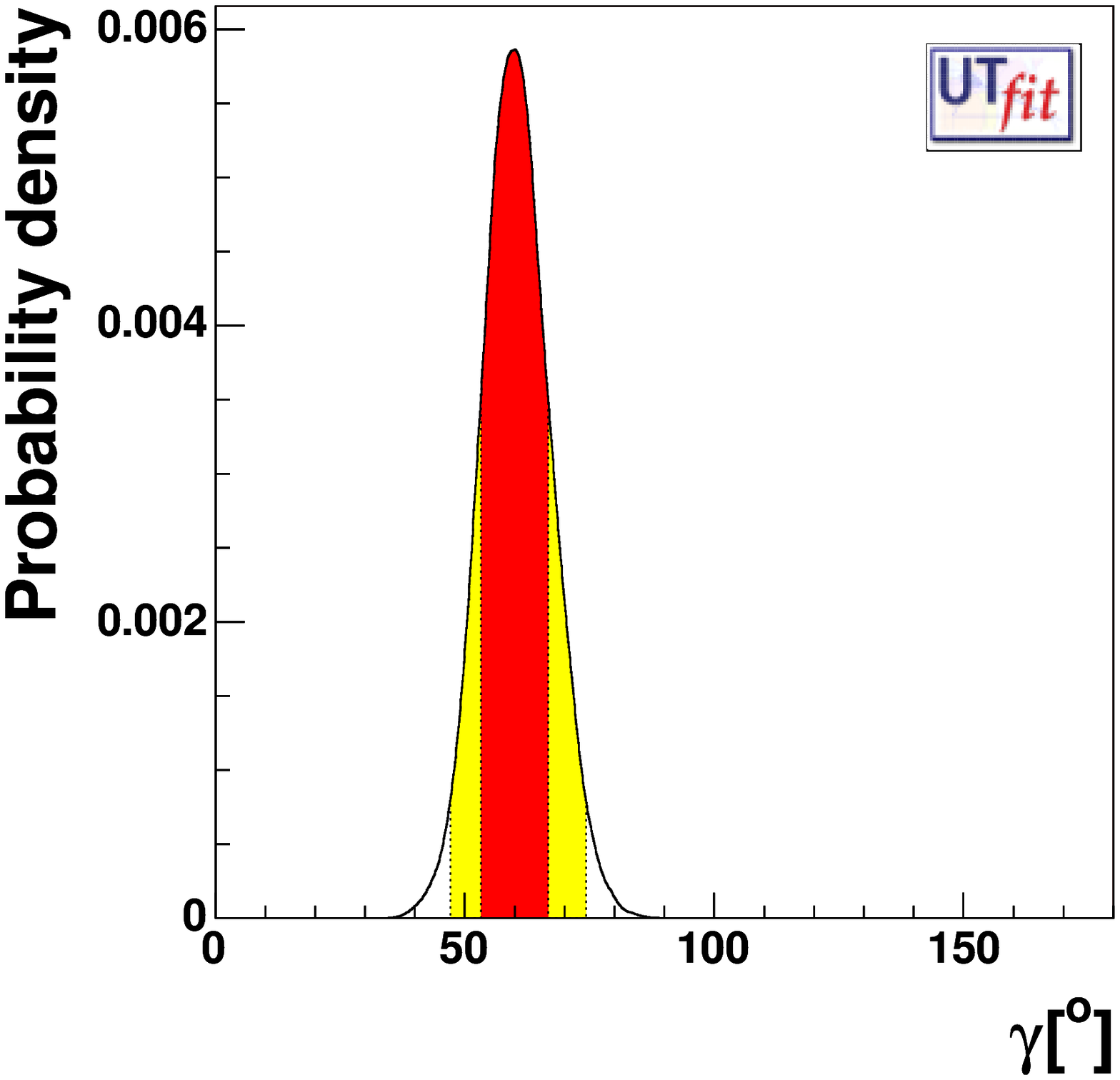}}
{\includegraphics[height=7.cm]{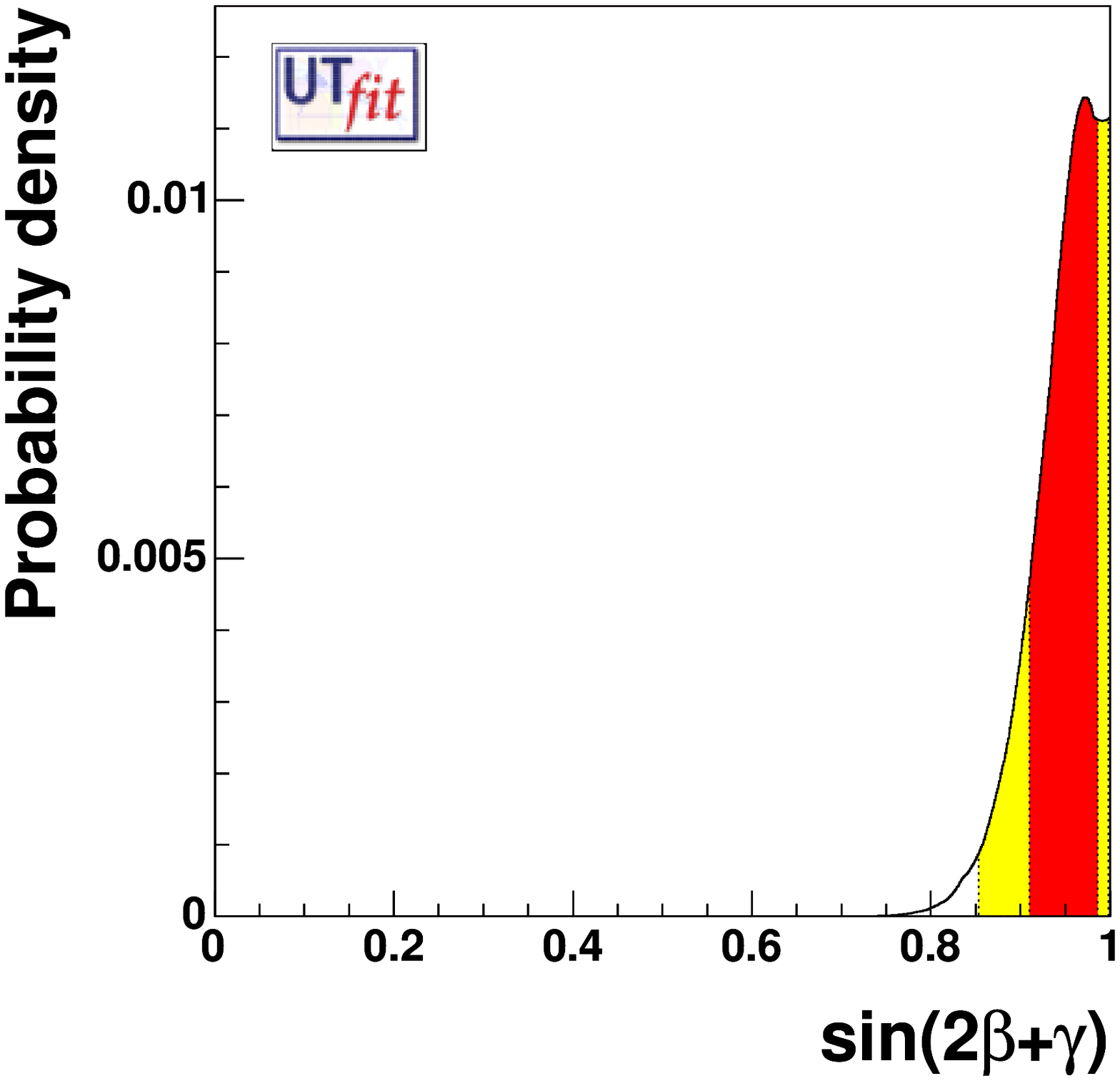}}
\caption{\it {From top to bottom and from left to right, the a-posteriori
    p.d.f.'s for $\rhobar$, $\etabar$, $\alpha$, $\snb$, $\gamma$ and
    $\sin(2\beta+\gamma)$. The red (darker) and the
    yellow (lighter) zones correspond respectively to 68\% and 95\% of
    the area. The following constraints have been used in the UT fit:
    $\left | V_{ub} /| V_{cb} \right |$, $\Delta m_d$,
    $\Delta m_s$, $\epsilonk$, and $\snb$ from the measurement of the
    CP asymmetry in the $J/\psi K^0$ decays.}}
\label{fig:1dim}
\end{center}
\end{figure}

\begin{figure}[htb!]
\begin{center}
\includegraphics[width=16cm]{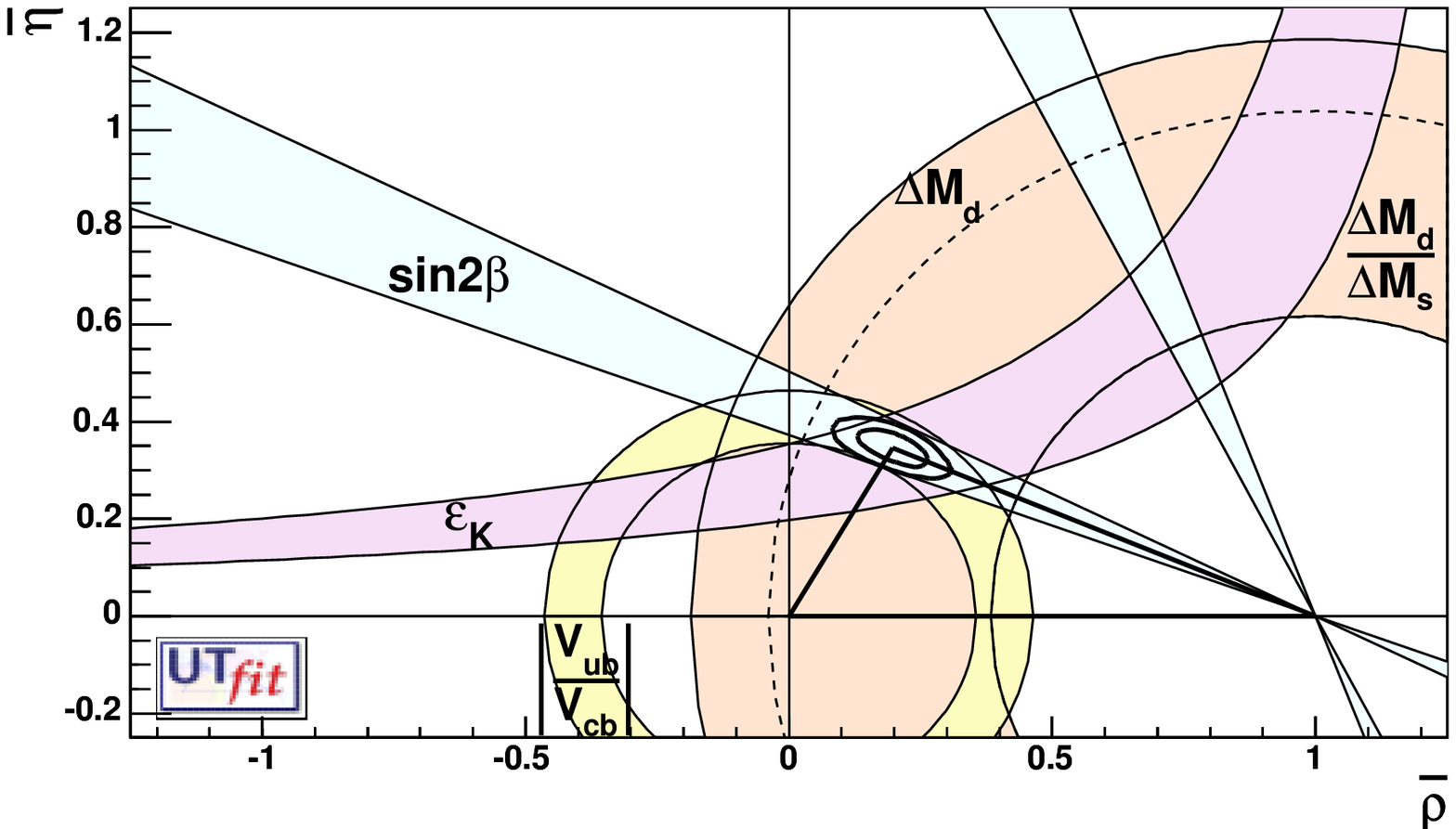}
\caption{ \it {Allowed regions for $\rhobar$ and $\etabar$ using the parameters
    listed in Table~\ref{tab:inputs}.  The closed contours at 68\% and
    95\% probability are shown. The full lines correspond to 95\%
    probability regions for each of the constraints, given respectively
    by the measurements of $\left | V_{ub} \right |/\left | V_{cb} \right |$,
    $\Delta m_d$, $\Delta m_s$, $\epsilonk$, and $\snb$ from the measurement
    of the CP asymmetry in the $J/\psi K^0$ decays.}}
\label{fig:rhoeta}
\end{center}
\end{figure}

Using the constraints from $\left | V_{ub} \right |/\left | V_{cb}
\right |$, $\Delta {m_d}$, $\Delta {m_s}$, $\epsilonk$
and $\snb$, we obtain the results given in Table~\ref{tab:1dim}.
The central value for each p.d.f. is calculated using the median 
and the error corresponds to $34\%$ probability regions
on each side of the median. Asymmetric errors are symmetrized 
changing the quoted central value.~\footnote{
For the p.d.f. with multiple solutions, e.g. for the UT angles in
Section \ref{sec:newinputs}, we consider the range corresponding to
$68\%$ probability and delimited by the intercept of the distribution
with an horizontal line.}
Figures~\ref{fig:1dim} and \ref{fig:rhoeta} show, respectively, the
probability density functions (p.d.f.'s) for some UT parameters and
the selected region in the $\rhobar-\etabar$ plane.

\begin{table*}[h]
\begin{center}
\begin{tabular}{@{}llllll}
\hline\hline
         Parameter               &   ~~~~~68$\%$        & ~~~~~95$\%$     &~~~~~
99$\%$    \\ \hline
~~~~~$\overline {\rho}$          & 0.196  $\pm$ 0.045   & [0.104,\,0.283]   & [0.073,\,0.314] \\
~~~~~$\overline {\eta}$          & 0.347  $\pm$ 0.025   & [0.296,\,0.396]   & [0.281,\,0.412] \\\hline
~~~~$\alpha [^{\circ}]$          & 96.1   $\pm$ 7.0     & [82.1,\,110.0]    & [77.7,\,114.8]  \\
~~~~$\beta [^{\circ}]$          & 23.4  $\pm$ 1.5     &  [20.8,\,26.1]   & [20.2,\,27.1]       \\
~~~~$\gamma[^{\circ}$]           & 60.3   $\pm$ 6.8     & [47.0,\,74.2]     & [42.5,\,78.9]   \\\hline
    ~$\sin 2\alpha$       & -0.21  $\pm$ 0.24     &  [-0.65,\,0.27]   & [-0.77,\,0.41]       \\
    ~$\snb$          & 0.726  $\pm$ 0.028   & [0.670,\,0.780]   & [0.651,\,0.797] \\
    ~$\sin(2\beta+\gamma)$    & 0.947  $\pm$ 0.038     &  [0.852,\,0.996]   & [0.813,\,0.998]       \\\hline
$\rm{Im} {\lambda}_t$[$10^{-5}$] & 13.3   $\pm$ 0.9     & [11.5,\,15.1]     & [10.9,\,15.6]   \\
\hline\hline
\end{tabular}
\end{center}
\caption {\it {Values and probability ranges for the UT parameters obtained 
from the UT fit using the following constraints: $\left | V_{ub} \right |/\left | V_{cb} \right |$,
$\Delta {m_d}$, $\Delta {m_s}$, $\epsilonk$ and $\snb$. The value of 
$\rm{Im} {\lambda}_t = \rm{Im} {V^*_{ts} V_{td}}$ is also given. }}
\label{tab:1dim}
\end{table*}

\subsection{Fundamental test of the Standard Model in the quark sector}
\label{sec:lati}

\begin{figure}[htb!]
\begin{center}
\includegraphics[width=16cm]{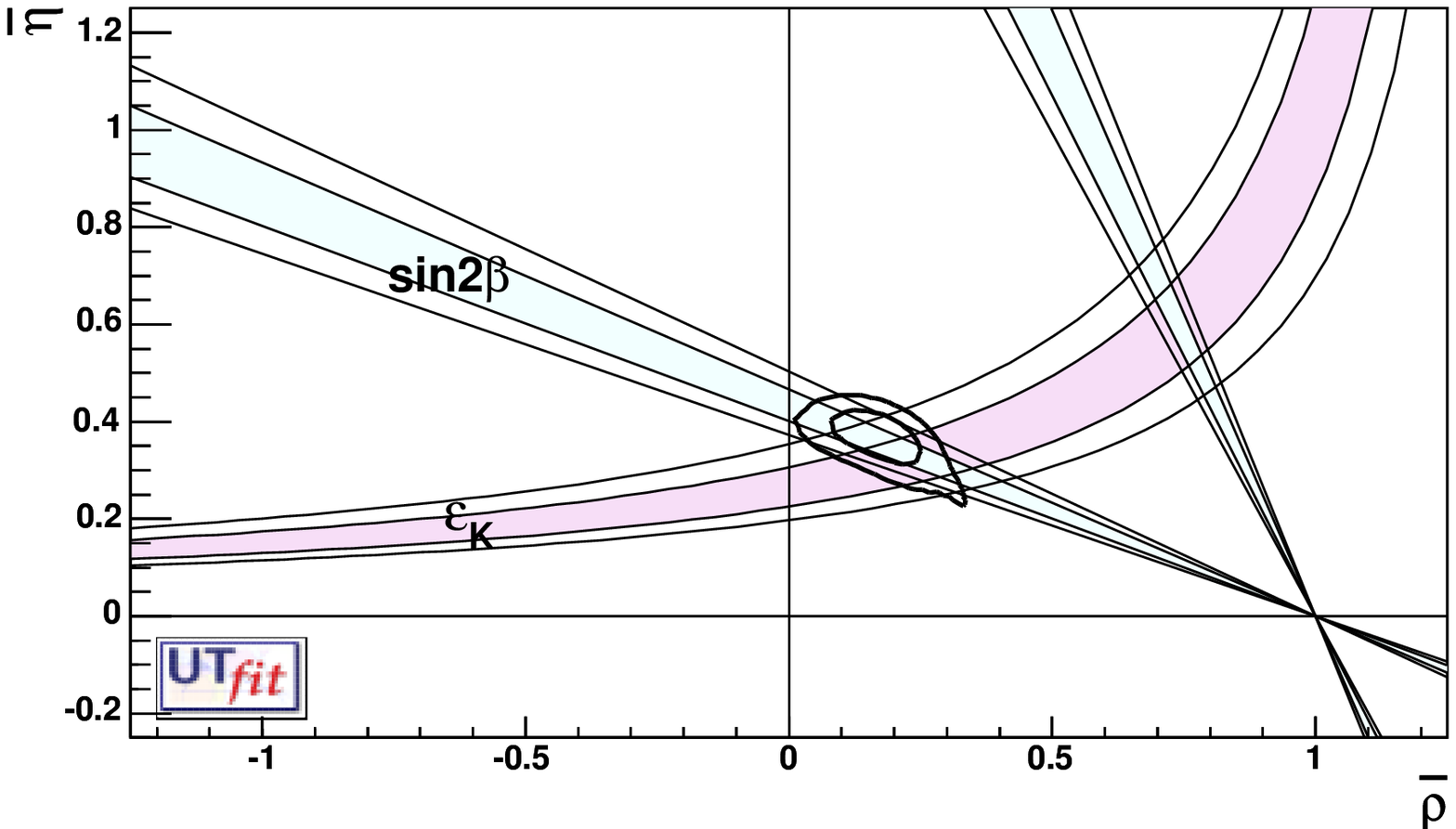}
\caption{\it {The allowed regions for $\overline{\rho}$ and $\overline{\eta}$
    (close contours at 68\%, 95\% probability ranges), as selected by the
    measurements of $\left | V_{ub} \right |/\left | V_{cb} \right |$,
    $\Delta {m_d}$, and by the limit on $\Delta {m_s}$,
    are compared with the bands (at 68\% and 95\% probability ranges)
    from the measurements of CP-violating quantities in the kaon
    ($\epsilonk$) and in the B ($\snb$) sectors.}}
\label{fig:testcp}
\end{center}
\end{figure}

The standard fit, illustrated in Figure~\ref{fig:rhoeta}, gives a clear picture of the
success of the SM. A crucial test consists in establishing CP violation by using the
sides of the UT, {\it i.e.} CP-conserving processes such as the semileptonic $B$ decays and $B_{d,s}-{\overline  B}_{d,s}$ 
oscillations. The comparison of the region selected by these constraints
and the one selected by the direct measurements of CP violation in the kaon ($\epsilonk$) or in the $B$ ($\snb$) sectors is shown in Figure~\ref{fig:testcp} and gives a picture of the success of the SM in the flavour sector.
This pictorial agreement is quantified through the comparison between the value of $\rhobar$ and $\etabar$ computed from
CP-conserving and CP-violating observables:~\footnote{The second allowed region for $\rhobar$ and $\etabar$ from the CP-violating
measurements (see Figure~\ref{fig:testcp}) has been discarded
since it is ruled out by the measurement of $\cos 2\beta$, see
Section~\ref{sec:c2b}.}
\begin{eqnarray}
&&\left.\begin{array}{l}
\rhobar=0.169\pm 0.057 ([0.055,\,0.310]~{\rm at} ~95\%)\\
\etabar=0.364\pm 0.037 ([0.252,\,0.430]~{\rm at} ~95\%)
\end{array}\right\} ~~\rm {UT~sides~only}\nonumber\\
&&\left.\begin{array}{l}
\rhobar=0.241\pm 0.070 ([0.098,\,0.363]~{\rm at} ~95\%)\\
\etabar=0.311\pm 0.030 ([0.260,\,0.371]~{\rm at} ~95\%)
\end{array}\right\} ~~\rm {S(J/\psi K^0)+\epsilonk}
\end{eqnarray}

\begin{figure}[htb!]
\begin{center}
\includegraphics[height=12cm]{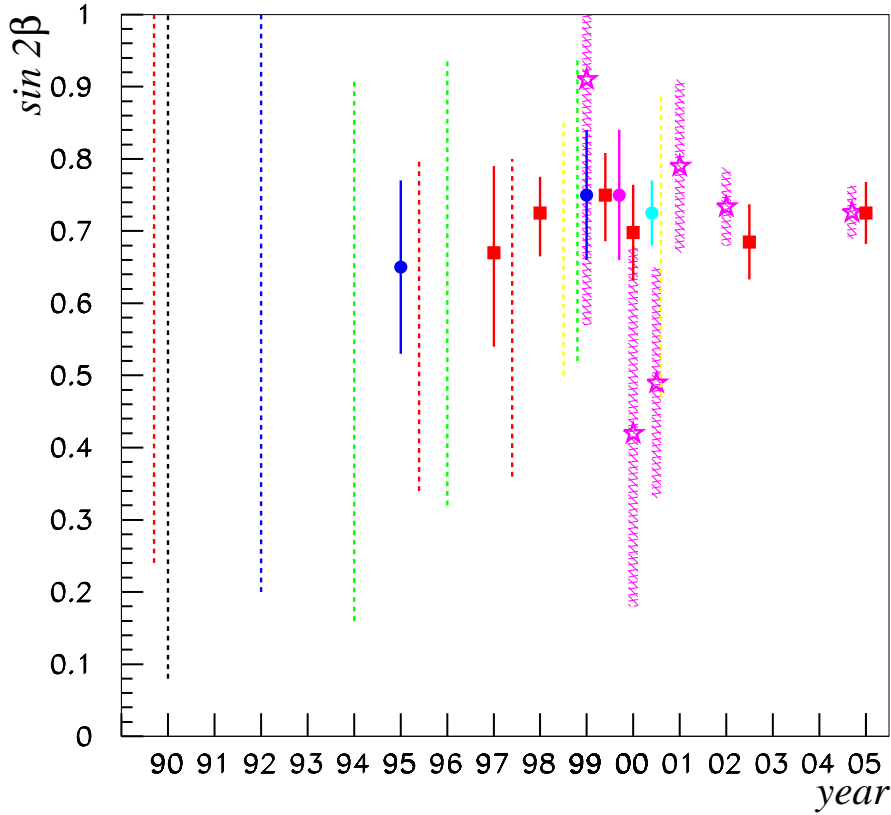}
\caption{\it {Evolution of the ``indirect'' determination of $\snb$ over the
    years (until 2002).  From left to right, they correspond to the
    following papers~\cite{ref:loro,ref:allrhoeta}: dRLMPPPPS90, DDGN90, LMMR92, AL94,
    CFMRS95, BBL95, AL96, PPRS97, BF97, BPS98, PS98, AL99, CFGLM99,
    CPRS99, M99, CDFLMPRS00, B.et.al.00, HLLL00 and
    CFLPSS~\cite{ref:noi, ref:noi2}. The dotted lines correspond to the 95$\%$
    C.L. regions (the only information given in those papers). The
    larger hatched bands (from year '99) correspond to values of $\snb$ from
    direct measurements ($\pm 1\sigma$). The most recent values contained in this paper are 
    also shown.}}
\label{fig:storiasin2beta}
\end{center}
\end{figure}

Another test can be performed by comparing the value of $\snb$ from $S(J/\psi K^0)$ and the one determined from ``sides" measurements
\begin{eqnarray}
\snb = & 0.734 \pm 0.043 ~([0.616, 0.811]~{\rm at} ~95\%)  & ~~\rm {UT~sides~only}      \nonumber \\
\snb = & 0.726 \pm 0.037 ~([0.652, 0.800]~{\rm at} ~95\%) & ~~S(J/\psi K^0)
\label{eq:sin2beta}
\end{eqnarray}
For completeness, we also give the value of $\snb$ obtained by using 
all the constraints but the direct determination:
\begin{equation}
\snb = 0.725 \pm 0.043 ~([0.634, 0.804]~{\rm at} ~95\%) ~~\rm {UT~sides}+\epsilon_K.
\end{equation}
As a matter of fact, the value of $\snb$ was predicted, before its
first direct\footnote{In the following, for
simplicity, we will denote as ``direct'' (``indirect'') the
determination of any given quantity from a direct measurement (from
the UT fit without using the measurement under consideration).}
measurement was obtained, by using all other available
constraints ($\left | V_{ub} \right |/\left | V_{cb} \right |$,
$\Delta m_d$, $\Delta m_s$, and $\epsilonk$). The ``indirect''
determination has improved regularly over the years, as shown in
Figure~\ref{fig:storiasin2beta}, where the direct measurement is 
also reported, as a reference.

The agreement of these determinations confirms the validity of CKM mechanism in the SM. This test relies on several non-perturbative 
techniques, such as the Operator Product Expansion for computing $B$ decay rates, the Heavy Quark Effective Theory and LQCD, 
which are used to extract the CKM parameters from the experimental 
measurements. The overall consistency of the UT fit gives confidence on the theoretical tools. Assuming the validity of the SM, it is possible to perform a more quantitative test of the non-perturbative techniques as discussed in following.

\subsection{Determination of other quantities}

In the previous sections we have obtained
the {\it a-posteriori} p.d.f.'s for all the UT parameters. It is also instructive to
remove from the fitting procedure the external information on the
value of one (or more) of the constraints.

In this section we study the distributions of $\dms$ and of the
hadronic parameters.
In the first case we do not include in the analysis the experimental
information on $B_s$--${\overline B}_s$ mixing coming from LEP and
SLD. In the case of the hadronic parameters, we remove from the fit the
constraints on their values coming from lattice calculations, and use them as
free parameters of the fit. In this way we can compare the
uncertainty obtained on a given quantity through the UT fit to the
present theoretical error on the same quantity.

\subsubsection{The expected distribution for \boldmath$\dms$}

\begin{figure}[htb!]
\begin{center}
\includegraphics[height=8cm]{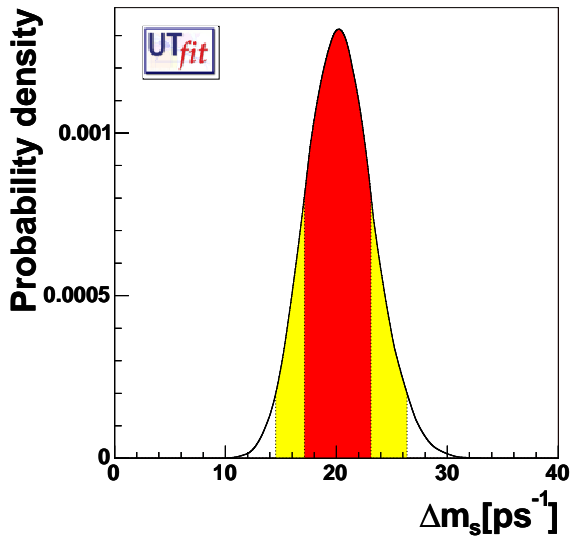} 
\end{center}
\caption{ \it {
    $\dms$ probability distribution, obtained without using the 
    experimental information
    from $B_s$--${\overline B}_s$ mixing.}}
\label{fig:dms}
\end{figure}

The p.d.f.~for $\dms$ obtained by removing the experimental information
coming from $B_s$--${\overline B}_s$ mixing is shown in Figure~\ref{fig:dms}.
The result of this exercise is given in Table~\ref{tab:dmsresults} (upper line).
The present experimental analyses of $B_s$--${\overline B}_s$ mixing 
at LEP and SLD have established a sensitivity 
of 18.3~ps$^{-1}$ and they show an evidence at about 2$\sigma$ for a positive 
signal at around 17.5 ps$^{-1}$, well compatible with the range of the $\dms$ 
distribution from the UT fit (Figure~\ref{fig:dms}). The inclusion of this information 
in the UT analysis has a large impact on the determination of $\dms$, as shown by the comparison between 
the permitted range obtained when this information is either 
included or not (see Table~\ref{tab:dmsresults}).
Accurate measurements of $\dms$, expected from the TeVatron in the near future, will provide an ingredient of the utmost importance for testing the SM.

\begin{table*}[h]
\begin{center}
\begin{tabular}{@{}lllll}
\hline\hline
      ~~~~~~~~~~~~~~Parameter                       & ~~~68$\%$       & ~~~~95$\%$    & ~~~~99$\%$    \\ \hline
~~$\dms$(\rm{without~$\dms$}) [{\rm ps}$^{-1}$]     &  21.2 $\pm$ 3.2 &  [15.4,\,27.8]  &  [13.8,\,30.0]  \\
~~$\dms$(\rm{including ~$\dms$}) [{\rm ps}$^{-1}$]  &  18.5 $\pm$ 1.6 &  [15.6,\,23.1]  &  [15.1,\,27.3]  \\
\hline\hline
\end{tabular} 
\end{center}
\caption {\it Central values and ranges for $\dms$ corresponding to defined
levels of probability, obtained by including (or not) the experimental 
information on $\dms$.}
\label{tab:dmsresults} 
\end{table*}

\subsubsection{Determination of \boldmath$\fbssqbs$, $\hat{B}_K$ and $\xi$}
\label{sec:2dpdfs}

\begin{figure}[htbp!]
\begin{center}
{\includegraphics[height=5.3cm]{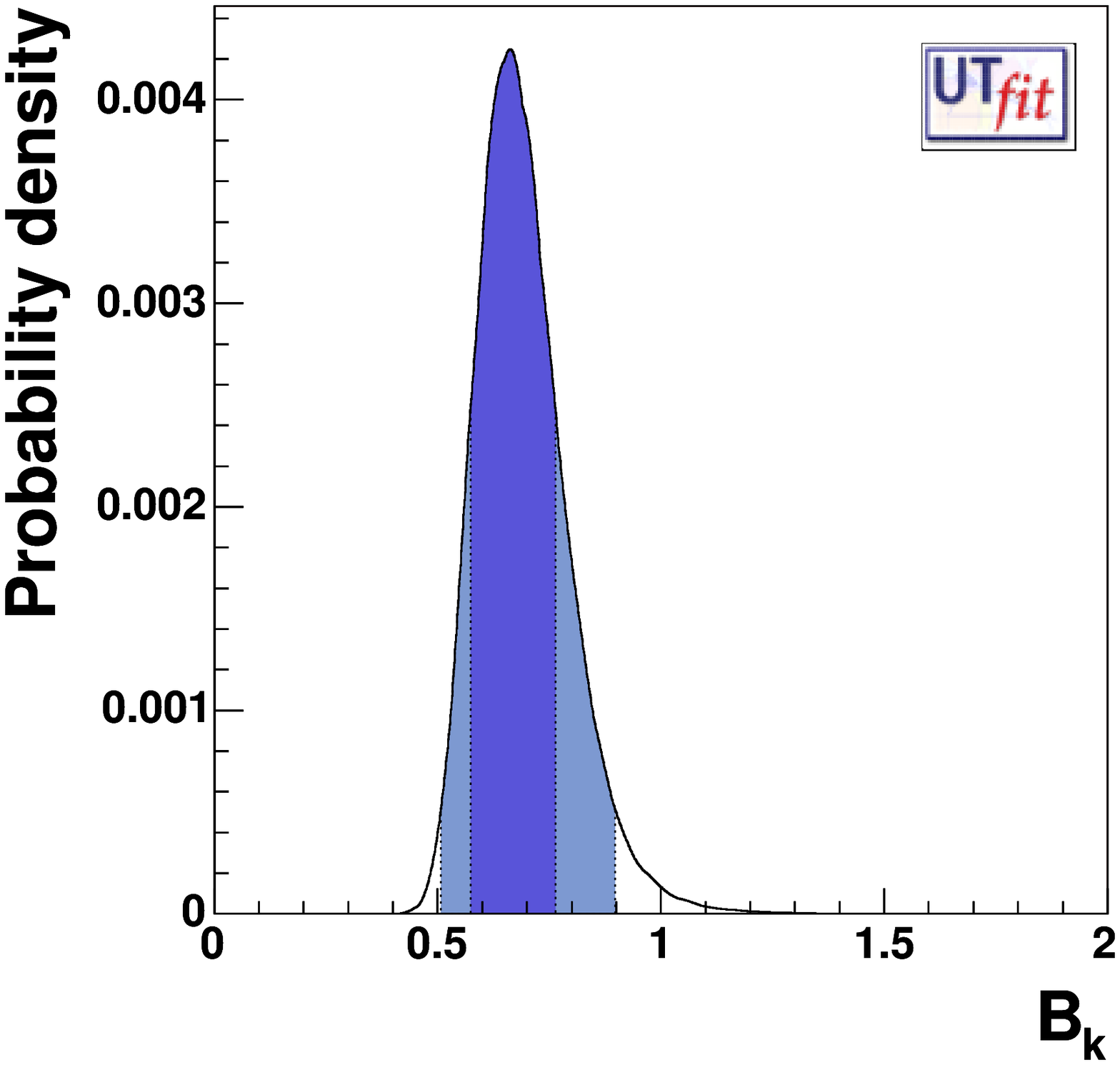}}
{\includegraphics[height=5.3cm]{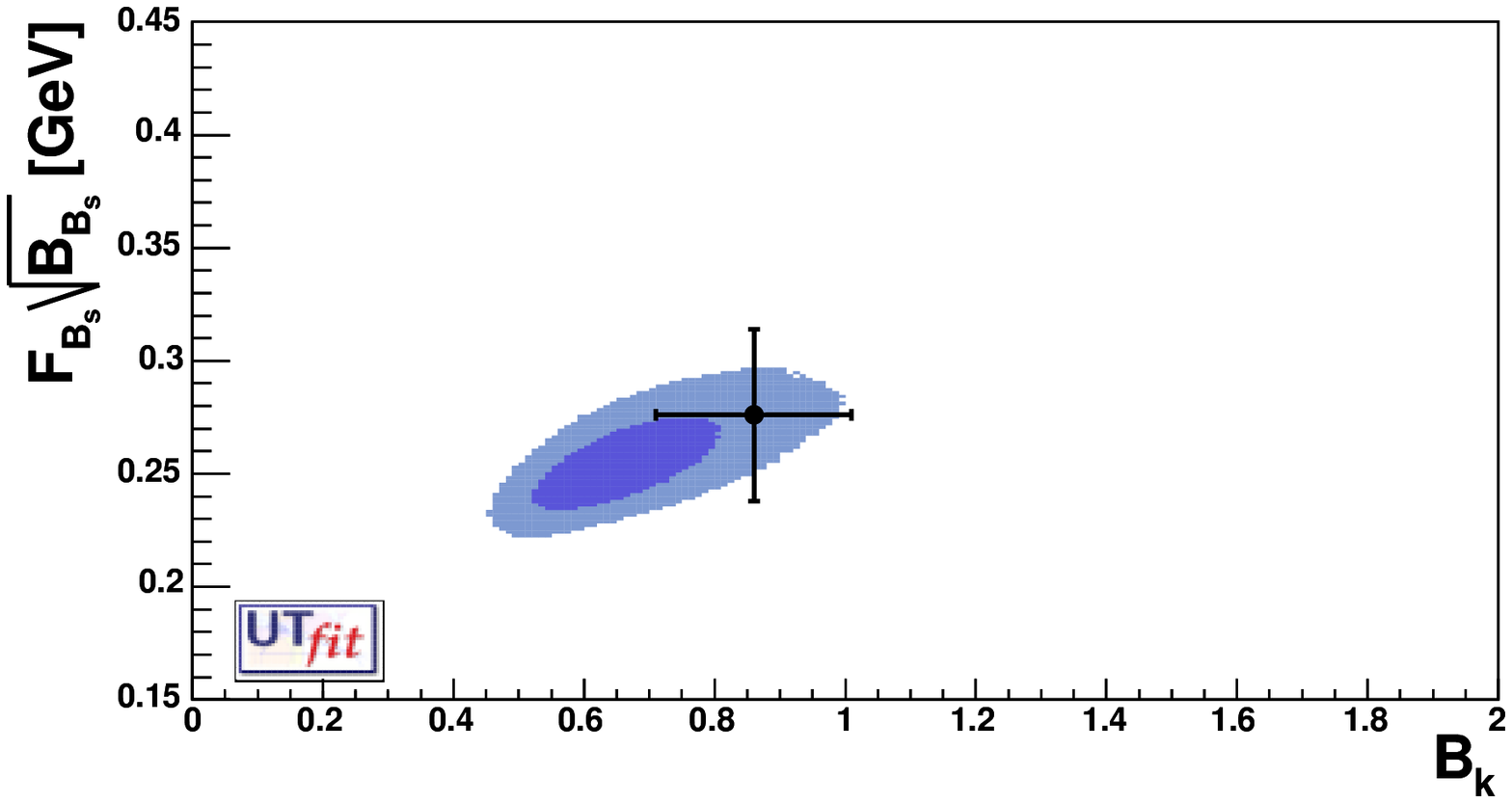}}  \\
{\includegraphics[height=5.3cm]{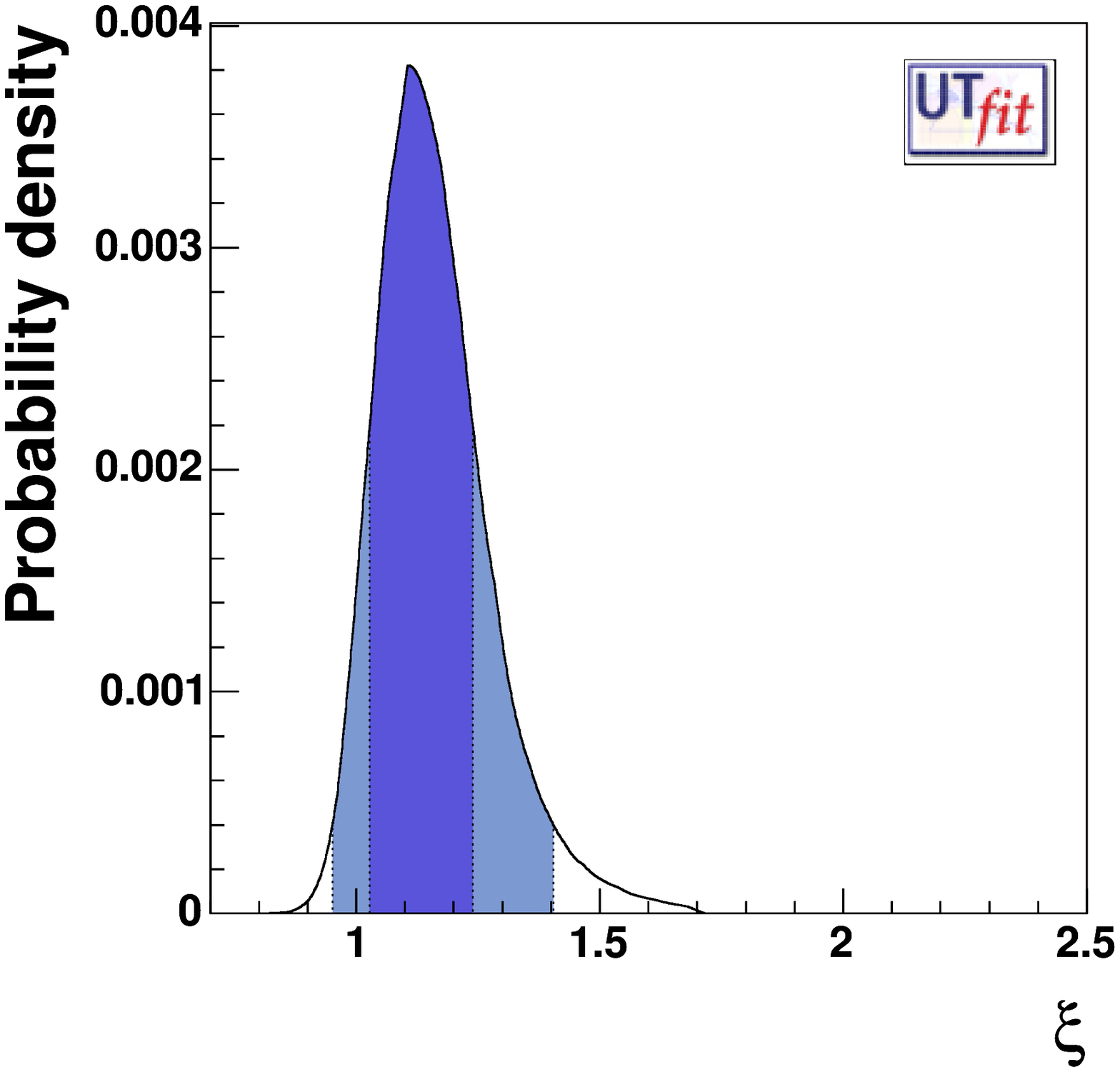}} 
{\includegraphics[height=5.3cm]{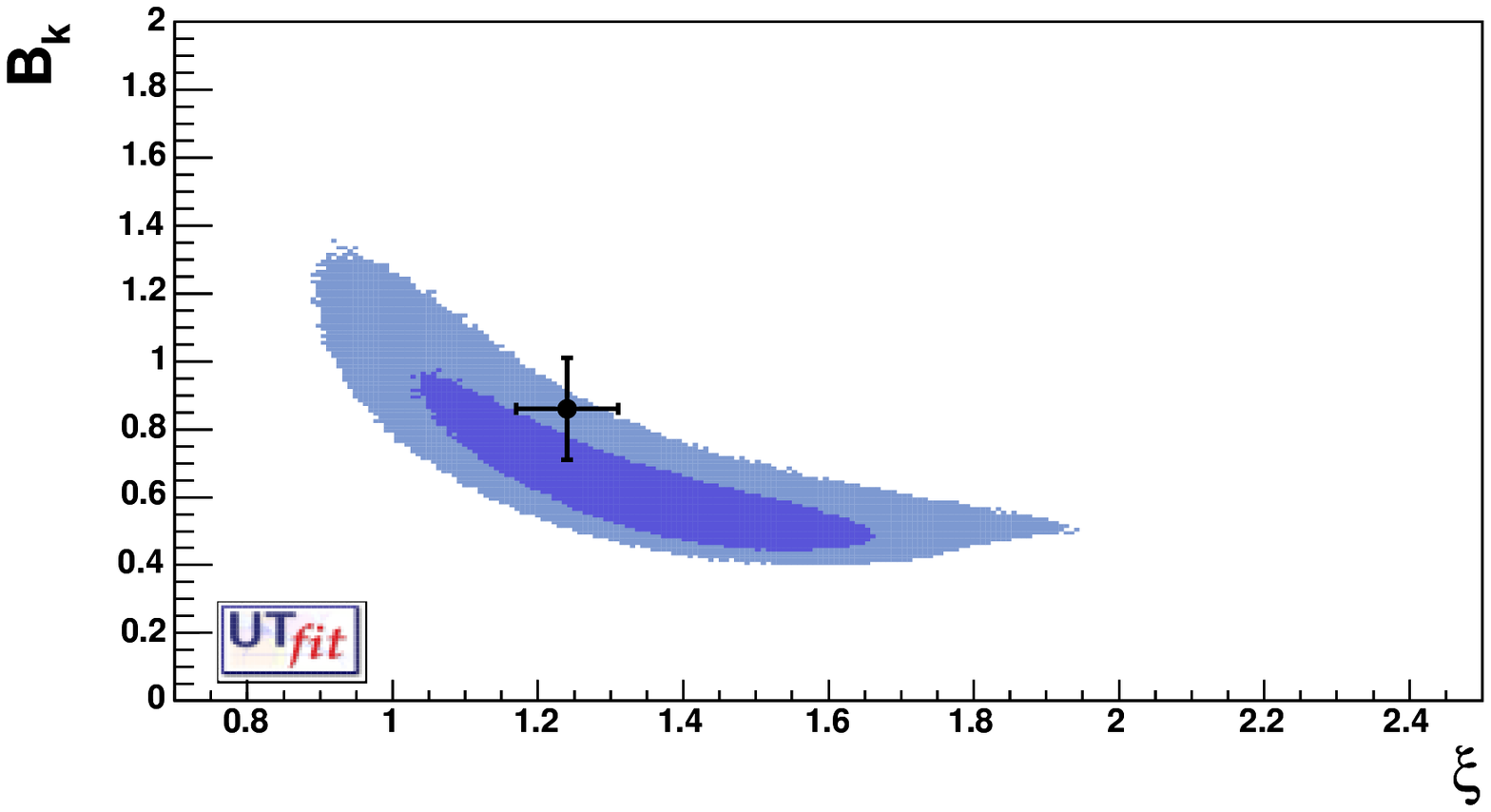}}\\
{\includegraphics[height=5.3cm]{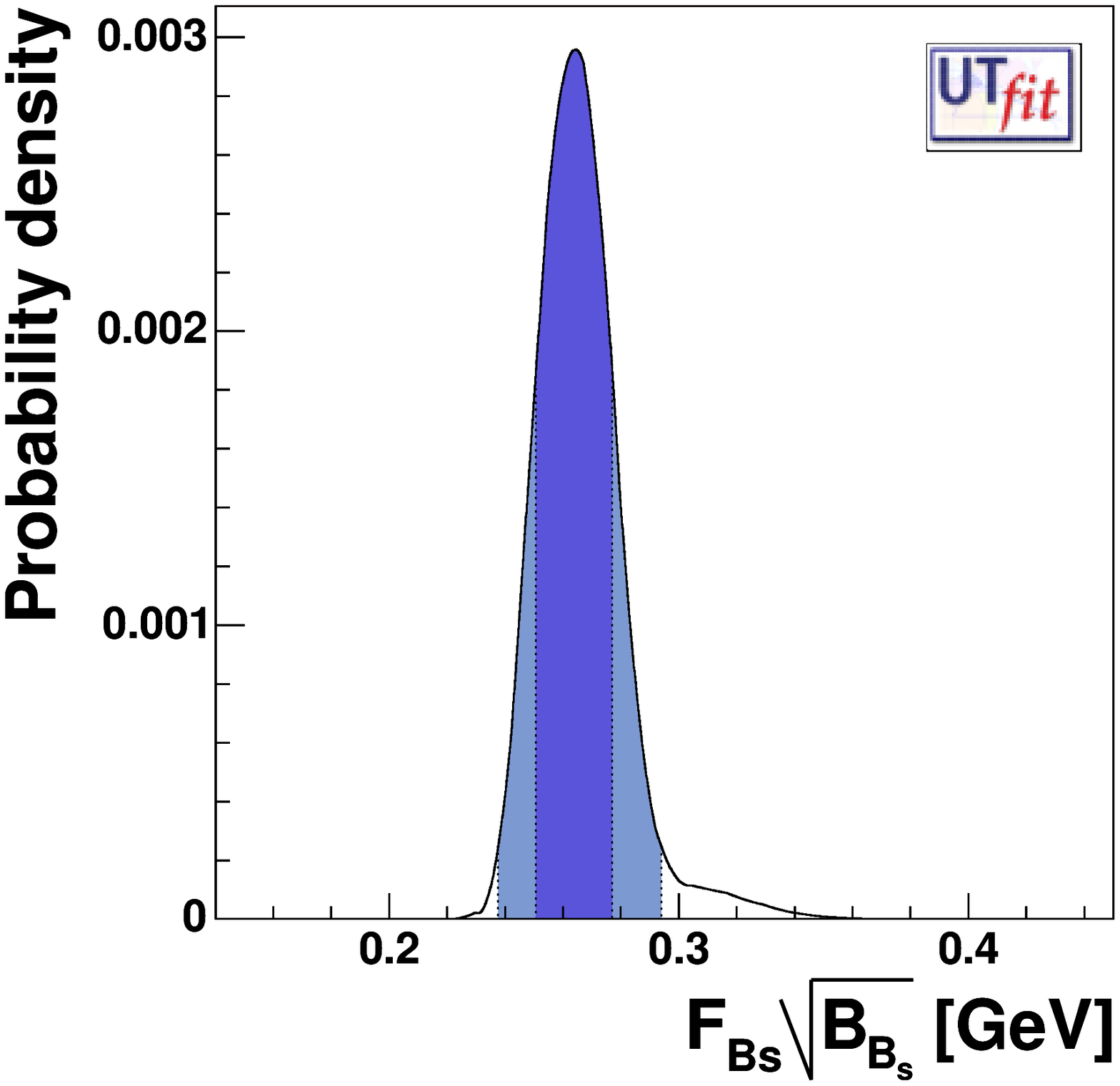}}      
{\includegraphics[height=5.3cm]{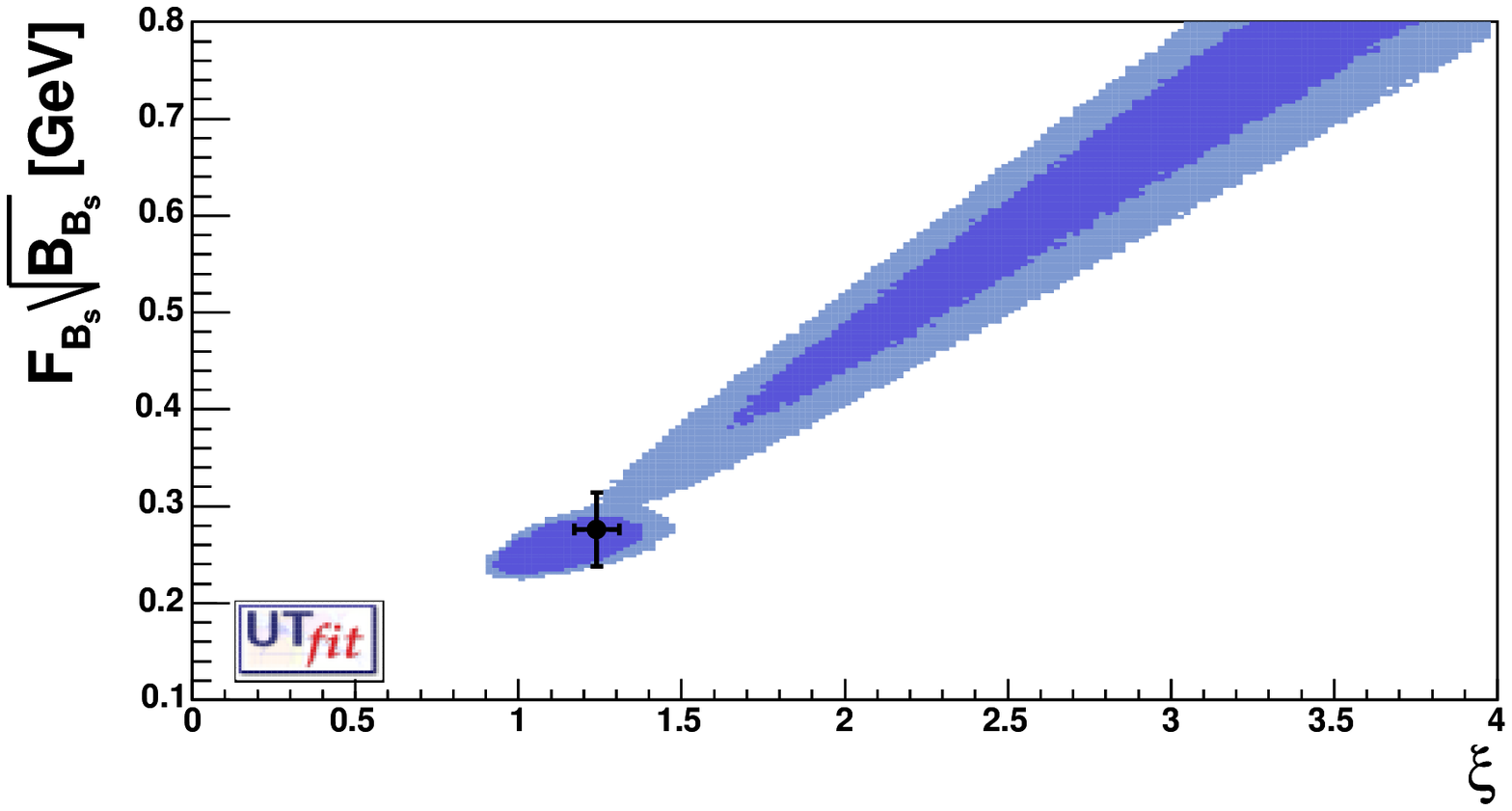}}  \\
\caption{\it {One- and two-dimensional probability distributions
for $\hat B_K$, $\xi$ and $\fbssqbs$ (see Section~\protect\ref{sec:2dpdfs}). The dark and the light (blue) zones correspond respectively to 68\% and 95\% of the area.
The 1$\sigma$ ranges for these quantities, as calculated from LQCD, are indicated with error bars. }} 
\label{fig:bkfb}
\end{center}
\end{figure}

To obtain the {\it a-posteriori} p.d.f. for a given
hadronic quantity, we perform the
UT fit imposing as input a uniform distribution in a range much larger
than the expected interval of values assumed for the quantity itself.
Table~\ref{tab:nonptsumm} and Figure~\ref{fig:bkfb} (left column)
show the results when one single parameter
is taken out of the fit with the above procedure.
The central value and the error of each of these quantities have to be compared
to the current evaluations from LQCD, given in Table~\ref{tab:inputs}
and also shown in Table~\ref{tab:nonptsumm} for the reader's convenience.

\begin{table}[h]
\begin{center}
\begin{tabular}{c|c|c|c}
\hline\hline
    Parameter          &     68$\%$             &      95$\%$  &  99$\%$     \\
\hline
$\xi$  -{\utfit}       & 1.15 $\pm$ 0.11       &  [0.97, 1.44] & [0.93, 1.59] \\
$\xi$  -LQCD           & 1.24 $\pm$ 0.06       &  [1.13, 1.34] & [1.10, 1.37] \\
\hline
$\fbssqbs$(MeV) -{\utfit} & 265  $\pm$ 13          &  [242, 305]   & [236, 330]   \\
$\fbssqbs$(MeV) -LQCD     & 276  $\pm$ 38          &  [201, 350]   & [179, 373]   \\
\hline
$\hat{B}_K$     -{\utfit} & 0.69 $\pm$ 0.10        &  [0.53, 0.93] & [0.49, 1.07] \\
$\hat{B}_K$     -LQCD     & 0.86 $\pm$ 0.11        &  [0.67, 1.05] & [0.62, 1.10] \\
\hline\hline
\end{tabular} 
\caption {\it Values and probability ranges for the non-perturbative QCD
  parameters, if the external information (input) coming from the
  theoretical calculation of these parameters is not used in the fit. The LQCD values are given with errors obtained 
by convoluting
  the Gaussian and flat contributions reported separately in
  Table~\ref{tab:inputs}.}
\label{tab:nonptsumm} 
\end{center}
\end{table}

Some conclusions can be drawn. The precision on $\fbssqbs$ obtained
from the fit has an accuracy which is better than the current
evaluation from LQCD. This proves that the standard UT fit
is, in practice, weakly dependent on the assumed theoretical
uncertainty on $\fbssqbs$.

The result on $\hat{B}_K$ indicates that values of $\hat{B}_K$ smaller
than $0.49$ are excluded at $99\%$ probability, while large values of
$\hat{B}_K$ are compatible with the prediction of the fit obtained
using the other constraints.  The present estimate of $\hat{B}_K$ from
LQCD, which has a 15$\%$ relative error
(Table~\ref{tab:inputs}), is as precise as the indirect
determination from the UT fit.  

The present best determination of the parameter $\xi$ comes from LQCD.

In the above exercise we have removed from the fit individual
quantities one by one.  It is also interesting to see what can be
obtained taking out two of them simultaneously.
Figure~\ref{fig:bkfb} shows the regions selected in the planes
($\fbssqbs$,~$\hat B_K$), ($\hat B_K$,~$\xi$) and ($\fbssqbs$,~$\xi$).
The corresponding results are summarized in
Table~\ref{tab:nonptsumm-two}. For ($\fbssqbs$,~$\xi$), the 2-dimensional
distribution is not limited by the fit. In such a case, the probability  
attached to a given range depends completely on the ranges of the {\it a-priori}
distributions chosen for the two variables. For this reason, we do not quote
any range in Table~\ref{tab:nonptsumm-two}. Still, from the plot in
Figure~\ref{fig:bkfb}, a combined lower bound $\fbssqbs\gtrsim 200$~MeV and
$\xi\gtrsim 0.9$ emerges.

\begin{table}[h]
\begin{center}
\begin{tabular}{c|c|c|c}
\hline\hline
    Parameter         &     68$\%$        &      95$\%$  &  99$\%$     \\ 
\hline 
$\hat{B}_K$           & 0.68$\pm$ 0.10   &  [0.52, 0.97] & [0.48, 1.21] \\ 
$\fbssqbs$(MeV)       & 257 $\pm$ 15     &  [232, 304]   & [221, 348]   \\
\hline
$\hat{B}_K$           & 0.68 $\pm$ 0.18 & [0.45,1.14]  & [0.41,1.33] \\ 
$\xi$                 & 1.31 $\pm$ 0.21        & [0.96,1.72]  & [0.89,1.98] \\
\hline\hline
\end{tabular} 
\caption {\it Values and probability ranges for the non-perturbative QCD
  parameters, if two external pieces of information (inputs) coming
  from the theoretical calculations of these parameters are not used in the fit.
}
\label{tab:nonptsumm-two}
\end{center}
\end{table}

\section{New Constraints from UT angle measurements}
\label{sec:newinputs}

The values for $\alpha$, $\gamma$, $\sin(2\beta+\gamma)$, and $\sin2\beta$
given in Table~\ref{tab:1dim} have to be taken as predictions for
future measurements. A strong message is given for instance for the
angle $\gamma$. Its indirect determination is known with an accuracy of about $10\%$.  

Thanks to the huge statistics collected at the $B$ factories, new
CP-violating quantities have been recently measured allowing for the
direct determination of $\alpha$, $\gamma$, $2\beta$+$\gamma$, and
$\cos2\beta$.
In the following, we study the constraints induced by
these new measurements in the $\rhobar-\etabar$
plane and their impact on the global UT fit.

\subsection{Determination of the angle \boldmath$\gamma$ using \boldmath$DK$ events}
\label{ref:gamma}

Various methods related to $B \to DK$ decays have been proposed to
determine the UT angle
$\gamma$~\cite{ref:GLW}-\cite{ref:DKDalitz}, using the fact that a charged 
$B$ can decay into a $D^0(\overline{D}^0) K$ final state via a
$V_{cb}$($V_{ub})$ mediated process. CP violation occurs if the $D^0$
and the $\overline{D}^0$ decay to the same final state.  These processes are thus sensitive to the
phase difference $\gamma$ between $V_{ub}$ and $V_{cb}$. The same argument
can be applied to $B \to D^{*}K$ and $B \to D^{(*)} K^*$ decays.

Three methods have been proposed:

\begin{itemize}
\item \underline{Gronau-London-Wyler method} (GLW)~\cite{ref:GLW}. It consists in 
reconstructing the neutral $D$ meson in a CP eigenstate:
$B^{\pm} \to D^0_{CP^{\pm}} K^{\pm}$, where $D^0_{CP^{\pm}}$ are the CP eigenstates 
of the $D$ meson. In this case, one can define four quantities,
sensitive to the value of the angle $\gamma$:
\begin{eqnarray}
&&R_{C\!P^{\pm}} = \frac{\Gamma(B^+ \to D^0_{CP^{\pm}}K^+)+\Gamma(B^- \to D^0_{CP^{\pm}}K^-)}
                    {\Gamma(B^+ \to D^0K^+)+\Gamma(B^- \to \overline D^0K^-)} =    
                    1 + r_B^2 \pm 2 r_B \cos \gamma \cos \delta_B  \nonumber \\
&&A_{C\!P^+} =      \frac{\Gamma(B^+ \to D^0_{CP^{\pm}}K^+)-\Gamma(B^- \to D^0_{CP^{\pm}}K^-)}
                    {\Gamma(B^+ \to D^0_{CP^{\pm}}K^+)+\Gamma(B^- \to D^0_{CP^{\pm}}K^-)} =    
                    \frac{\pm 2 r_B \sin \gamma \sin \delta_B}{R_{C\!P^{\pm}}}  
\label{eq:GLW}
\end{eqnarray}
where $r_B$ is the absolute value of the ratio of the Cabibbo-suppressed
over the Cabibbo-allowed amplitudes:
\begin{equation}
 r_B(DK) =  \left\vert \frac{{\cal A} (B^- \to \overline{D}^0 K^-)}{{\cal A} (B^- \to {D}^0 K^-)}
 \right\vert.
\label{eq:rbdef}
\end{equation}
The phase $\delta_B$ in
Eq.~(\ref{eq:GLW}) is the strong phase difference between the $b \to u$ and 
the $b \to c$ decays. The main limitation of this method 
comes from the dilution of the interference effect, 
since $r_B$ is expected to be small ($\sim |V_{ub}/V_{cb}|$). 
One can nevertheless repeat this measurement for several $f_{CP}$ 
final states of $D$ meson.
Moreover, this method (and the ones introduced below) can be generalized
to $D^*K$, $D K^*$ and $D^* K^*$ final states, having different values
for $r_B$ and $\delta_B$, but the same functional dependence on 
$\gamma$.
In case of $B \to D^*K$ decays, there is an effective strong-interaction phase 
difference of $180^\circ$ whenever the $D^*$ is reconstructed as 
$D^* \to D \gamma$ or $D^* \to D \pi$~\cite{Bondar:2004bi}.

\item \underline{Atwood-Dunietz-Soni method} (ADS)~\cite{ref:ADS}. It consists in 
forcing the $\bar D^0$ ($D^0$) meson, coming from the Cabibbo-suppressed
(Cabibbo-allowed) $b \to u$ ($b \to c$) transition to decay 
into the Cabibbo-allowed (Cabibbo-suppressed) $K \pi$ final state.
In this way, one can look at the interference between two amplitudes 
having the same order of magnitude. Two quantities are defined:
\begin{eqnarray}
R_{ADS}  &=& \frac{\Gamma(B^- \to [K^+\pi^-]_DK^-)+\Gamma(B^+ \to [K^-\pi^+]_DK^+)}
                 {\Gamma(B^- \to [K^-\pi^+]_DK^-)+\Gamma(B^+ \to [K^+\pi^-]_DK^+)} \nonumber \\
                &=&  r_{DCS}^2 + r_B^2 + 2 r_B r_{DCS} \cos \gamma \cos (\delta_B + \delta_D) \nonumber \\[2mm]
A_{ADS}  &=& \frac{\Gamma(B^- \to [K^+\pi^-]_DK^-)-\Gamma(B^+ \to [K^-\pi^+]_DK^+)}
                 {\Gamma(B^- \to [K^+\pi^-]_DK^-)+\Gamma(B^+ \to [K^-\pi^+]_DK^+)}  \nonumber \\
                &=& 2 r_B r_{DCS} \sin \gamma \sin (\delta_B + \delta_D) / R_{ADS}
\label{eq:ADS}
\end{eqnarray}
which are functions (as in the case of the GLW method) of $\gamma$, $r_B$
and $\delta_B$. Since in this case $D^0$ and $\bar D^0$ are forced to
decay into different final states, the expressions in Eq.~(\ref{eq:ADS}) 
also depend on the ratio of the Cabibbo-suppressed over the
Cabibbo-allowed decays of the $D$,
\begin{equation}
r_{DCS} = \left\vert\frac{{\cal A}(D^0 \to K^+ \pi^-)}{{\cal A}(D^0 \to K^-\pi^+)}\right\vert ,
\end{equation}
and on the additional strong phase shift $\delta_D$ introduced by these decays. 
One can also use $D^*K$, $D K^*$ and $D^* K^*$, having different values of
$r_B$ and $\delta_B$, but the same values of $\gamma$, $r_{DCS}$ and $\delta_D$.

\item \underline{Dalitz method}~\cite{ref:DKDalitz}. It consists in studying the
interference between the $b \to u$ and the $b \to c$ transitions 
using the Dalitz plot of $D$ mesons reconstructed into three-body 
final states (such as, for instance, $D^0 \to K_s \pi^- \pi^{+}$).
The advantage of this method is that the 
full sub-resonance structure of the three-body decay is considered, 
including interferences such as those used for GLW and ADS methods 
plus additional interferences due to the overlap between broad resonances 
in some regions of the Dalitz plot.  
The same analysis is also performed using $B \to D^{*0}K$ 
decays. The Dalitz analysis has 
only a twofold discrete ambiguity ($\gamma,\gamma -\pi$) and not a 
fourfold ambiguity as in the case of the GLW and ADS methods.  
One can fit the experimental data to extract a 3D likelihood as
a function of $r_B$, $\gamma$ and $\delta_B$.
\end{itemize}

\begin{figure}[htb!]
\begin{center}
\begin{tabular}{cc}
\includegraphics[height=4.9cm]{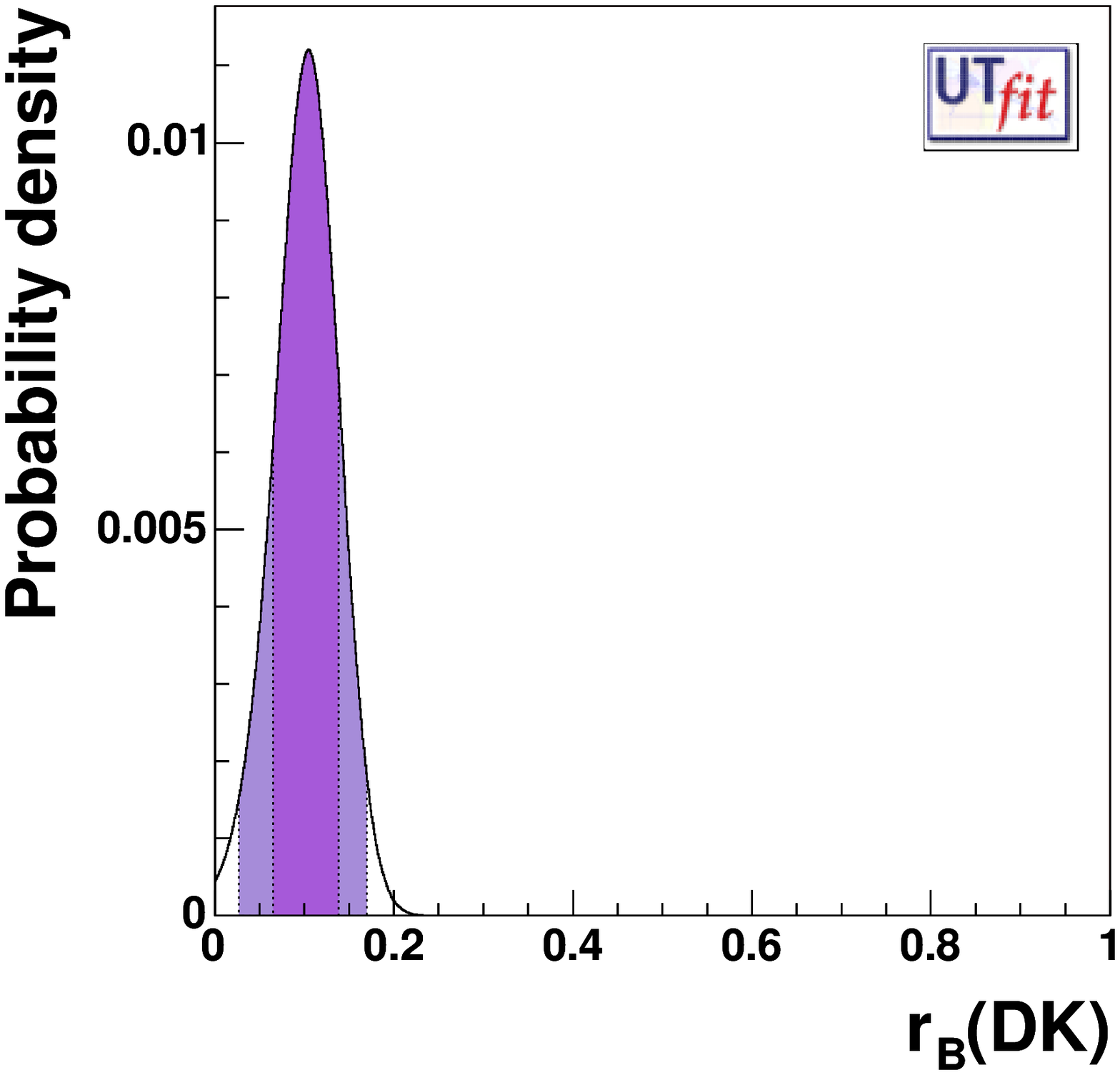}
\includegraphics[height=4.9cm]{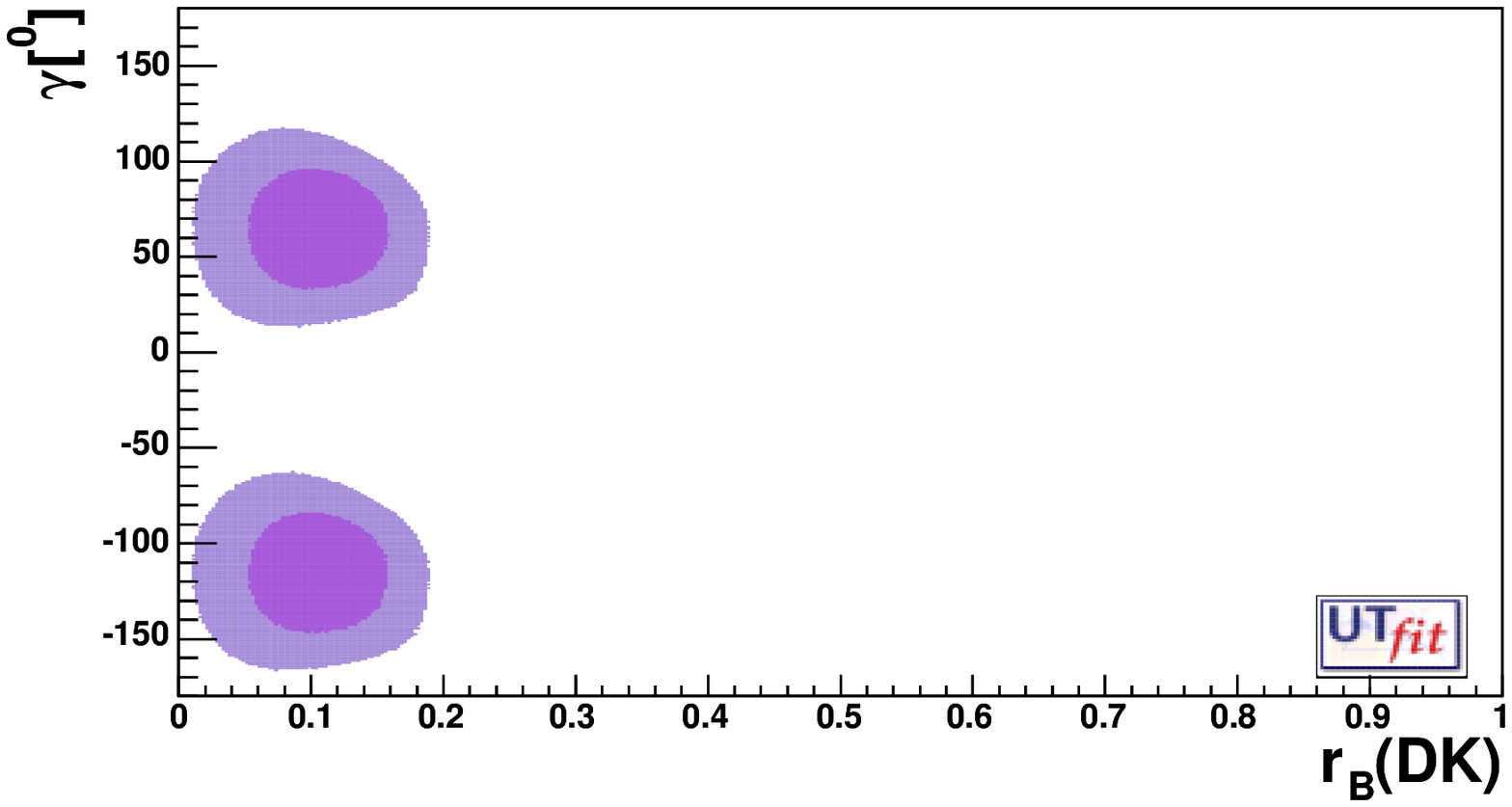} \\
\includegraphics[height=4.9cm]{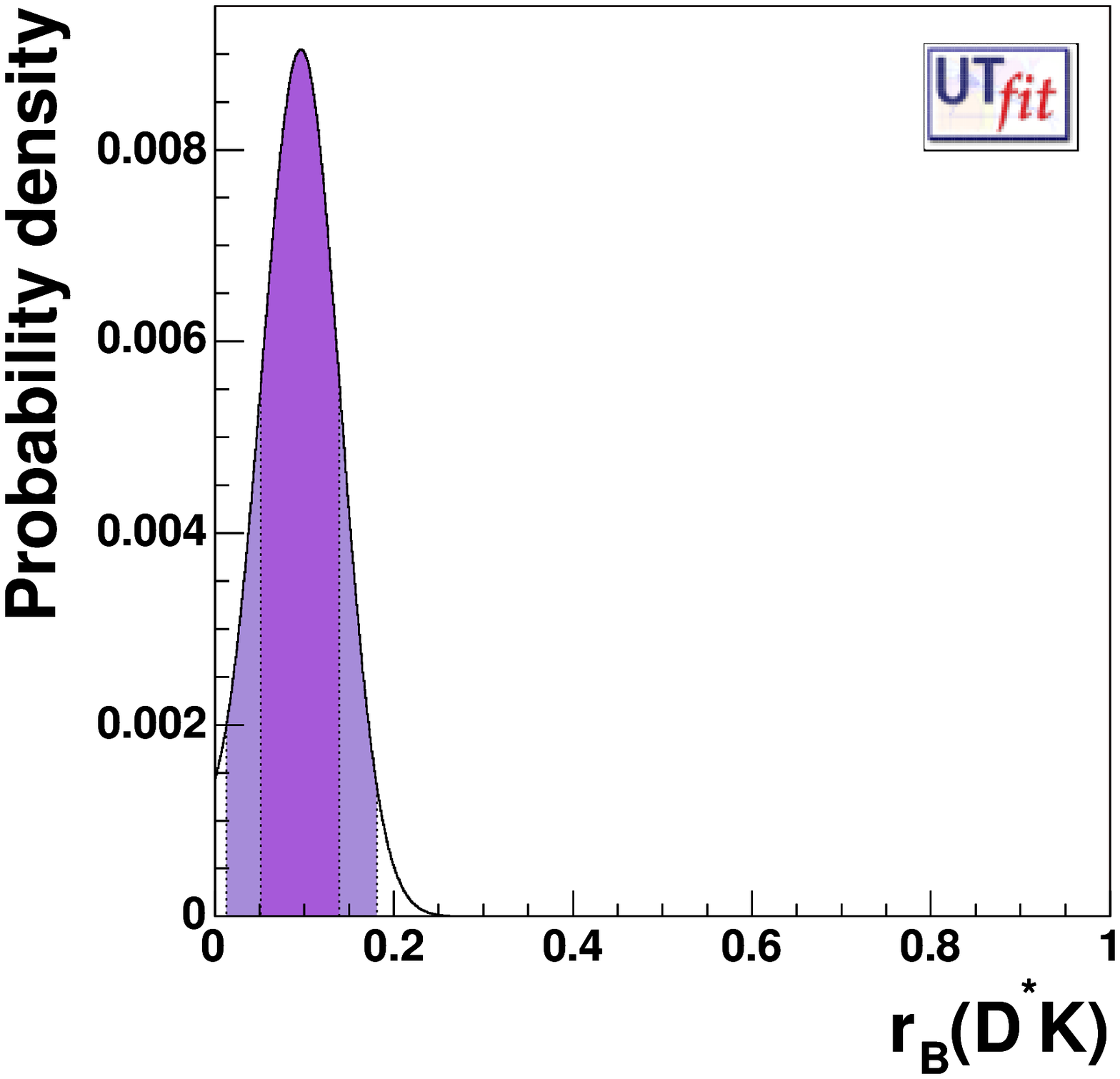} 
\includegraphics[height=4.9cm]{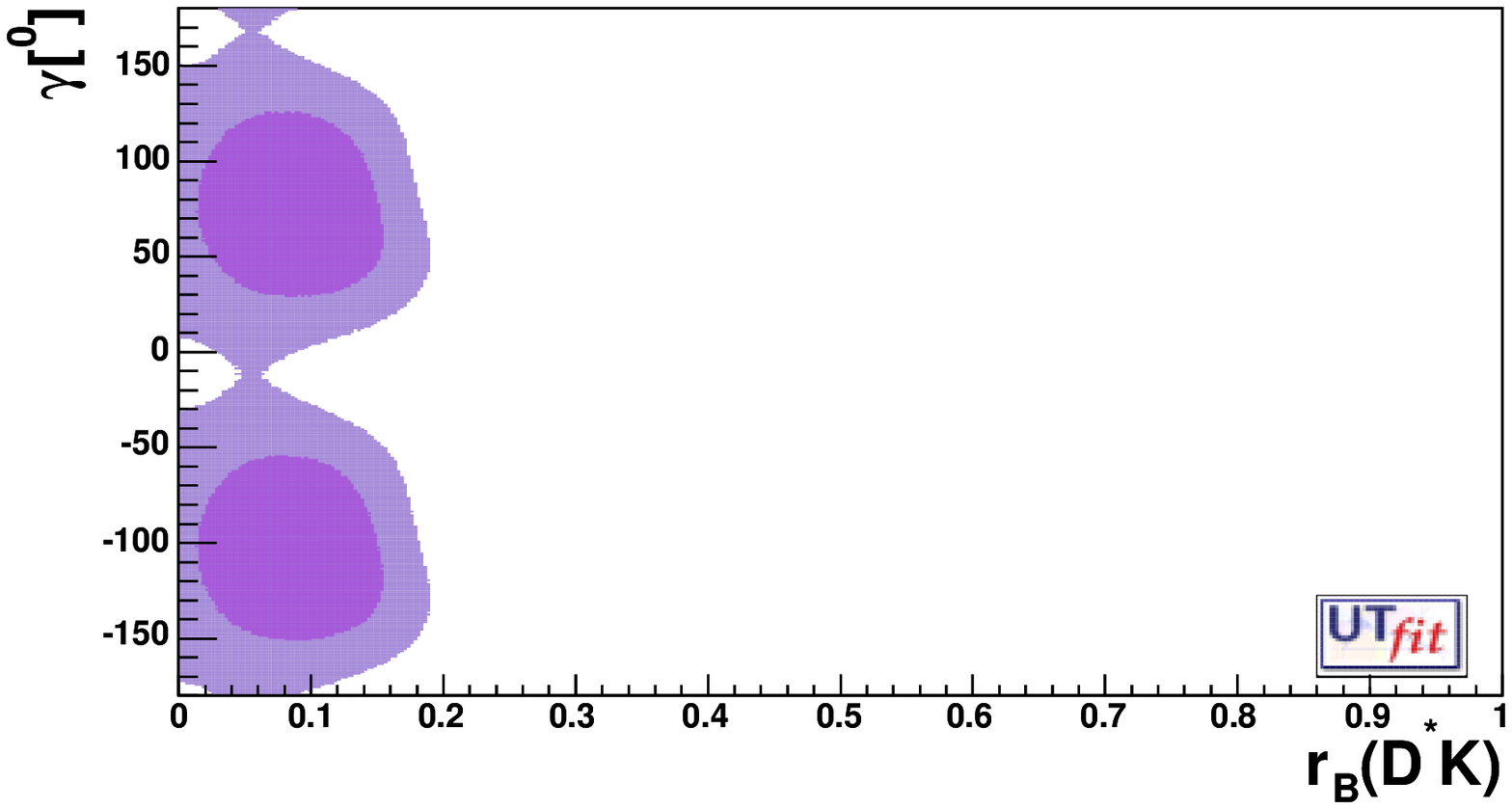} \\
\includegraphics[height=4.9cm]{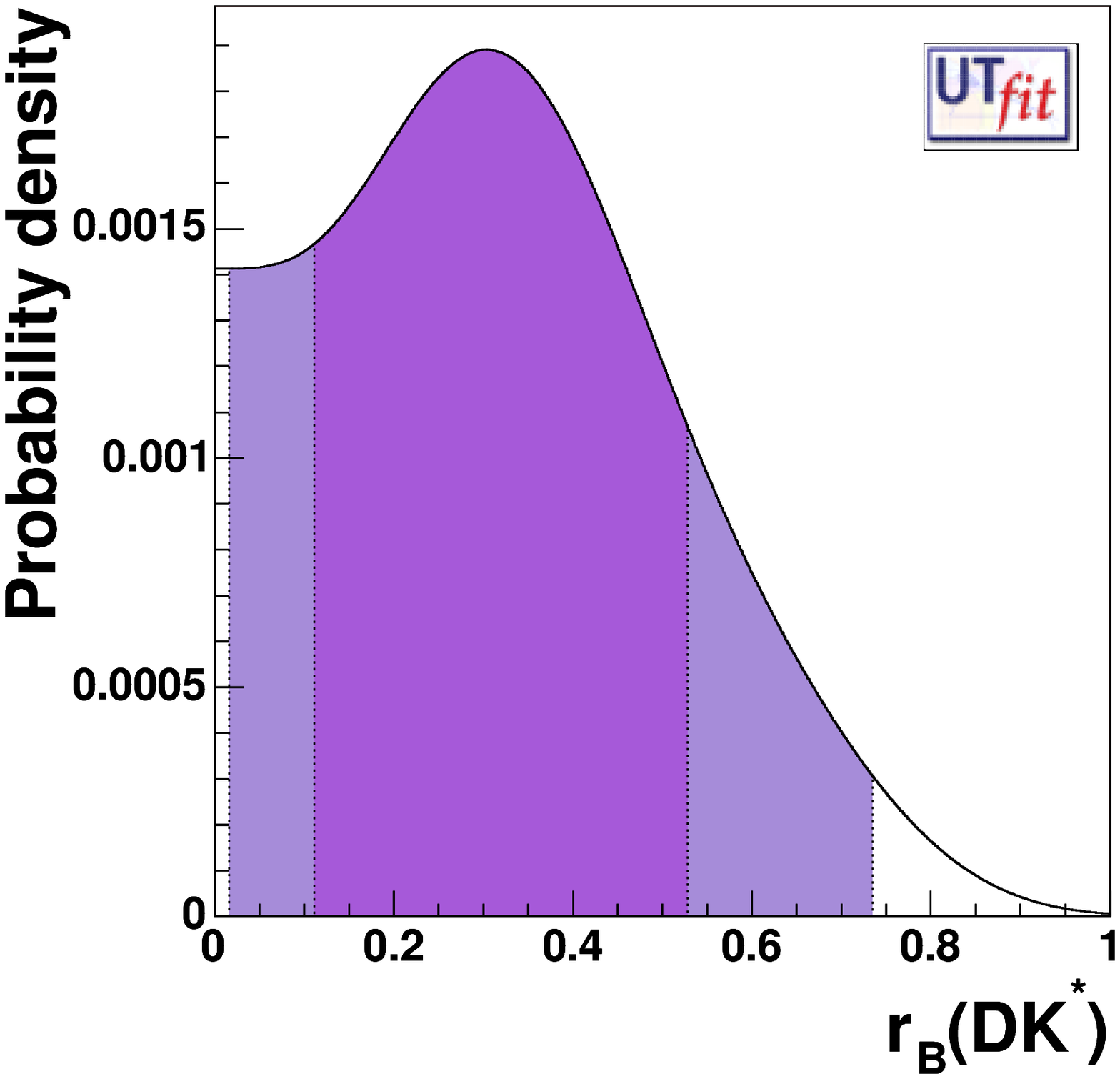} 
\includegraphics[height=4.9cm]{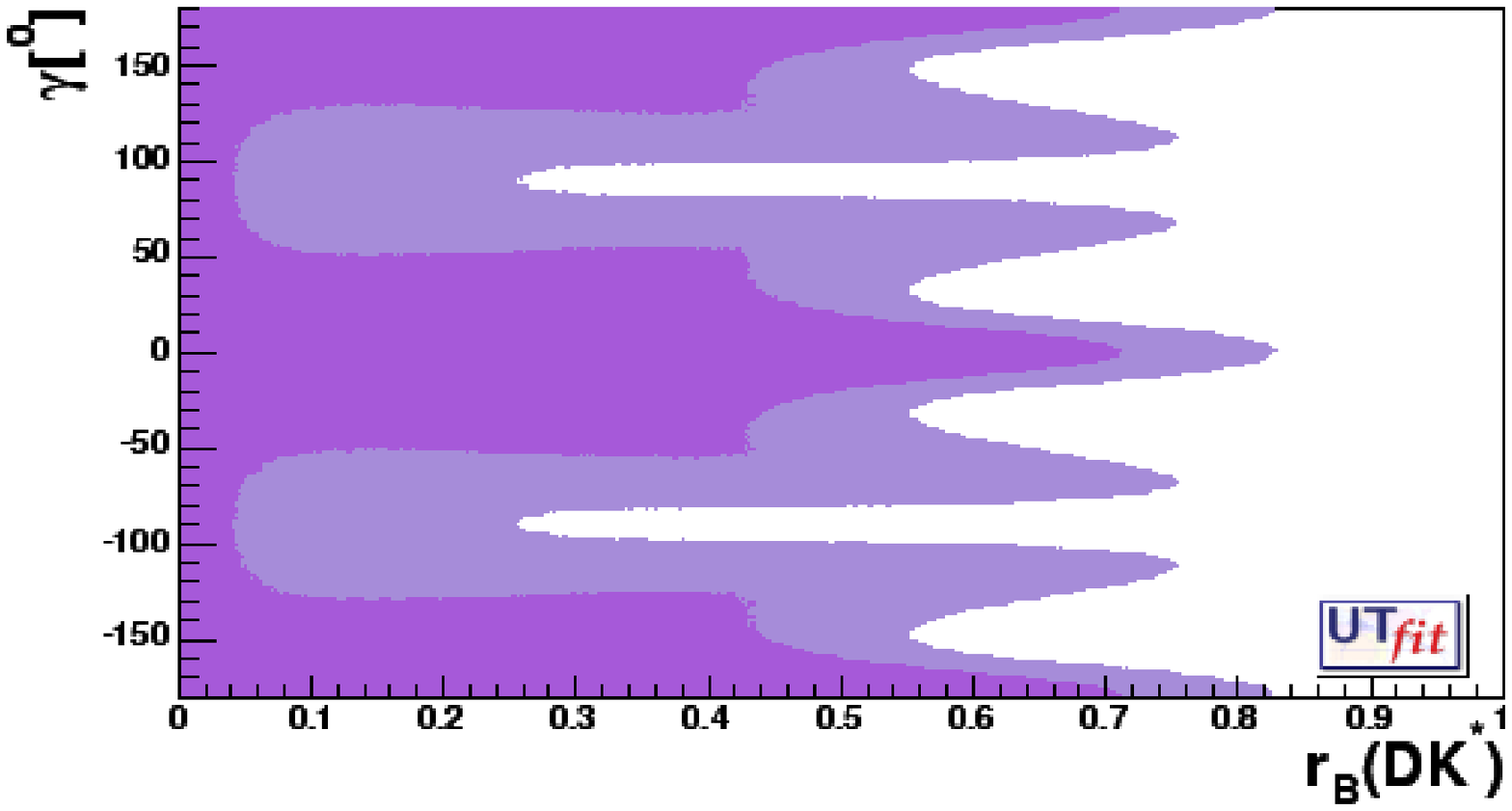} \\
\includegraphics[height=4.9cm]{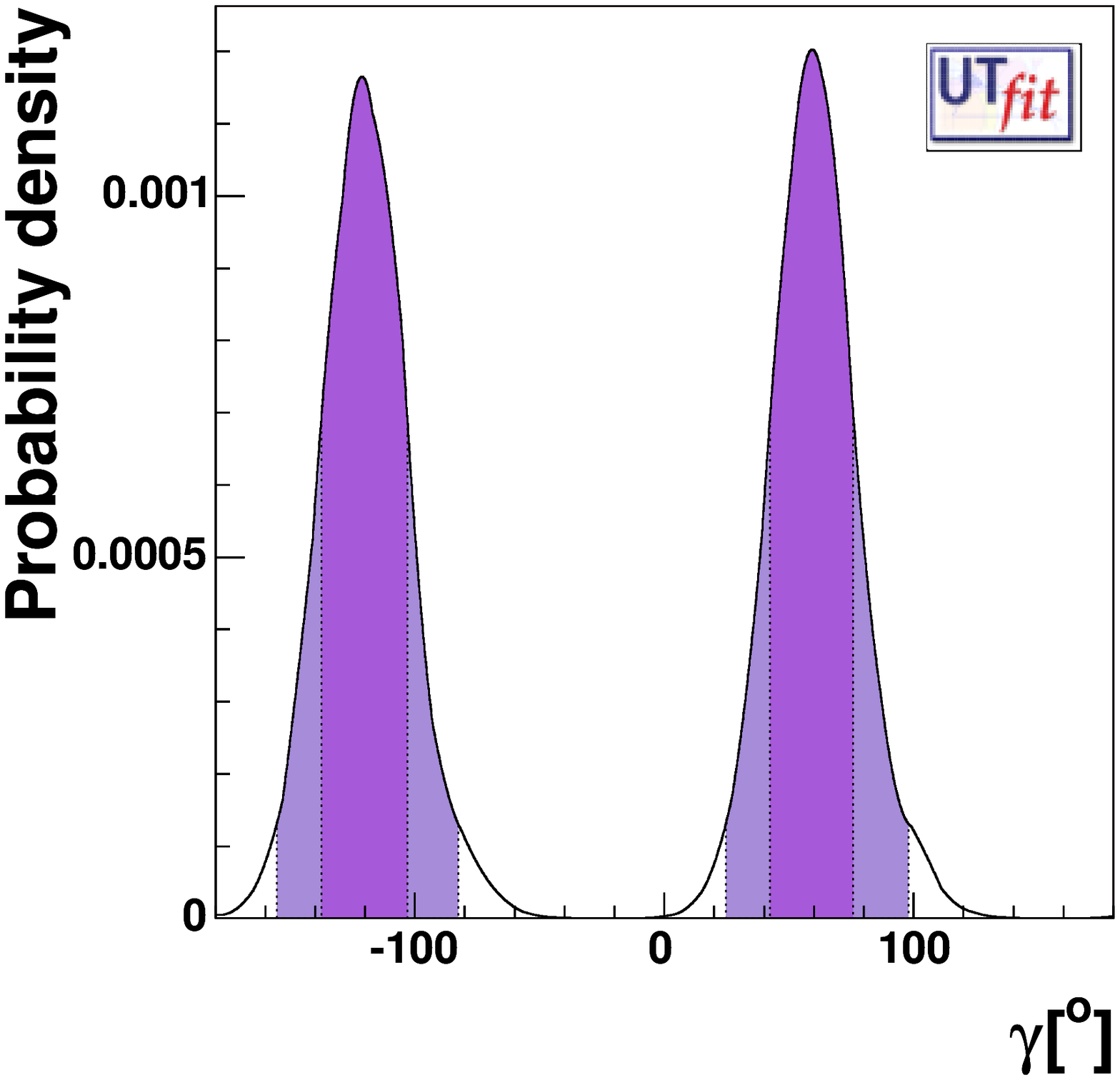} 
\includegraphics[height=4.9cm]{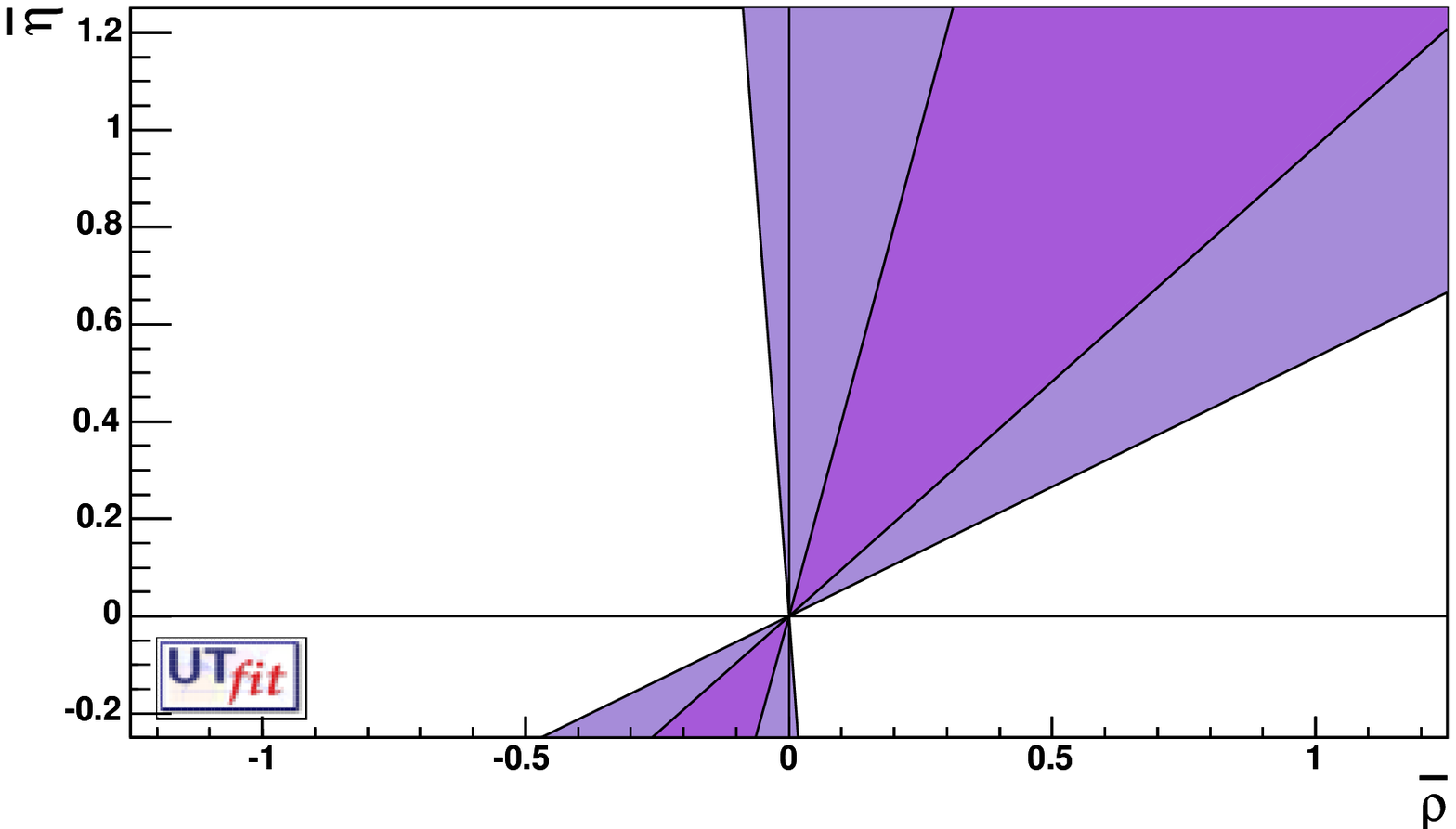} 
\end{tabular}
\end{center}
\caption{ {\it Left column: a-posteriori one-dimensional p.d.f.'s
    (from top-left to bottom-left) for $r_B (DK)$, $r_B(D^*K)$, 
    $r_B(DK^*)$ and $\gamma$, using the results
    from $B \to D^{(*)}K^{(*)}$ decays.  The plots on the right show
    the measurement of the angle $\gamma$ in the $\gamma-r_B$
    (first three rows) and $\rhobar-\etabar$ planes (bottom).}}
\label{fig:gammadir}
\end{figure}

Both BaBar and Belle have presented results applying the 
three methods to several final states. It is important
to notice that in the case of ADS and Dalitz measurements 
we used directly the experimental likelihood because of the presence of non-Gaussian effects:
For Dalitz analyses, there are non-trivial correlations between $r_B$ and $\gamma$ and the
error on $\gamma$ is proportional to $1/r_B$; for ADS analyses, the likelihood of $R_{ADS}$
is bounded to be positive.

\begin{table}[htbp!]
\begin{center}
\begin{tabular}{ccccc} \hline\hline 
Observable         &     $DK$               &   $D^*K$                   &   $DK^*$                \\  \hline
$A_{C\!P^+}$(GLW)  & 0.22  $\pm$ 0.11       & -0.14 $\pm$ 0.18           & -0.07 $\pm$ 0.18        \\    
$A_{C\!P^-}$(GLW)  & 0.02  $\pm$ 0.12       &  0.26 $\pm$ 0.26           & -0.16 $\pm$ 0.29        \\    
$R_{C\!P^+}$(GLW)  & 0.91  $\pm$ 0.12       &  1.25 $\pm$ 0.20           &  1.77 $\pm$ 0.39        \\      
$R_{C\!P^-}$(GLW)  & 1.02  $\pm$ 0.12       &  0.94 $\pm$ 0.29           &  0.76$^{+0.30}_{-0.33}$ \\  \hline  
$R_{ADS}$          & 0.017 $\pm$ 0.009      &   $<0.16$@90$\%$ C.L.                 &         -               \\  
$A_{ADS}$          & 0.49$^{+0.53}_{-0.46}$ &          -                 &         -               \\ \hline
$r_{B}$(Dalitz)-Belle  &  $0.21 \pm 0.08 \pm 0.03 \pm 0.04$ & $0.12^{+0.16}_{-0.11} \pm 0.02 \pm 0.04$  &  -  \\
$\gamma$$[^{\circ}]$(Dalitz)-Belle &  $68 \pm 15 \pm 13 \pm 11$         & $75 \pm 57 \pm 11 \pm 11$                 &  -  \\
$r_{B}$(Dalitz)-BaBar  &  $<0.19~@90\%~{\rm C.L.}$ & $0.155^{+0.070}_{-0.077} \pm 0.04
0 \pm 0.020$  &  -  \\
$\gamma$$[^{\circ}]$(Dalitz)-BaBar & \multicolumn{2}{c}{$70 \pm 26 \pm 10 \pm 10$}  &  -  \\
\hline
\hline
\end{tabular}
\caption{ \it {Summary of the experimental results obtained at the B factories from $B \to D^{(*)}K^{(*)}$
    decays using the GLW, ADS and Dalitz methods~\cite{ref:hfag,ref:expgamma}.  
    The last two lines give the results from the Dalitz analysis, for
    which the ambiguity $\gamma \to \pi - \gamma$ is implicit.}}
\label{tab:DK}
\end{center}
\end{table}

The p.d.f.'s of $\gamma$ and $r_B$, and the selected region in
the $\gamma-r_B$ plane are shown in Figure~\ref{fig:gammadir},
together with the impact of this measurement on the $\rhobar-\etabar$ plane.

\begin{figure}[htb!]
\begin{center}
\includegraphics[height=8.9cm]{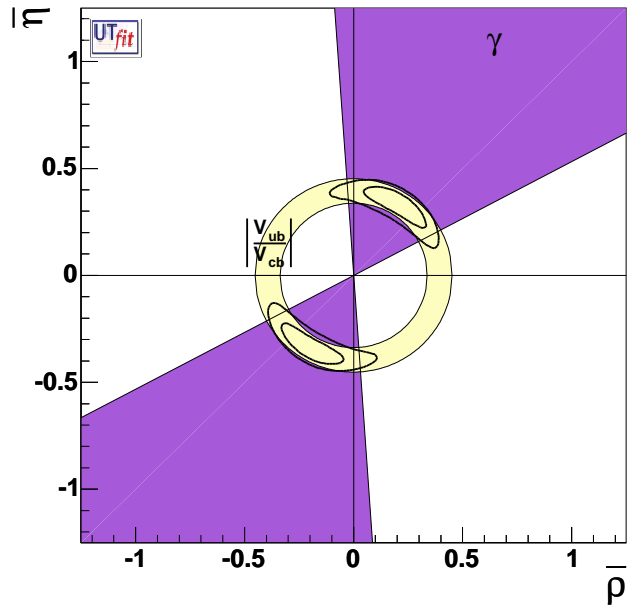}
\end{center}
\caption{ {\it Regions of the $\rhobar$--$\etabar$ plane selected by the constraints imposed by the determination of
$\vert V_{ub}/V_{cb}\vert$ and $\gamma$ from tree-level processes.}}
\label{fig:gammavub}
\end{figure}

The comparison between the direct and the indirect determination is:
\begin{eqnarray}
\gamma[^{\circ}] & = & ~~~~~~~~60.3 \pm 6.8 ~~~~([47.0,\, 74.2]~{\rm at~95\%})~~~~~~~~~~~\rm {indirect-{\mbox{\utfit}}} \\
\gamma[^{\circ}] & = &\left\{
\begin{array}{l} 
59.1 \pm 16.7 ~~\cup~~ -120.3 \pm 17.2 \\
{}([24.7,\, 97.9] ~~\cup~~ [-155.4,\, -82.7]~{\rm at~95\%})
\end{array}\right.
~~\,\rm {direct-{\it D^{(*)}K^{(*)}}} \nonumber  
\label{eq:gamma}
\end{eqnarray}
where the two values of the direct determination approximately correspond to
the discrete ambiguity of the Dalitz method, which dominates the determination of 
$\gamma$.
An important result of this analysis is also the distributions for the different $r_B$'s:
\begin{eqnarray}
r_B (DK)     & = & 0.10  \pm 0.04 ~([0.03,\,0.17]~{\rm at} ~95\%~ C.L.) \\ \nonumber
r_B (D^*K)   & = & 0.09  \pm 0.04 ~([0.01,\,0.18]~{\rm at} ~95\%~ C.L.) \\ \nonumber
r_B (DK^*)   & = & 0.32  \pm 0.21 ~([0.02,\,0.74]~{\rm at} ~95\%~ C.L.)  
\label{eq:rbrb}
\end{eqnarray}
The small values for $r_B (DK)$ and $r_B (D^*K)$ are limiting the precision
on the present determination of $\gamma$.

It is interesting to observe that the determination of $\gamma$
discussed in this section is not affected by new physics (NP) under
the mild assumption that NP does not change tree-level processes. 
Therefore, together with the measurement of $\vert V_{ub}/V_{cb}\vert$, it
already provides a constraint on the $\rhobar$--$\etabar$ plane which must be fulfilled by
any NP model. 
The regions selected by these two constraints are shown in Figure~\ref{fig:gammavub}. We obtain 
\begin{equation}
\bar \rho = \pm (0.21 \pm 0.10),
~ ~ ~ \bar \eta = \pm (0.36 \pm 0.06).
\end{equation}

A more detailed and quantitative analysis of the impact of the UT fit on specific NP models will be presented in a forthcoming paper.

\subsection{Determination of $2\beta$+$\gamma$ using $D^{(*)}\pi(\rho)$ events}
\label{ref:sin2bg}

The interference effects between the $b \to c$ and $b \to u$ decay amplitudes in the 
time-dependent asymmetries of $B$ decaying into $D^{(*)}\pi$ 
and $D^{(*)}\rho$ final states allow for the determination of
$2\beta$+$\gamma$.
The time-dependent rates, in the case of $D\pi$ final states, 
can be written as
\begin{eqnarray}
R(B^0 \to D^{-} \pi^+) = N~e^{-\Gamma t}~{(1+C~\cos(\Delta m_d t) + S~\sin(\Delta m_d t)~)}  \\ \nonumber
R(\overline{B}^0 \to D^{-} \pi^+) = N~e^{-\Gamma t}~ {(1-C~\cos(\Delta m_d t) -S~\sin(\Delta m_d t)~)}   \\ \nonumber
R(B^0 \to D^{+} \pi^-) = N~e^{-\Gamma t}~ {(1+C ~\cos(\Delta m_d t)-\overline{S}~\sin(\Delta m_d t)~)}   \\ \nonumber
R(\overline{B}^0 \to D^{+} \pi^-) = N~e^{-\Gamma t}~ {(1-C~\cos(\Delta m_d t) +\overline{S}~\sin(\Delta m_d t)~)} 
\label{eq:betagamma}
\end{eqnarray}
where $S$, $\overline{S}$ and $C$ are defined as 
\begin{eqnarray}
S \equiv S(D^-\pi^+) &=& 2\frac{r}{1+r^2} \sin(2 \beta +\gamma-\delta)              \\ \nonumber
\overline{S} \equiv S(D^+\pi^-) &=& 2\frac{r}{1+r^2} \sin(2 \beta+\gamma+\delta)    \\ \nonumber
C &=& \frac{1-r^2}{1+r^2} 
\end{eqnarray}
and $r$ and $\delta$ are the absolute value and the strong phase 
of the amplitude ratio
${\cal A}(\overline{B}^0 \to D^- \pi^+) / {\cal A}(B^0 \to D^-\pi^+)$.
All these expressions can be generalized to the case of $D^* \pi$ and $D \rho$ final states,
which have in principle different values of $r$ and $\delta$.
The ratio $r$ is expected to be rather small being of the order of 
$\lambda |V_{ub}/V_{cb}| \simeq$ 0.02. 

The extraction of the weak phase is made even more difficult by the presence of a 
correlation between the tag side and the reconstruction
side in time-dependent CP measurements at $B$ factories~\cite{ref:owen}.
This is related to the possibility that the interference between $b \to c$ and $b \to u$ transitions in $B \to D X$ decays
occurs also on the tag side. In such a case, it is useful to replace $S$ and $\overline{S}$ by 
two new parameters $a$ and $c$ which can be written as~\cite{ref:hfag}
\begin{eqnarray}
\label{eq:ac}
a &\equiv& \frac{S + \overline{S}}{2} \simeq 2 r \sin(2 \beta +\gamma) \cos(\delta)      \\ \nonumber
c &\equiv& -\frac{S - \overline{S}}{2} \simeq 2 \cos(2\beta +\gamma)(r \sin(\delta)-r^\prime \sin(\delta^\prime))
\end{eqnarray}
retaining only linear terms in $r$ and $r^\prime$,
where $r^\prime$ and $\delta^\prime$ are the analogue of $r$ and $\delta$
for the tag side. It is important to stress that the interference in
the tag side cannot occur when $B$ mesons are tagged using semileptonic
decays.  In other words, $r^{'}$= 0 when only semileptonic decays are used. 
In the following we will consider the observable $a$,
$c_{\ell}$ (denoting $c$ evaluated for lepton-tagged events),
which are functions of $r$, $\delta$ and $2\beta$+$\gamma$.
The experimental situation is summarized in Table ~\ref{tab:Dpi}.

\begin{table}[htbp!]
\begin{center}
\begin{tabular}{cccc} 
\hline\hline
  Parameter          & {{$D\pi$}}           &      {{$D^*\pi$}}    &    {{$D\rho$}}     \\ \hline
a                    &  -0.045 $\pm$ 0.027  &  -0.030 $\pm$ 0.014  & -0.005 $\pm$ 0.049   \\    
$c_{\ell}$           &  -0.035 $\pm$ 0.035  &   0.010 $\pm$ 0.021  & -0.147 $\pm$ 0.082  \\ 
\hline
\end{tabular}
\caption{ \it {Summary of the experimental results from BaBar and 
Belle Collaborations. World averages are calculated by HFAG~\cite{ref:hfag}
from~\cite{expdpi}.}}
\label{tab:Dpi}
\end{center}
\end{table}

\begin{figure}[htbp!]
\begin{center}
\includegraphics[width=0.58\textwidth]{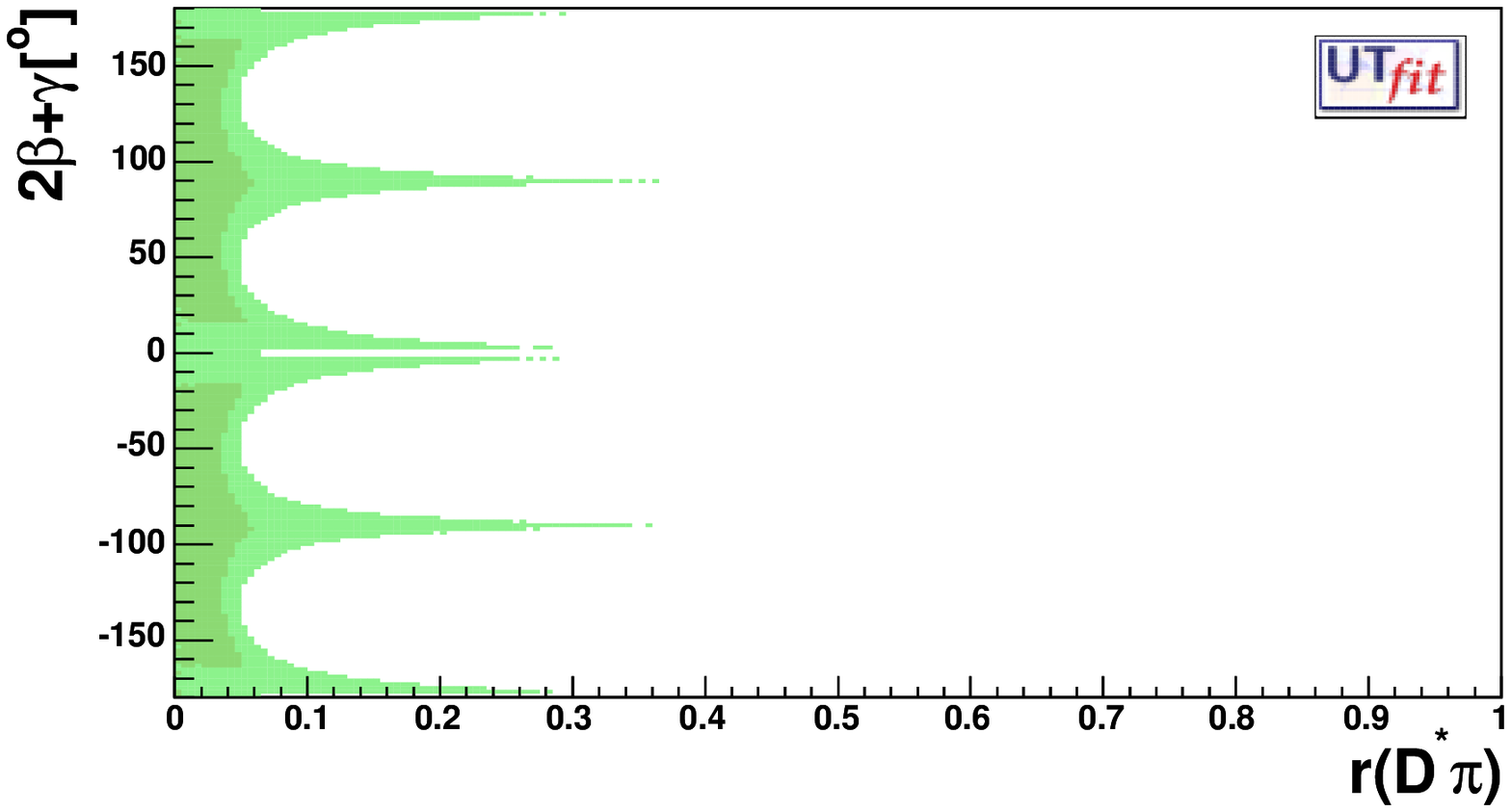} \\
\includegraphics[width=0.34\textwidth]{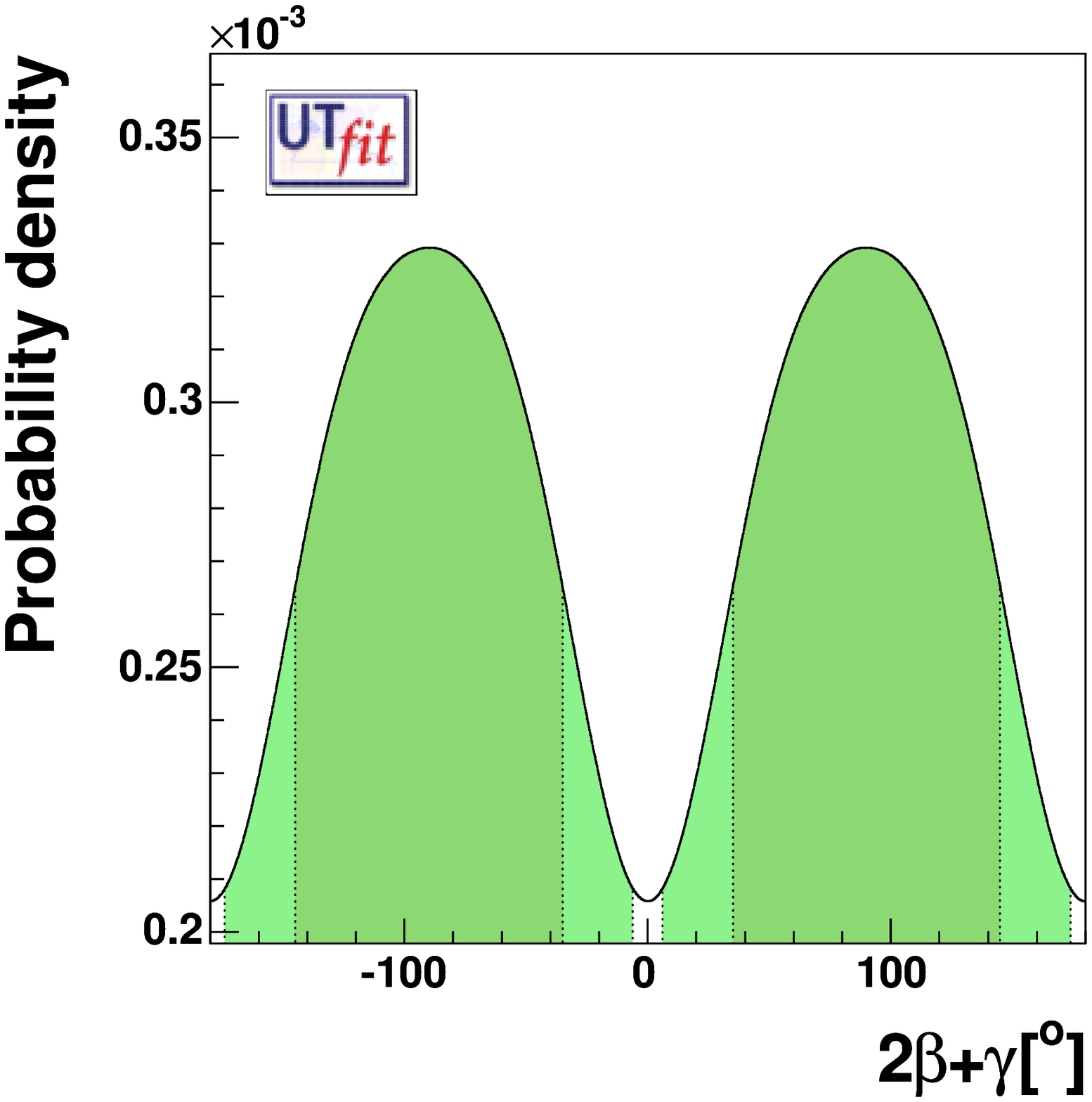}
\includegraphics[width=0.65\textwidth]{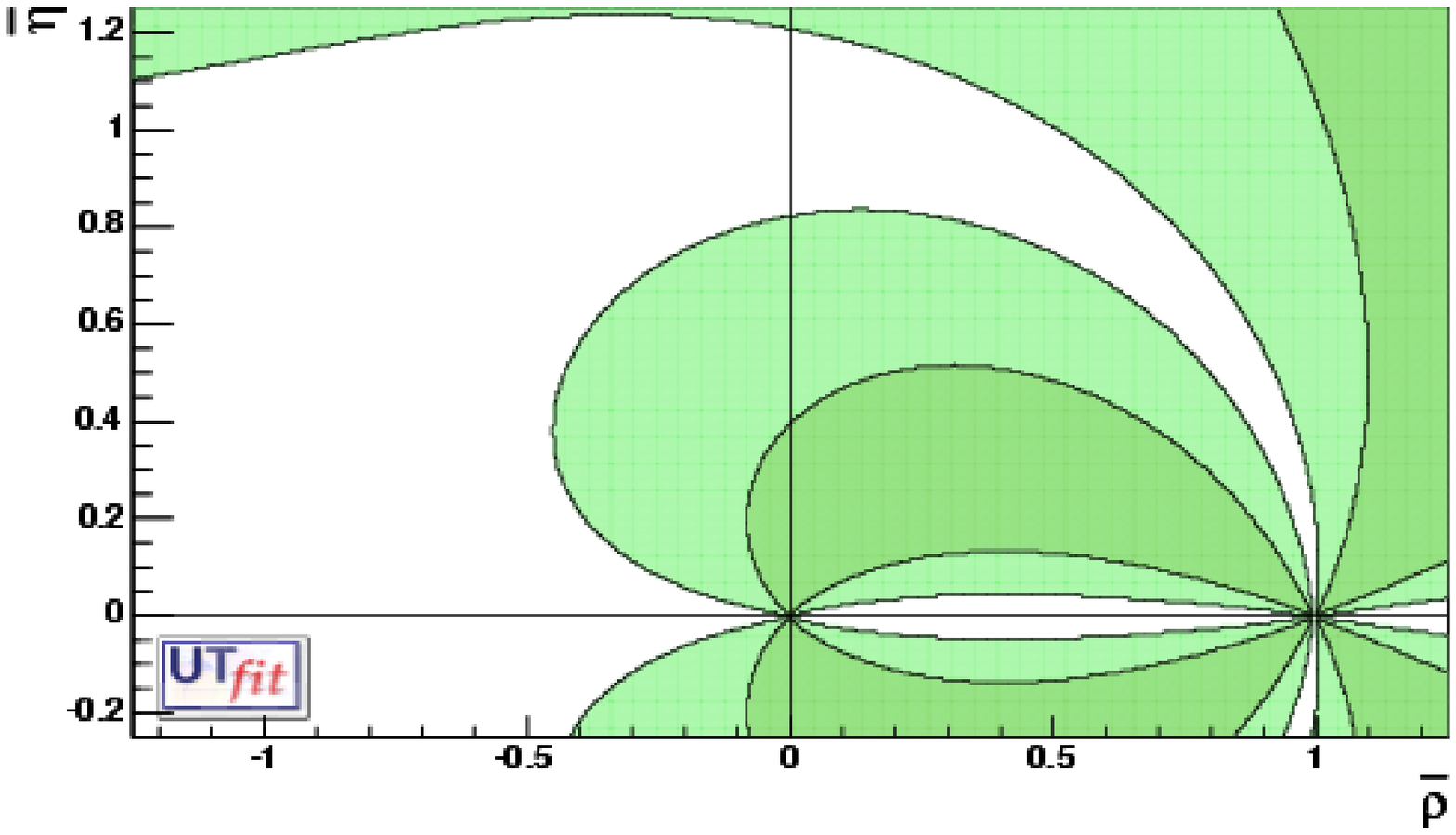}
\hspace*{-2.5cm}
\end{center}
\caption{ {\it Distribution of $r$ vs. $2\beta+\gamma$ for $D^*\pi$ decays (top). For illustration, we also show the p.d.f. of $2\beta+\gamma$ (bottom-left) and the constraint in the $\rhobar-\etabar$ plane
(bottom-right) obtained assuming $SU(3)$ flavour symmetry and neglecting the annihilation contribution in $A(\overline{B}^0 \to D^{*-} \pi^+)$.}}
\label{fig:2bgdir}
\end{figure}

With the present experimental data, a determination of $2\beta$+$\gamma$ cannot be obtained from 
$D^{(*)} \pi (\rho)$ modes alone. The number of free parameters exceed the available constraints 
as shown by Eqs.(\ref{eq:ac}). Without further input, one can only find correlations 
among $2\beta$+$\gamma$ and the hadronic parameters. For example, the correlation between $r(D^{*}\pi)$ 
and $2\beta$+$\gamma$ is shown in Figure~\ref{fig:2bgdir}. An independent information on the hadronic
parameters would allow a determination of $2\beta$+$\gamma$. For instance, assuming $SU(3)$ flavour symmetry 
and neglecting annihilation contributions, one can estimate 
$BR(\overline{B}^0 \to D^{*-} \pi^+)$ from $BR(\overline{B}^0 \to D_s^{*-} \pi^+)$, obtaining
$r(D^*\pi)= 0.015 \pm 0.006 \pm 0.005$ (where the first error is statistical and 
the second is a guessed theoretical error associated to the SU(3) breaking effect
and to the size of annihilation contributions~\cite{expdpi}).
Under these two assumptions, we get a constraint on $2\beta$+$\gamma$ as shown in 
Figure~\ref{fig:2bgdir}. The same can be done for the other two modes.
This strategy suffers from  theoretical uncertainties which cannot be
reliably estimated. For this reason, we do not use any bound on $2\beta$+$\gamma$ in 
the global fit. A more fruitful use of these data could be possible 
when the experimental analyses included the coefficient of the cosine term (in such a case the formulae in Eq.~(\ref{eq:ac}) should not be expanded in $r$ and $r^{\prime}$).
Additional decay modes, such as $B^0 \to D^0 K_S$ or the Dalitz analysis of
$B^0\to D^{\pm} K_S \pi^\mp$ decays~\cite{Aleksan:2003fm}, would also help.


\subsection{Determination of the angle \boldmath$\alpha$}
\label{ref:alpha}

The angle $\alpha$ can be obtained using the time-dependent analyses of $B^0 \to \pi^+ \pi^-$, 
$B^0 \to \rho^+ \rho^-$ and $B^0 \to (\rho \pi)^0$.                

In the absence of contributions from penguin diagrams, these decays
give a measurement of $\sin 2\alpha$.
Penguin diagrams with $c$ and $t$ quarks in the loop introduce an additional
amplitude with a different weak phase. In this case, the experimentally measured
quantity is $\sin 2\alpha_{e\!f\!f}$, which is a function of $\sin 2\alpha$ but also of 
unknown hadronic parameters.
Several strategies have been proposed to get rid of this so-called ``penguin pollution''.

\subsubsection{Isospin analysis of \boldmath$B \to \pi \pi $ and \boldmath$B \to \rho \rho$}

Assuming SU(2) flavour symmetry and neglecting electroweak penguins, the decay amplitudes of $B\to\pi\pi (\rho\rho)$ can be written as
\begin{eqnarray}
A^{+-} &=& -T e^{-i\alpha} + P e^{i\delta_P} \nonumber \\
A^{+0} &=& -\frac{1}{\sqrt{2}} \left [e^{-i\alpha} (T +
T_c~e^{i\delta_{T_c}})\right ] \nonumber \\
A^{00} &=& -\frac{1}{\sqrt{2}} \left [e^{-i\alpha}  T_c~e^{i\delta_{T_c}} +
P~e^{i\delta_P} \right ]\,,
\label{eq:su2analysis}
\end{eqnarray}
where $T$, $T_c$ and $P$ are real parameters~\footnote{These parameters are directly related to the RGI quantities of ref.~\cite{burassilv}. In particular, up to a trivial rescaling,
$P$ is the charming penguin parameter $P_1$ of \cite{burassilv},
while $T = E_1 - P_1^{GIM} + A_1$ and $T_c = E_2 + P_1^{GIM} - A_1$.} and $\delta_P$ and $\delta_{T_c}$ are strong phases (the strong phase of the $T$ term is conventionally set to zero).
It should be noted that these parameters are different for $B
\to \pi \pi$ and $B \to \rho \rho$ decays. For $B \to \rho \rho$,
following the experimental results, only longitudinally polarized final states 
have been considered. 

\begin{figure}[htb!]
\begin{center}
\includegraphics[width=0.48\textwidth]{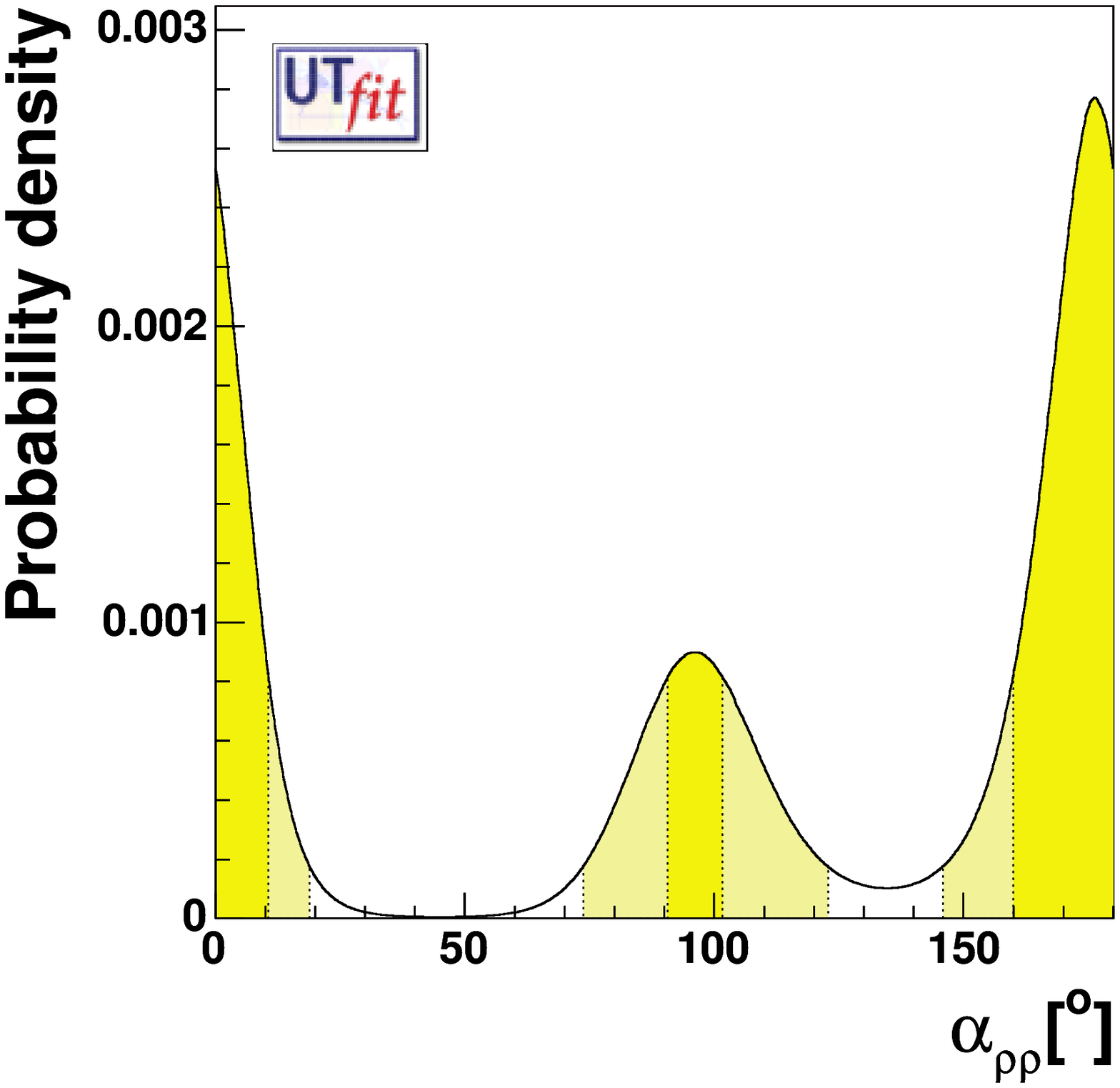}
\includegraphics[width=0.48\textwidth]{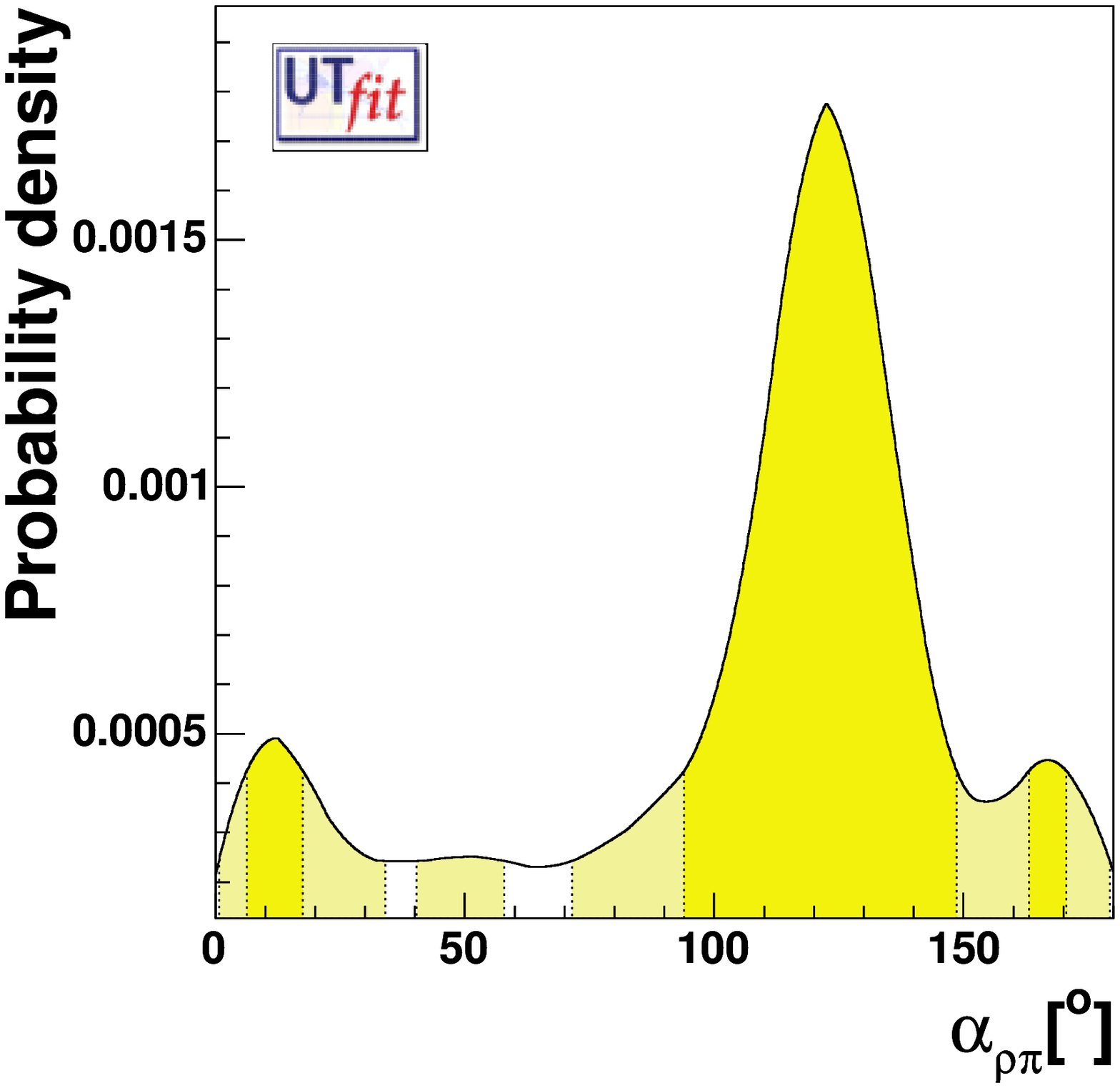} \\
\hspace*{-0.6cm}
\includegraphics[width=0.48\textwidth]{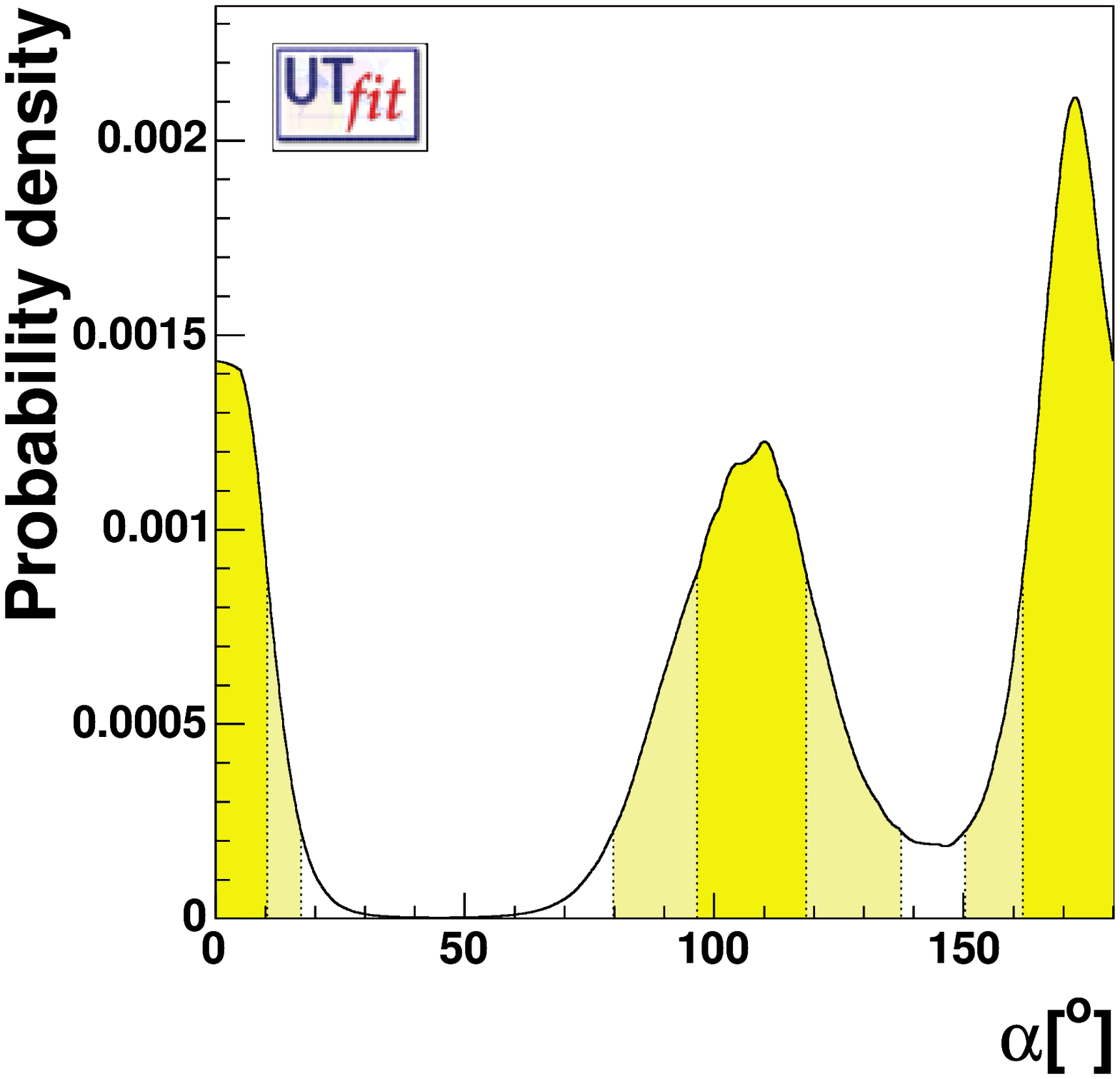} 
\hspace*{-1.0cm}
\includegraphics[width=0.6\textwidth]{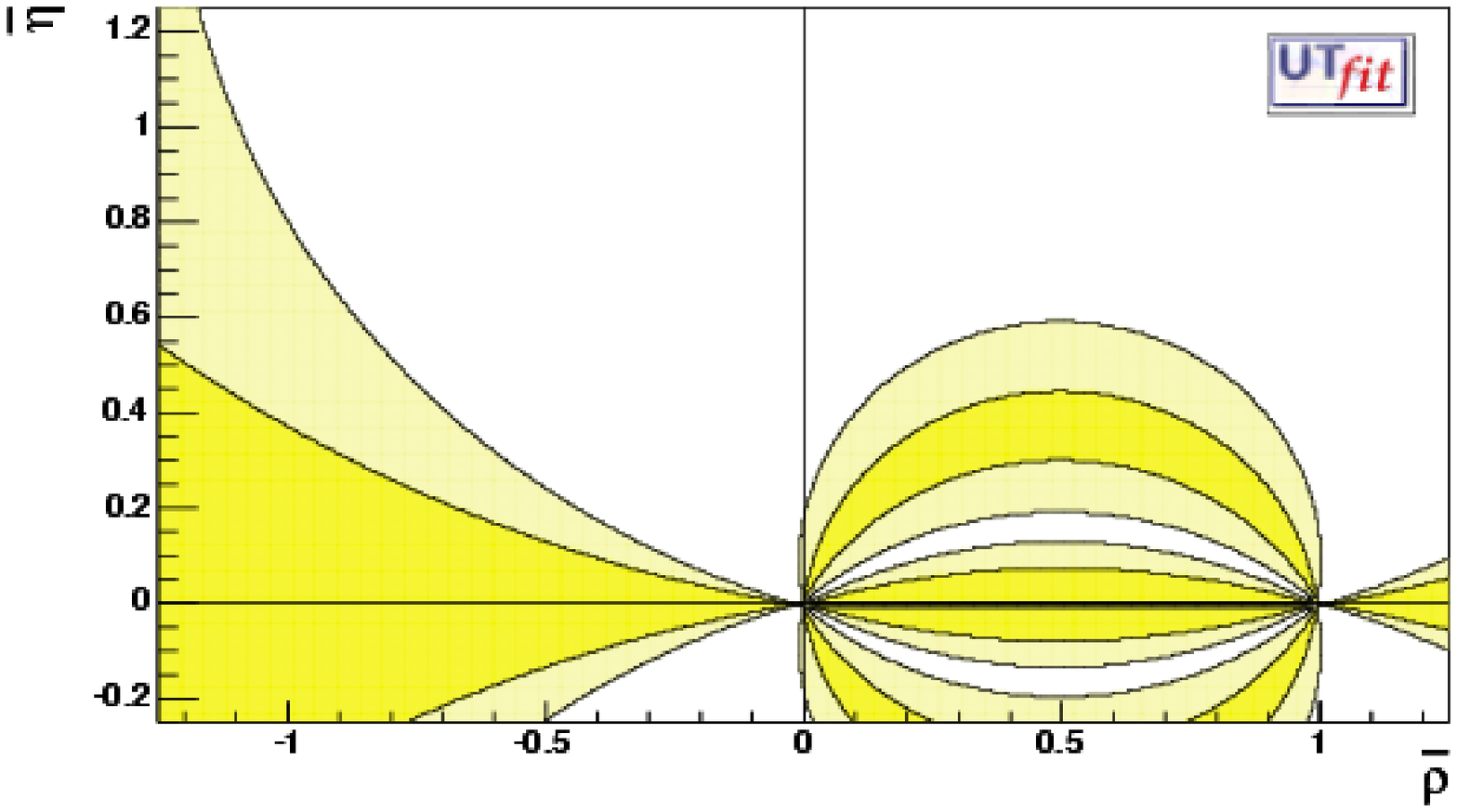}
\hspace*{-1.0cm}
\end{center}
\caption{{\it A-posteriori one-dimensional p.d.f.'s of $\alpha$ from $\rho\rho$ (top-left), 
$\rho\pi$ (top-right), and their combination (bottom-left).
On bottom right we show the impact of the $\alpha$ determination
on the $\rhobar-\etabar$ plane.}}
\label{fig:alpha}
\end{figure}

In Table~\ref{tab:alpha-exp}, we collect the experimental information on these modes.
It is important to remark the situation of
the measurements of $S$ and $C$ in $B \to \pi\pi$ decays. The BaBar and Belle Collaborations, even with 
increased statistics, confirmed the disagreement between their results (at about $3.2\sigma$) \cite{ref:pipi}.
For this reason, we exclude these measurements from the global fit waiting for a clarification 
of the experimental results in the near future and use only the experimental informations 
from $B \to \rho \rho$ decay modes.
The result on $\alpha$ from $\rho\rho$ is shown in Figure~\ref{fig:alpha} and given in Table~\ref{tab:isospin-pipi}.

\begin{table}[tb]
  \centerline{\small
    \begin{tabular}{c|ccc|ccc}
     \hline\hline
                  &      \multicolumn{3}{c}{$\pi \pi$}  &
\multicolumn{3}{c}{$\rho \rho$} \\ 
     \hline     
     Observable   &      BaBar      &   Belle      &     Average    &    BaBar  
   &      Belle   &   Average \\
     \hline     
     $C$                &  -0.09  $\pm$ 0.16  & -0.58 $\pm$ 0.17  & -                  & -0.23 $\pm$ 0.28  &   -           & -0.23  $\pm$ 0.28 \\
     $S$                &  -0.30  $\pm$ 0.17  & -1.00 $\pm$ 0.22  & -                  & -0.19 $\pm$ 0.35  &    -          & -0.19  $\pm$ 0.35 \\
     $BR^{+-}(10^{-6})$ &   4.7   $\pm$ 0.6   &  4.4 $\pm$ 0.7   &  4.6   $\pm$ 0.4   & 30.0  $\pm$ 6.0   &     -         &  30.0  $\pm$ 6.0 \\
     $BR^{+0}(10^{-6})$ &   5.8   $\pm$ 0.7   &  5.0 $\pm$ 1.3   &  5.5   $\pm$ 0.6   & 22.5  $\pm$ 8.1   &   31.7 $\pm$ 9.8 &  26.4  $\pm$ 6.4 \\
     $BR^{00}(10^{-6})$ &   1.17  $\pm$ 0.34  &  2.32 $\pm$ 0.53  &  1.51  $\pm$ 0.28  & 0.54  $\pm$ 0.41   &     -         &    0.54 $\pm$ 0.41 \\
     \hline\hline
    \end{tabular}}
  \caption{{\it Experimental inputs from isospin analyses in $B \to \pi \pi$ and
$B \to \rho \rho$ decays~\cite{ref:hfag}. The $B \to \rho \rho$ decays are assumed to be 
fully polarized, in agreement with available measurements.}}
  \label{tab:alpha-exp}
\end{table}

\begin{table}[htbp!]
\begin{center}
\begin{tabular}{cccc} 
\hline\hline
  & Output Value & $95\%$  & $99\%$  \\
$\rho \rho$ \\
\hline
$\alpha[^{\circ}$]& 96.2 $\pm$ 5.5 $\cup$ 175.3 $\pm$15.4 & $[73.7,122.7]$  $\cup$   $[153.4,192.5]$ &  $[68.7,202.7]$  \\[-1.2mm]
$BR(\rho^+\rho^-)$ & 32.3 $\pm$ 5.4       & $[21.7,43.0]$         &  $[18.5,46.2]$  \\  
$BR(\rho^+\rho^0)$ & 20.3 $\pm$ 4.4       & $[11.9,28.8]$         &  $[9.5,31.4]$  \\  
$BR(\rho^0\rho^0)$ &  0.6 $\pm$ 0.7       & $[0.1,1.4]$           &  $<1.6$  \\  
\hline \hline
\end{tabular}
\end{center}
\caption{{\it Output values for the main parameters and experimental
observable entering the isospin analysis. 
Branching ratios are quoted in units of $10^{-6}$.}}
\label{tab:isospin-pipi}
\end{table}

\begin{figure}[htb!]
\begin{center}
\includegraphics[height=4.2cm]{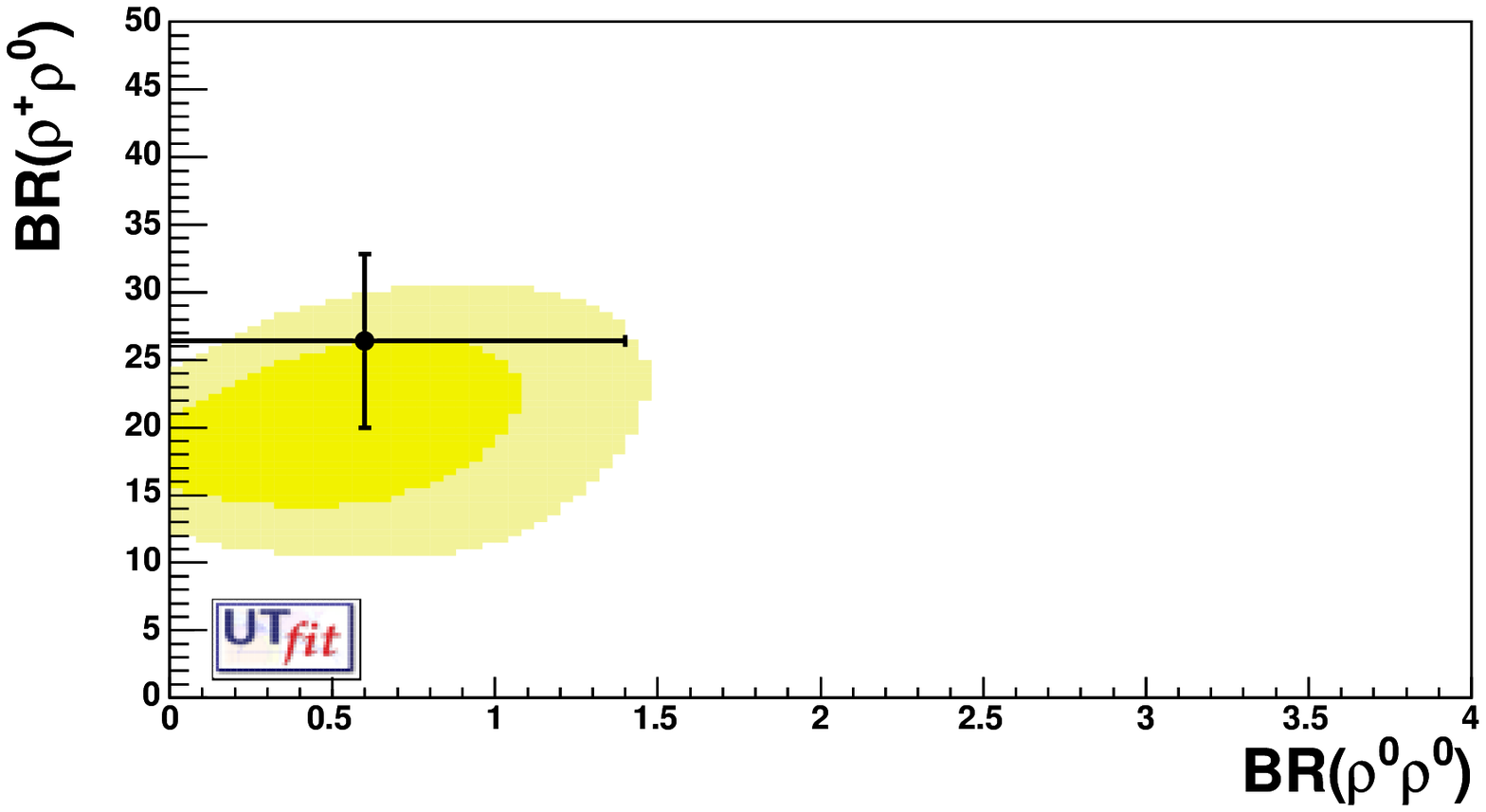}
\includegraphics[height=4.2cm]{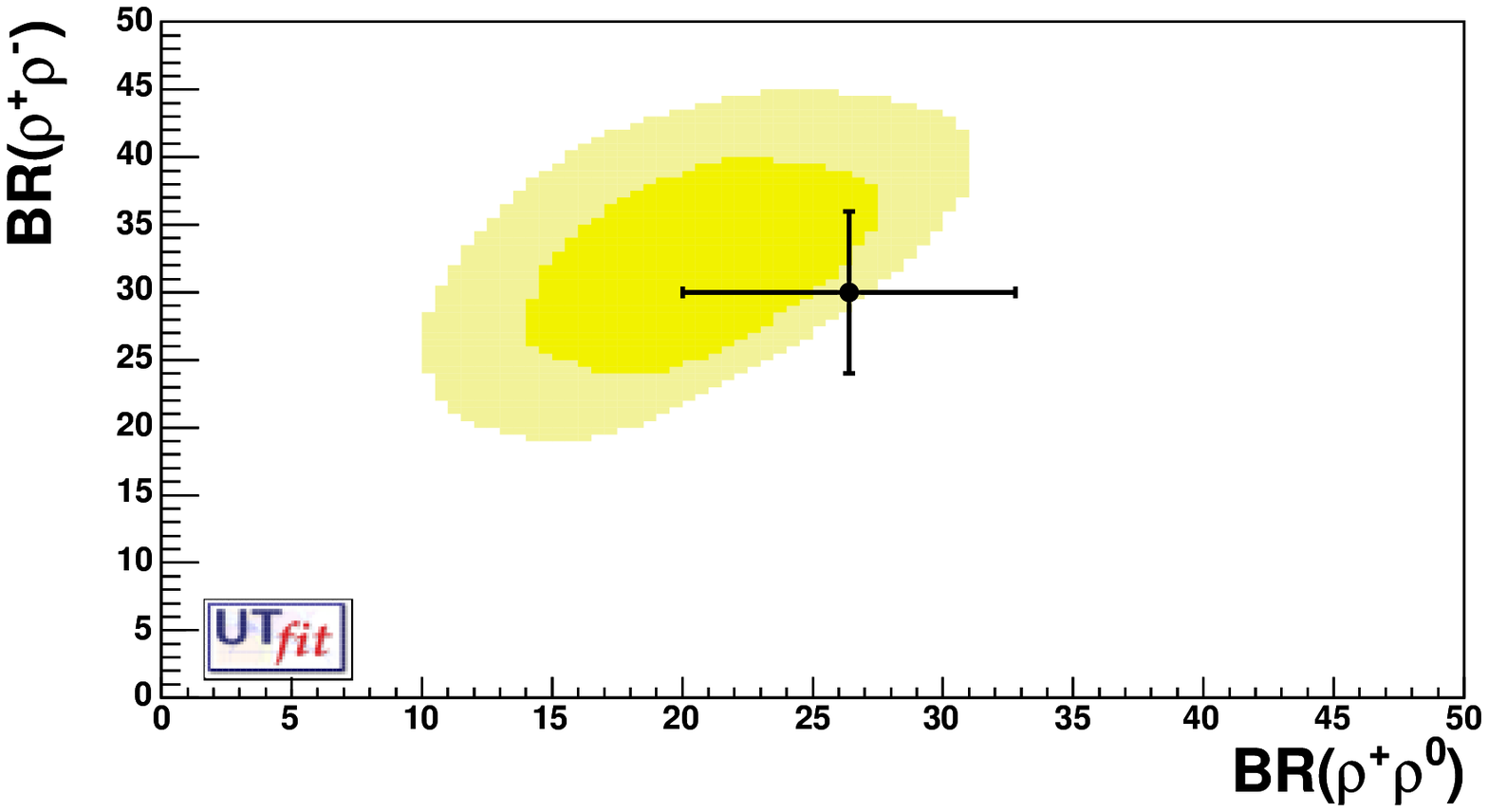} \\
\end{center}
\caption{{\it Plots showing the correlation of $BR(B^+ \to \rho^+ \rho^0)$ vs
    $BR(B^0 \to \rho^0 \rho^0)$(left) and $BR(B^0 \to \rho^+ \rho^-)$
    vs $BR(B^+ \to \rho^+ \rho^0)$(right). 
    The crosses indicate the 1$\sigma$ experimental 
    ranges for the branching ratios, quoted in units of $10^{-6}$.}}
\label{fig:brrhorhp2D}
\end{figure}

The different branching ratios can also be obtained {\it a-posteriori} and
compared with the measured input values.
The results of this test are presented in Table~\ref{tab:isospin-pipi}.
In Figure~\ref{fig:brrhorhp2D} we show the 2D distribution of the
$B \to \rho \rho$ case.

 
\subsubsection{Dalitz analysis of \boldmath$B^0 \to (\rho\pi)^0$}

The time-dependent study of $B^0 \to (\rho\pi)^0$ decays on the Dalitz plot is a powerful way to obtain $\alpha$~\cite{quinn-snyder}. 
The sensitivity of this analysis depends upon the measurements of the branching ratios and CP asymmetries for the various
intermediate resonances as well as from the interference between 
different amplitudes through mixing.
In the SM and without further theoretical assumptions,
one can write the amplitudes of the various intermediate decay 
modes of the $B$ meson as
\begin{eqnarray}
A^{+-} &=& T^{+-} e^{-i\alpha} + P^{+-} \nonumber \\
A^{-+} &=& T^{-+} e^{-i\alpha} + P^{-+} \nonumber \\
A^{00} &=& T^{00} e^{-i\alpha} + P^{00}
\label{eq:rhopi}
\end{eqnarray}
where $T$ and $P$ parameters are complex amplitudes, carrying their own
strong phase, and the first (second) superscript refers to the 
$\rho$ ($\pi$) charge. The amplitudes for the CP-conjugate decays can be parameterized in the same way.
The two sets of relations are characterized by 13 unknowns, one of which is
a global phase that can be removed and one is fixed by the normalization 
of the Dalitz plot. Moreover, using SU(2) flavour symmetry, $P^{00}$ can be written as a 
function of $P^{+-}$ and $P^{-+}$. This leaves 9 quantities to be determined
including $\alpha$.   
Recently, BaBar reported a result of this analysis~\cite{BaBar-rhopi}
which gives the values of 16 experimental observables that can be written as
functions of these 9 unknowns. The result for $\alpha$ is shown in the top--right plot
of Figure~\ref{fig:alpha}. \\

\subsubsection{Combined results on $\alpha$}

The results on $\alpha$ from $\rho\rho$, $\rho\pi$, their combination
and the bound on the $\rhobar-\etabar$ plane are shown in Figure \ref{fig:alpha}.
We obtain
\begin{equation}
\alpha = (107 \pm 11)^\circ \cup (176 \pm 14)^\circ ~ ~ ([80,138] \cup [150,197]~@95\%)
\end{equation}
Notice however that the interpretation of this result can be misleading. Indeed, if an independent information would allow discarding the second solution at $176^\circ$, the error associated to the first determination of $\alpha$ would be 
larger than $11^\circ$.

\subsubsection{Determination of \boldmath$\cos 2\beta$ from \boldmath$J\psi K^{*0}$ decays}
\label{sec:c2b}

From the time-dependent analysis of the decay $B^0 \to J/\psi K^{*0}$, 
it is possible to extract both $\sin 2\beta$ and $\cos 2\beta$~\cite{babar-book}.
The results obtained by Belle and BaBar are barely compatible (see Table \ref{tab:cos2b}) 
and we combined them using the skeptical approach of ref.~\cite{dago}, assuming
the values $\delta=1.3$ and $\lambda=0.6$ for the parameters of the model.
We also varied these two parameters without obtaining sizable deviations for 
the observed combined p.d.f.

\begin{table}[htbp!]
\begin{center}
\begin{tabular}{cc} 
\hline\hline
             &  $\cos 2\beta$         \\
\hline
BaBar        &   3.32$^{+0.76}_{-0.96}\pm$0.27 \\ 
Belle        &   0.31 $\pm$ 0.91 $\pm$ 0.11   \\
Skeptical  &   1.9 $\pm$ 1.3 \\ 
\hline \hline
\end{tabular}
\end{center}
\caption{{\it Experimental values for $\cos 2\beta$ and
our skeptical average.}}
\label{tab:cos2b}
\end{table}

\begin{figure}[htb!]
\begin{center}
\includegraphics[width=0.4\textwidth]{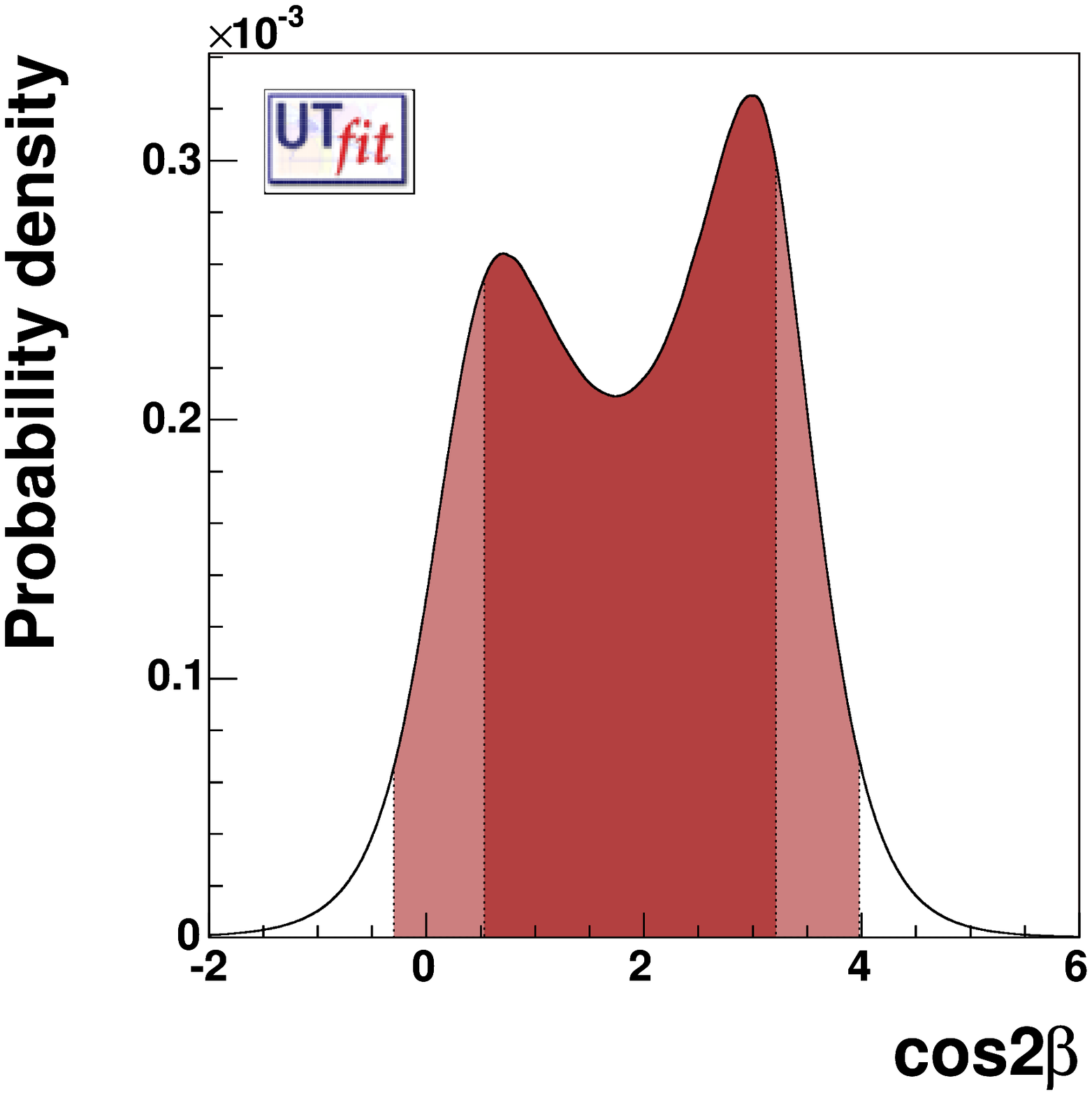}
\includegraphics[width=0.58\textwidth]{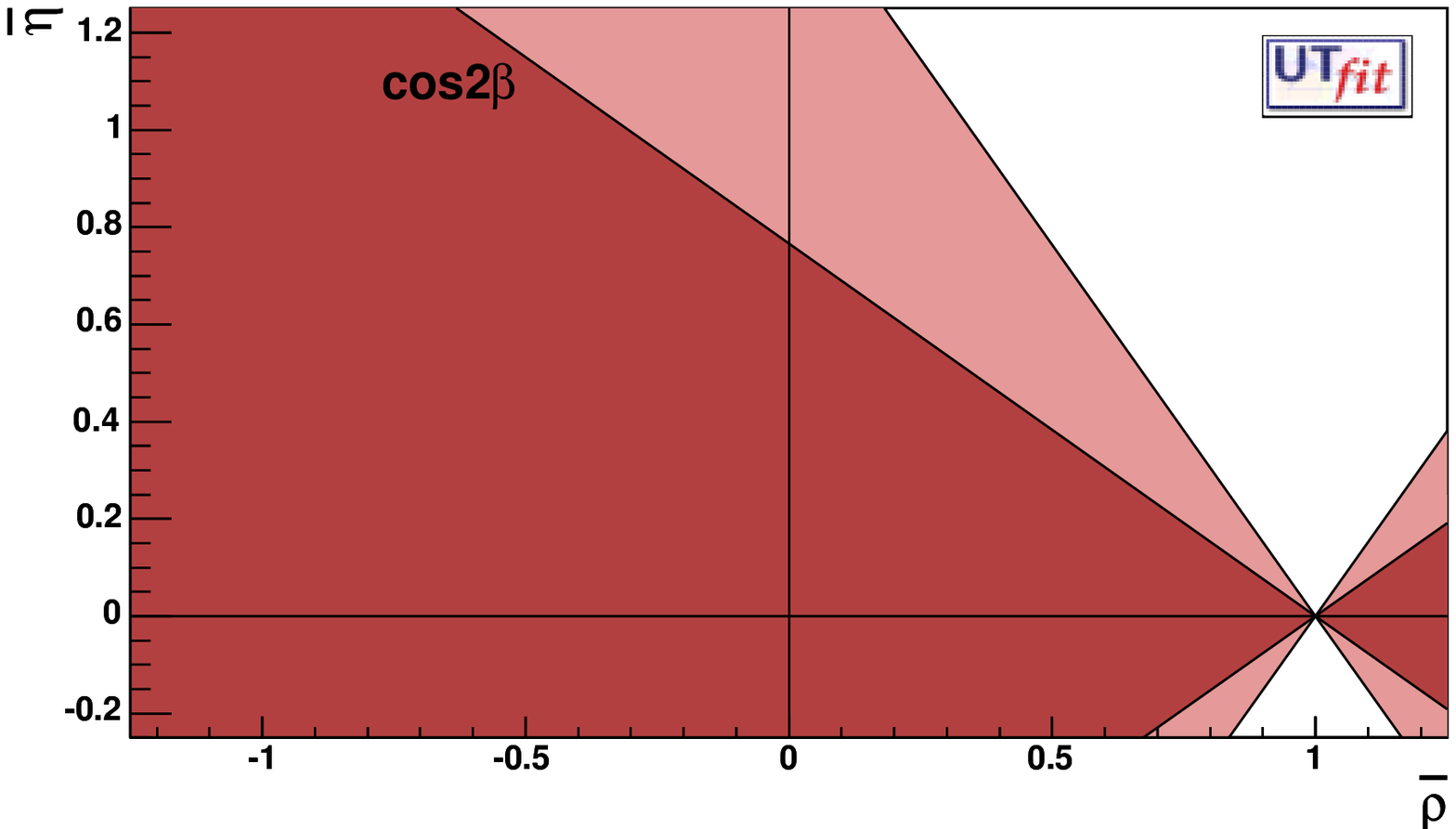} \\
\end{center}
\caption{{\it The skeptical likelihood of $\cos 2\beta$ (left) and the $\cos 2\beta$ constraint on the $\rhobar-\etabar$ plane (right).}}
\label{fig:cos2b}
\end{figure}

The skeptical likelihood of $\cos 2\beta$ and the impact of this determination
on the $\rhobar-\etabar$ plane, once the
{\it a-priori} bound $|\cos 2\beta|<1$ is imposed, are shown in
Figure~\ref{fig:cos2b}. 
The probability of $\cos 2\beta$ being less than zero is about $13\%$. This implies that this measurement 
removes the ambiguity associated to $\sin 2 \beta$, suppressing one of the two allowed bands for $\sin 2\beta$ in Figure~\ref{fig:rhoeta}.

\subsection{Determination of the Unitarity Triangle parameters using also the new UT angle measurements}
\label{sec:newresults}

\begin{figure}[tbp]
\begin{center}
\includegraphics[width=14cm]{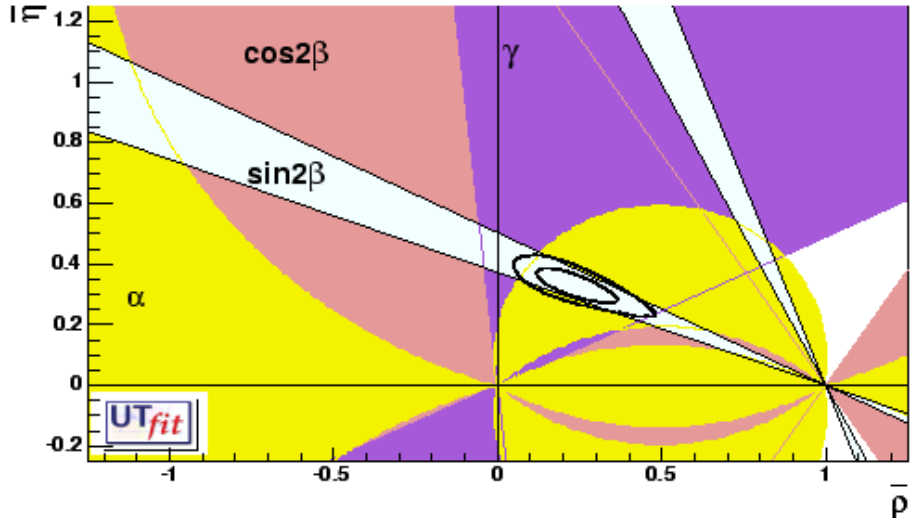}
\caption{ \it {Allowed regions for $\rhobar$ and $\etabar$ obtained using the
    measurements of the UT angles only: $\snb$, $\alpha$,
    $\gamma$, and $\cos 2\beta$. The closed contours at $68\%$
    and $95\%$ probability are shown. The full zones correspond to $95\%$
    probability regions from individual constraints.}}
\label{fig:soloangoli}
\end{center}
\end{figure}

It is interesting to see the selected region in $\rhobar-\etabar$
plane from the direct measurements of the UT angles: $\snb$, $\gamma$, $\alpha$, and $\cos 2\beta$. The plot is shown in Figure~\ref{fig:soloangoli}. In Table~\ref{tab-bfactory}
we report the results obtained using these constraints.

\begin{figure}[htbp!]
\begin{center}
\includegraphics[width=14cm]{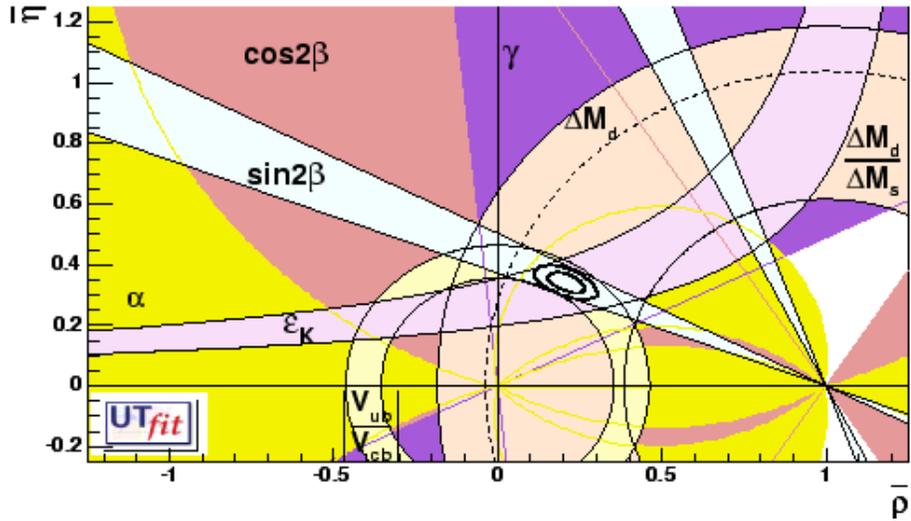}
\caption{ \it {Allowed regions for $\rhobar$ and $\etabar$ using the parameters
    listed in Table~\ref{tab:inputs} together with the new UT angle measurements.
    The closed contours at $68\%$ and $95\%$ probability are shown. The full lines
    correspond to $95\%$ probability regions for each of the constraints, given by the
    measurements of $\left | V_{ub} \right |/\left | V_{cb} \right |$,
    $\Delta m_d$, $\Delta m_s$, $\epsilonk$, $\snb$, $\gamma$,
    $\alpha$, and $\cos 2\beta$, respectively.}}
\label{fig:allall}
\end{center}
\end{figure}

The results given in Table~\ref{tab:1dimangoli} are obtained using all
the available constraints: $\left | V_{ub} \right |/\left | V_{cb}
\right |$, $\Delta {m_d}$, $\Delta {m_s}$, $\epsilonk$, $\snb$, $\gamma$, $\alpha$, and $\cos 2\beta$.
Figure~\ref{fig:allall} shows the corresponding selected region in the $\rhobar-\etabar$ plane.

\begin{table*}[h]
\begin{center}
\begin{tabular}{@{}llllll}
\hline\hline  
    Parameter               &     ~~~~~68$\%$             &      ~~~~~95$\%$     &    ~~~~~99$\%$      \\  \hline 
~~~~~$\overline {\rho}$          & 0.241  $\pm$ 0.081   & [0.090,\,0.441]   & [0.036,\,0.753] \\
~~~~~$\overline {\eta}$          & 0.328  $\pm$ 0.038   & [0.249,\,0.411]   & [0.182,\,0.468] \\ \hline
~~~~$\alpha [^{\circ}]$          & 102  $\pm$ 11     & [78,\,124]    & [66,\,134]  \\
~~~~$\beta [^{\circ}]$          & 23.6  $\pm$ 1.8     &  [19.9,\,26.9]   & [18.2,\,28.7]       \\
~~~~$\gamma[^{\circ}$]           & 54.1   $\pm$ 11.7     & [30.6,\,76.6]     & [21.9,\,85.6]   \\ \hline
    ~$\sin 2\alpha$       & -0.38  $\pm$ 0.36     &  [-0.94,\,0.38]   & [-0.99,\,0.66]       \\
          ~$\snb$    & 0.727  $\pm$ 0.037   & [0.654,\,0.800]   & [0.632,\,0.823] \\
    ~$\sin(2\beta+\gamma)$    & 0.955  $\pm$ 0.038     &  [0.805,\,0.998]   & [0.159,\,1.0]       \\\hline
$\rm{Im} {\lambda}_t$[$10^{-5}$] & 12.8   $\pm$ 1.6     & [9.6,\,16.0]     & [8.2,\,17.3]   \\
\hline\hline
\end{tabular} 
\end{center}
\caption {\it Values and probability ranges for the UT parameters obtained by using the constraints:
$\snb$, $\gamma$, $\alpha$, and $\cos 2\beta$.}
\label{tab-bfactory} 
\end{table*}

\begin{table*}[h]
\begin{center}
\begin{tabular}{@{}llllll}
\hline\hline  
    Parameter               &      ~~~~~68$\%$       &      ~~~~~95$\%$   &    ~~~~~99$\%$     \\   \hline
~~~~~$\overline {\rho}$          & 0.207  $\pm$ 0.038   & [0.129,\,0.282]   & [0.106,\,0.308] \\
~~~~~$\overline {\eta}$          & 0.341  $\pm$ 0.023   & [0.296,\,0.386]   & [0.282,\,0.400] \\\hline
~~~~$\alpha [^{\circ}]$             & 97.9   $\pm$ 6.0     & [86.0,\,109.7]    & [82.5,\,113.7]  \\
 ~~~~$\beta [^{\circ}]$          & 23.4  $\pm$ 1.5     &  [20.7,\,26.1]   & [20.2,\,27.1]       \\
~~~~$\gamma[^{\circ}$]           & 58.5   $\pm$ 5.8     & [47.3,\,70.2]     & [43.5,\,73.8]   \\ \hline
    ~$\sin 2\alpha$       & -0.27  $\pm$ 0.20     &  [-0.64,\,0.13]   & [-0.75,\,0.25]       \\
         ~$\snb$                & 0.725  $\pm$ 0.028   & [0.669,\,0.779]   & [0.651,\,0.795] \\
    ~$\sin(2\beta+\gamma)$    & 0.958  $\pm$ 0.030     &  [0.884,\,0.997]   & [0.853,\,0.998]       \\\hline
$\rm{Im} {\lambda}_t$[$10^{-5}$] & 13.2   $\pm$ 0.9     & [11.5,\,14.8]     & [11.0,\,15.3]   \\
\hline\hline
\end{tabular} 
\end{center}
\caption {\it Values and probability ranges for the Unitarity Triangle
parameters obtained by using all 
the available constraints: $\left | V_{ub} \right |/\left | V_{cb} \right |$, 
$\Delta {m_d}$, $\Delta {m_s}$, $\epsilonk$, $\snb$, $\gamma$, $\alpha$, and $\cos 2\beta$.}
\label{tab:1dimangoli} 
\end{table*}

Given the present experimental uncertainties, the new measurements of UT angles,
taken individually, would loosely constrain the values of $\bar \rho$ and $\bar \eta$. In addition, 
some dependence of the results on the choice of the {\it a-priori} distributions is
present, particularly for those quantities which are poorly constrained by the
experiments (as it should be in the Bayesian approach). 
On the other hand, when combined with the $\snb$ measurement, they select an area in 
the $\rhobar$--$\etabar$ plane comparable to the one available in the 
pre-$B$-factory era and largely independent of the chosen {\it a-priori} distributions.
The agreement between this area and the output of the standard UT fit is an important
demonstration of the consistency of the CKM mechanism in describing non-leptonic
$B$ decays and CP asymmetries.

\section {Compatibility plots}
\label{sec:pull}

In this section we discuss the interest of measuring with a better precision the various
physical quantities entering the UT analysis.
We investigate, in particular, to which extent future and improved
determinations of the experimental constraints, such as
$\sin{2\beta}$, $\dms$ and $\gamma$, could allow us to possibly
invalidate the SM, thus signalling the presence of NP effects.

\subsection{Compatibility between individual constraints. The pull
distributions.}
\label{sec:compa}

In CKM fits based on a $\chi^2$ minimization, a conventional
evaluation of compatibility stems automatically from the value of the
$\chi^2$ at its minimum.  The compatibility between constraints in the
Bayesian approach is simply done by comparing two different p.d.f.'s.

Let us consider, for instance, two p.d.f.'s for a given quantity
obtained from the UT fit, $f(x_1)$, and from a direct measurement,
$f(x_2)$: their compatibility is evaluated by constructing the p.d.f. of
the difference variable, $x_2-x_1$, and by estimating the distance of
the most probable value from zero in units of standard deviations. The
latter is done by integrating this p.d.f. between zero and the most
probable value and converting it into the equivalent number of standard
deviations for a gaussian distribution~\footnote{In the case of Gaussian
distributions for both $x_1$ and $x_2$, this quantity coincides with the
pull, which is defined as the difference between the central values of
the two distributions divided by the sum in quadrature of the
r.m.s of the distributions themselves.}.
The advantage of this approach is that no approximation is made
on the shape of p.d.f.'s. In the following analysis, $f(x_1)$ is the
p.d.f. predicted by the UT fit while the p.d.f of the measured quantity,
$f(x_2)$, is taken Gaussian for simplicity. The number of standard
deviations between the measured value, $\bar{x}_2 \pm \sigma(x_2)$,
and the predicted value (distributed according to $f(x_1)$) is plotted
as a function of $\bar{x}_2$ (x-axis) and $\sigma(x_2)$ (y-axis). The
compatibility between $x_1$ and $x_2$ can be then directly
estimated on the plot, for any central value and error of the
measurement of $x_2$.


If two constraints turn out to be incompatible, further investigation
is necessary to tell if this originates from a ``wrong'' evaluation of
the input parameters or from a NP contribution.

\subsection{Pull distribution for \boldmath$\sin{2\beta}$}
\begin{figure}[htbp!]
\hbox to\hsize{\hss
\includegraphics[width=0.5\hsize]{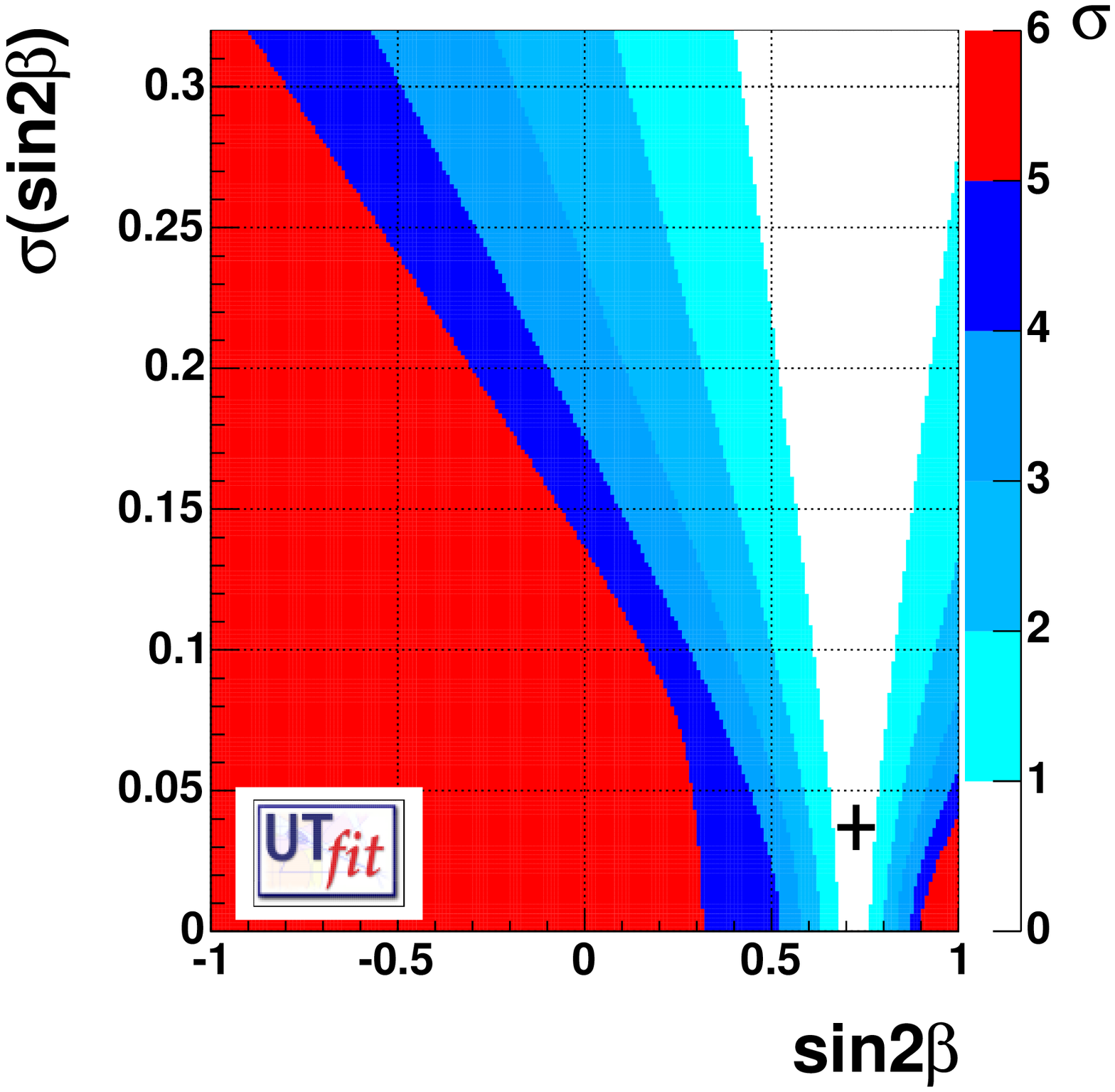}
\includegraphics[width=0.5\hsize]{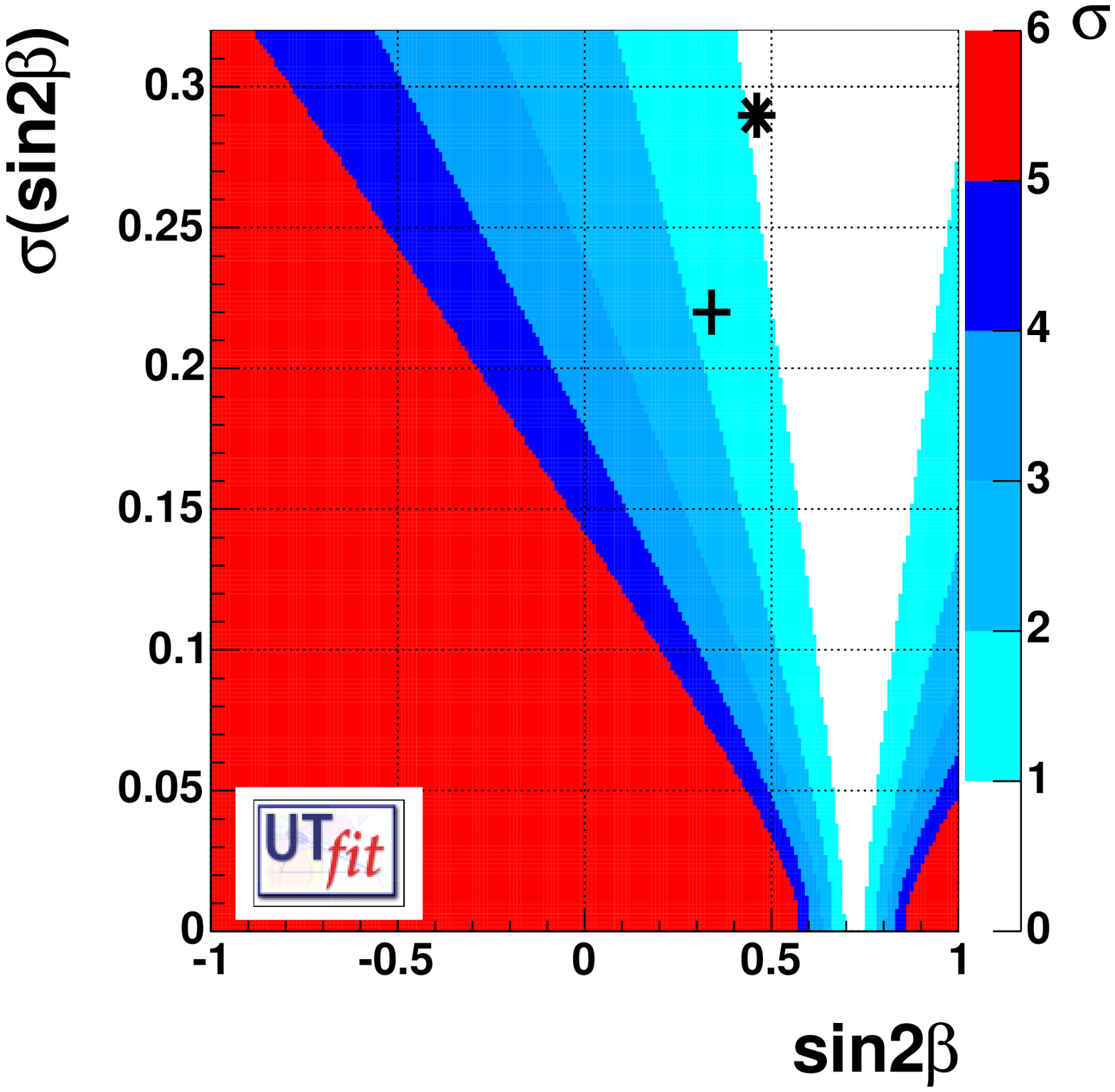}
\hss}
\caption{\it {The compatibility (``pull'') between the direct and indirect 
    determination of $\snb$ as a function of the 
    measured values and errors.
    On the left, the indirect distribution of $\snb$ is computed
    from the UT fit including $\vert V_{ub}/V_{cb}\vert$, $\dmd$, $\dms$, and $\epsilonk$, without using the direct measurement from $B^0 \to J/\psi K^0$. On the right,
    the indirect distribution includes also the constraint from the measurements of $\snb$ from $B^0 \to J/\psi K^0$. The compatibility regions from $1\sigma$ to $6\sigma$
    are displayed. The crosses display the position (value/error) 
    of the measurements of $\snb$ from $B^0 \to J/\psi K^0$(left plot) and from $B\to\phi K^0$ (right plot), extracted from the HFAG average for $S(\phi K^0)$. We also mark with an asterisk  the value of $\snb$ obtained using instead our skeptical combination for $S(\phi K^0)$.}}
\label{fig:pull_sin2b}
\end{figure}

We start this analysis by considering the measurement
of sin2$\beta$. 
The left plot in Figure~\ref{fig:pull_sin2b} shows the compatibility (``pull'') between the measurement of $\sin{2\beta}$ and its indirect
determination, obtained in the SM using the constraints $\left | V_{ub} /| V_{cb} \right |$, $\Delta m_d$, $\Delta m_s$, and $\epsilonk$ (but excluding $S(J/\psi K^0)$), as
function of the measured value (x-axis) and error (y-axis) 
of $\snb$. The cross indicates the present experimental average 
of $S(J/\psi K^0)$.

The plot shows
that, considering the present precision of $0.037$ on the measured
value of $\sin{2\beta}$, the 3$\sigma$ compatibility region is in the range
[0.51,\,0.88].  Values outside this range would be, therefore, not
compatible with the SM prediction at more than $3\sigma$ level. To get
these values, however, the presently measured central value should
shift by at least $6\sigma$.

The conclusion that can be derived from Figure~\ref{fig:pull_sin2b} is
the following: although the improvement on the error on sin2$\beta$
has an important impact on the accuracy of the UT parameter determination, 
it is very unlikely that in the near future we will be
able to use this measurement to detect any failure of the SM, unless
the other constraints entering the fit improve substantially or, of
course, in case the central value of the direct measurement moves away
from the present one by several standard deviations.

\subsubsection{R\^ole of \boldmath$\snb$ from Penguin processes}

It was pointed out some time ago that the comparison of the
time-dependent CP asymmetries in various $B$ decay modes could provide
evidence of NP in $B$ decay amplitudes~\cite{thphi}. Since
$\sin{2\beta}$ is known from $S(J/\psi K^0)$, a significant deviation of
the time-dependent asymmetry parameters of penguin dominated channels
from their expected values would indicate the presence of NP.

From this point of view, the pure penguin $b \to s \bar s s$ 
processes as $B \to \phi K^0$ are the cleanest probes of NP. The only 
SM uncertainty in the extraction of $\snb$ from $S(\phi K^0)$ comes 
from the penguin matrix elements of current-current operators containing 
up-type quarks. This is expected to be small and can be estimated in a given 
model of hadron dynamics and constrained using the experimental data.
For example, using the model of ref.~\cite{strike}, we obtain 
\begin{equation}
\snb = \left\{
\begin{array}{l l r}   
 0.34 \pm 0.20 \pm 0.08& {\rm from~} S_{\phi K^0}=0.34 \pm 0.20& {\rm (HFAG)}\\
 0.46 \pm 0.28 \pm 0.08& {\rm from~} S_{\phi K^0}=0.46 \pm 0.28&{\rm (skeptical)}
\end{array}
\right.
\label{eq:sphik}
\end{equation}
In the previous equation we have used both the HFAG average for $S(\phi K^0)$ and our estimate 
obtained using the skeptical approach of ref.~\cite{dago} (with $\delta=0.6$ and $\lambda=1.3$).
The average value of $S(\phi K^0)$ is obtained using four measurements having $\chi^2/NDF$=2.6. 
As this result is used to find evidence for NP it has been considered that systematic uncertainties may have been
underestimated, so errors have been inflated using the
approach~\cite{dago}. If the present discrepancy were instead due
to a statistical fluctuation, it should disappear in the future and the two approaches will converge to the same result.

The compatibility of $\snb$ shown in the right plot of Figure~\ref{fig:pull_sin2b} has been obtained using 
all the constraints of the standard analysis, namely $\vert V_{ub} / V_{cb}\vert$, $\Delta m_d$, $\Delta m_s$, 
$\epsilonk$, including $\snb$. The cross and star on this plot correspond to the values of $\snb$ in 
Eq.~(\ref{eq:sphik}), extracted from the HFAG and the skeptical average respectively. Depending on the 
chosen average procedure, the compatibility of $\snb$ from $B\to\phi K^0$ with the indirect determination ranges from 
about $1\sigma$ to $2\sigma$.

The extraction of $\snb$ could be extended to channels receiving additional
contributions from $b \to u \bar u s$  transitions, such as $B \to \eta^\prime K^0$,
$B^0 \to f_0 K^0$ or $B \to \pi^0 K^0$. However, in these cases, hadronic uncertainties are more difficult to estimate and expected to be channel-dependent. For this reason,
CP asymmetries in these channels can be different and should not be na\"{\i}vely averaged.
A detailed analysis of these modes goes beyond the
goal of this work and will be presented elsewhere.

\subsection{Pull distribution for $\dms$}
\begin{figure}[htbp!]
\hbox to\hsize{\hss
\includegraphics[width=0.5\hsize]{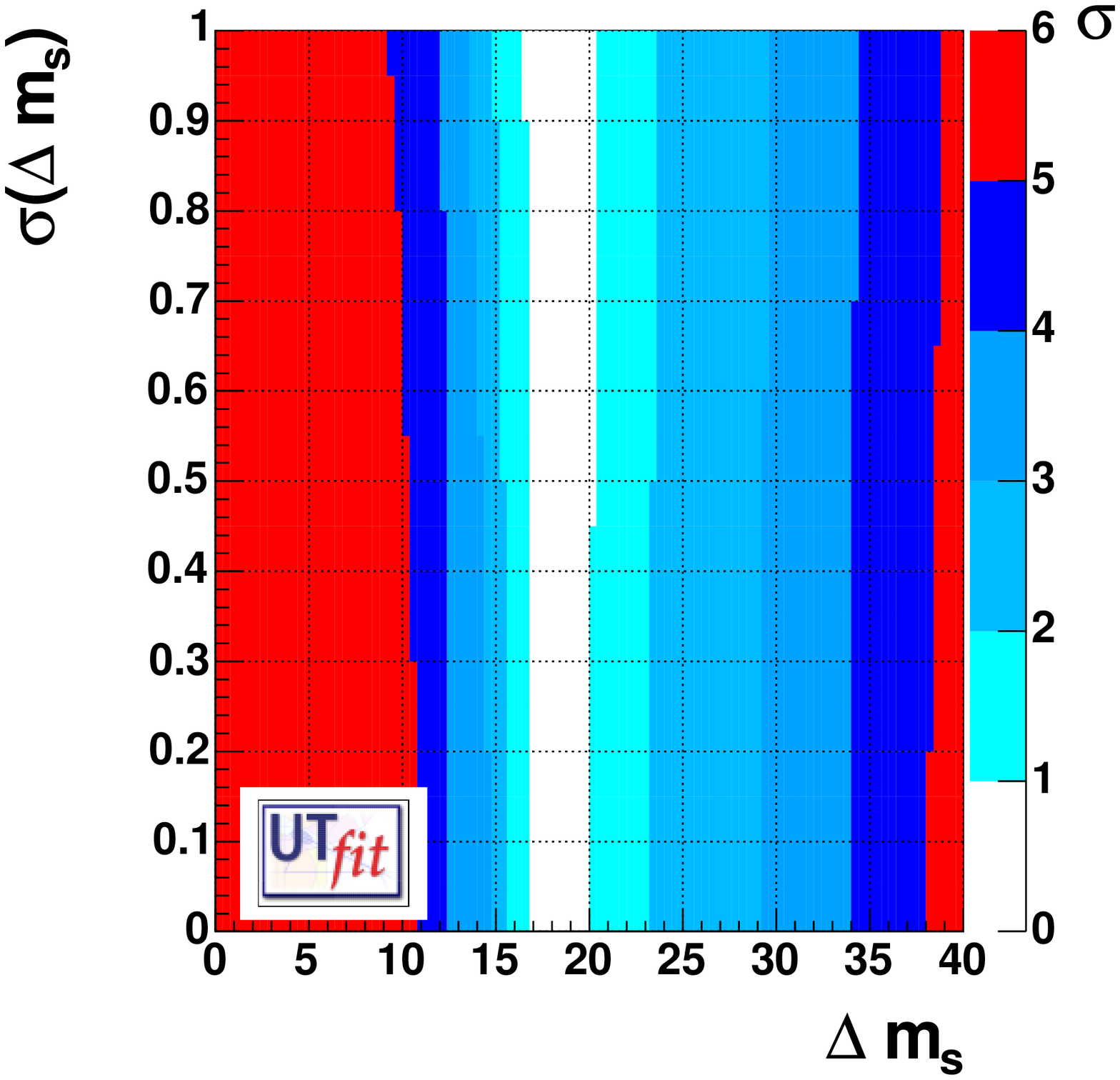}
\includegraphics[width=0.5\hsize]{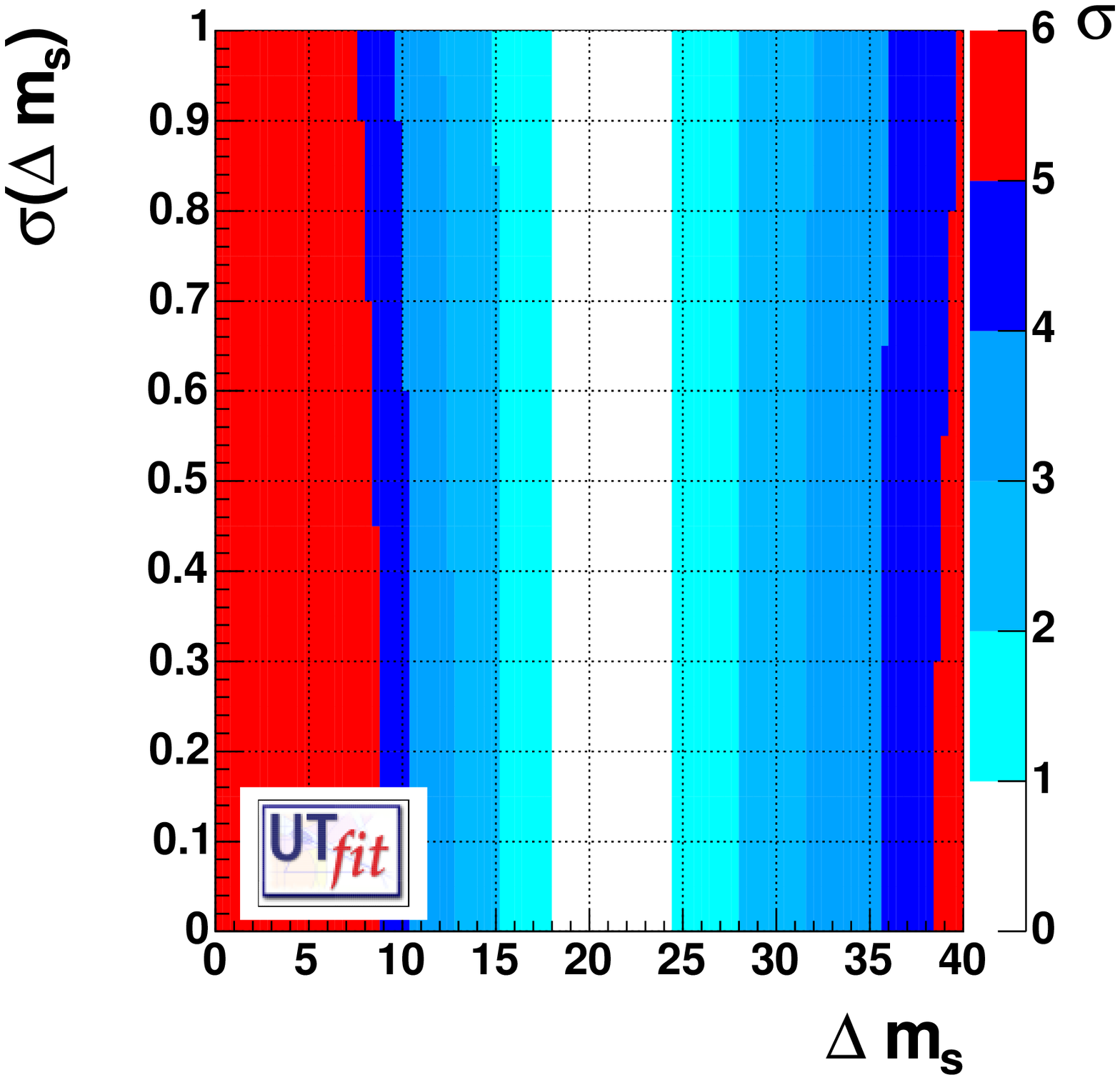}
\hss}
\caption{\it {The compatibility between the direct and indirect determination of
    $\dms$, as a function of the value of $\dms$($ps^{-1}$), using (left) or
    ignoring (right) the present experimental bound.}}
\label{fig:pull_dms}
\end{figure}

The plots in Figure~\ref{fig:pull_dms} show the compatibility of the
indirect determination of $\dms$ with a future determination of the
same quantity, obtained using or ignoring the experimental information
coming from the present bound.

From the plot in Figure~\ref{fig:pull_dms} we conclude that,
once a measurement of $\dms$ with an expected accuracy of $\sim 1$
ps$^{-1}$ is available, a value of $\dms$ greater than $32$ ps$^{-1}$
would imply NP at $3\sigma$ level or more.

\subsection{Pull distribution for the angles $\gamma$ and $\alpha$}

\begin{figure}[htbp!]
\hbox to\hsize{\hss
\includegraphics[width=0.5\hsize]{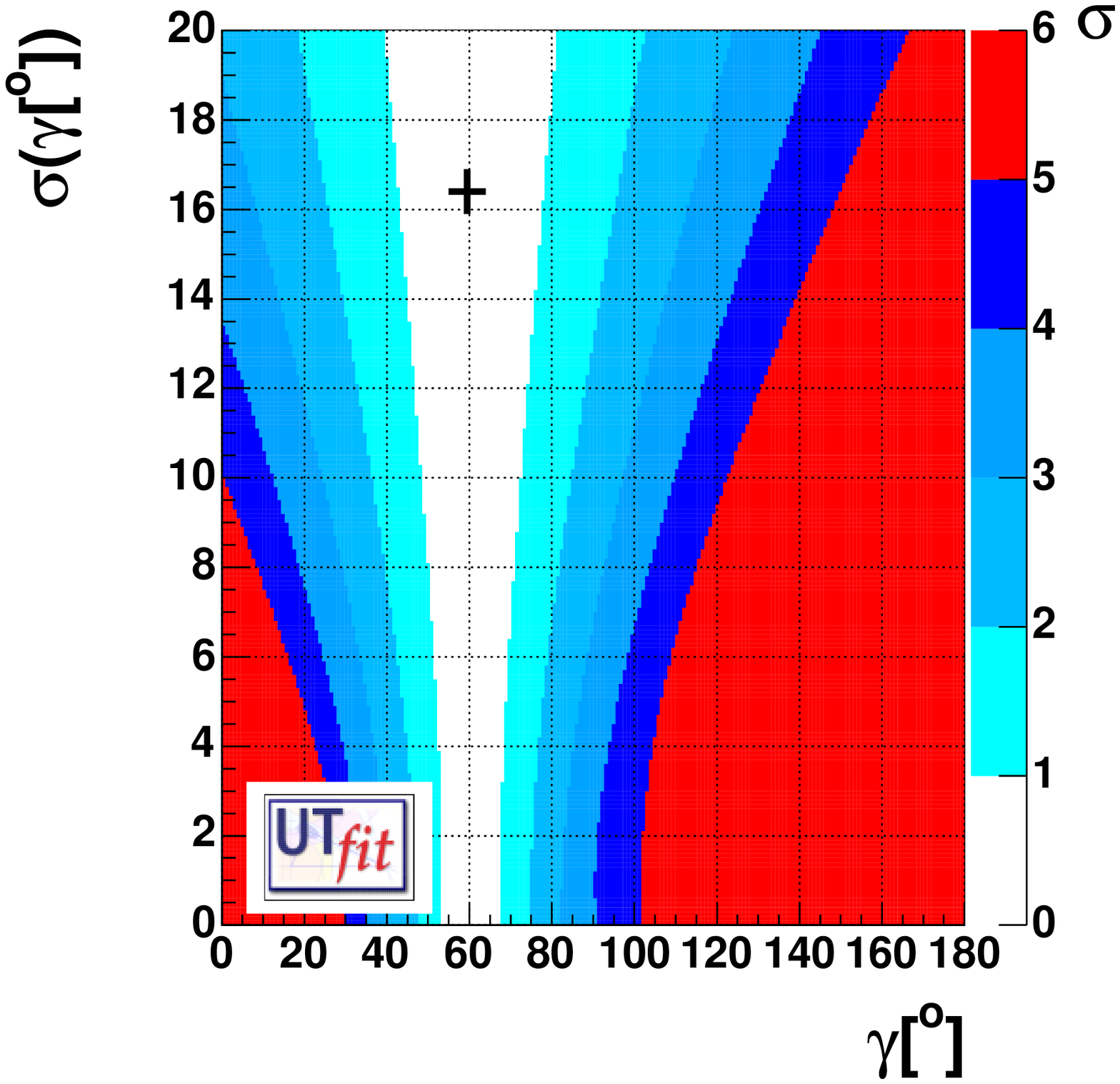}
\includegraphics[width=0.5\hsize]{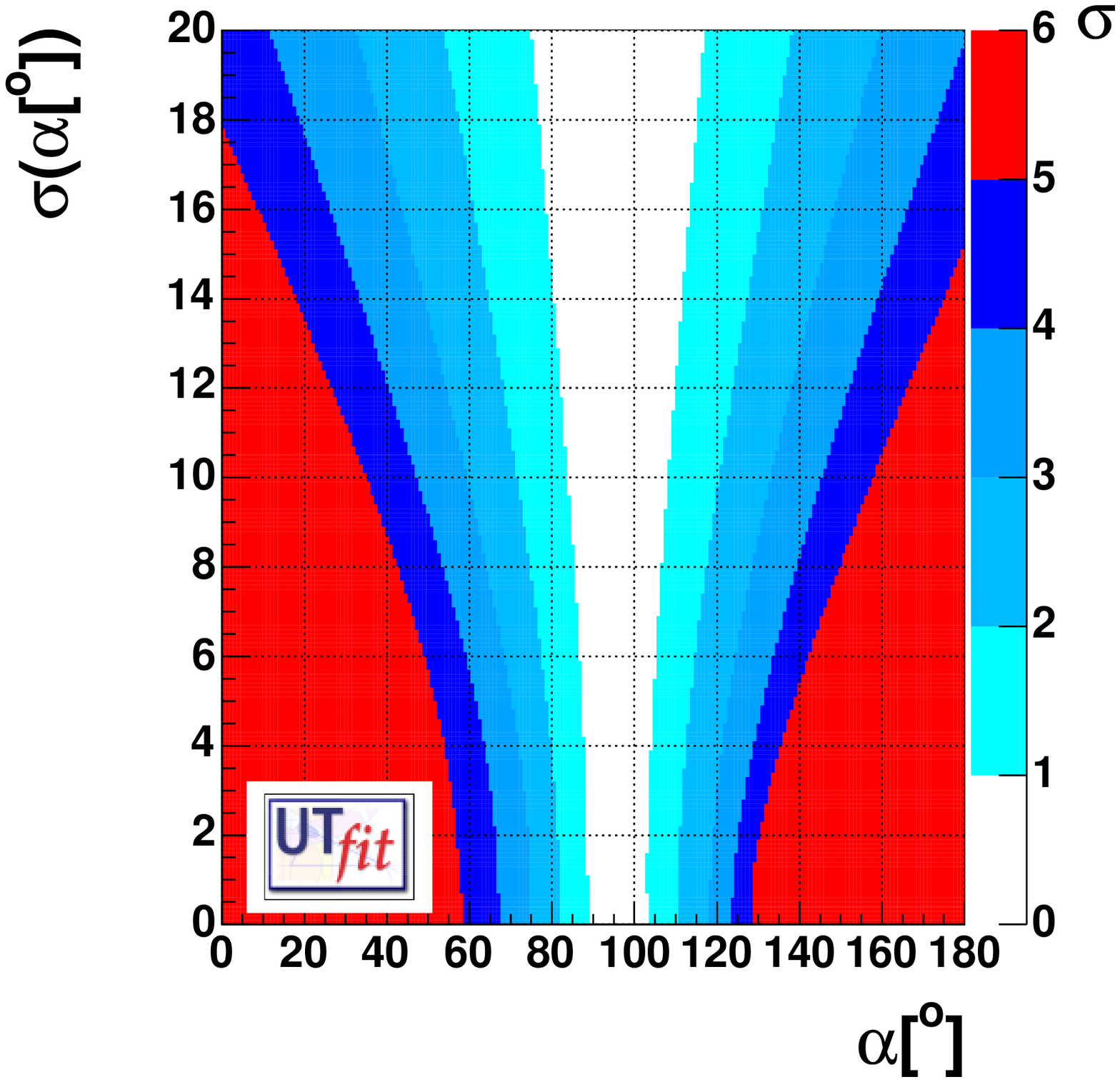}
\hss}
\caption{\it {The compatibility between the direct and indirect determination of
    $\gamma$ (left) and $\alpha$ (right), 
    as a function of their measured value and corresponding uncertainty using the {\utfit} results.
    The cross displays the position (value/error) of the present measurements 
    of $\gamma$ from charged B mesons decaying into $D^{*}K^{*}$ final states.}}
\label{fig:pull_gamma}
\end{figure}

The left plot in Figure~\ref{fig:pull_gamma} shows the compatibility of the
direct and indirect determination of $\gamma$. It can be noted that, even in case
the angle $\gamma$ can be measured with a precision of $10^{\circ}$
from $B$ decays, the predicted $3\sigma$ region is still rather large,
corresponding to the interval [25,/,95]$^{\circ}$. Values larger than
100$^{\circ}$ would clearly indicate physics beyond the Standard Model. 

The present direct determination of the angle $\gamma$, being 
in perfect agreement with UT fit, cannot provide 
any evidence of NP independently of the precision of the measurement.
To a lesser extent, the same conclusion can be drawn for
$\alpha$, see right plot in Figure~\ref{fig:pull_gamma}.
Assuming again a precision of $10^{\circ}$, NP at $3\sigma$ shows up
outside the range [58,/,132]$^{\circ}$.

\section{Conclusions}

Flavour physics in the quark sector has entered its mature age.  Today
the Unitarity Triangle parameters are known with good precision.  A
crucial test has been already done: the comparison between the
Unitarity Triangle parameters, as determined with quantities sensitive
to the sides of the triangle (semileptonic $B$ decays and oscillations),
and the measurements of CP violation in the kaon ($\epsilon_K$) and in
the $B$ (sin2$\beta$) sectors. The agreement is ``unfortunately''
excellent. The Standard Model is ``Standardissimo'': it is also
working in the flavour sector. This test of the SM has been 
allowed by the impressive improvements achieved on non-perturbative
methods which have been used to extract the CKM parameters.

Many $B$ decay branching ratios and CP asymmetries have been
measured at $B$ factories.  The outstanding result is the determination
of sin 2$\beta$ from $B$ hadronic decays into charmonium-$K^0$ final
states. On the other hand many other exclusive hadronic rare $B$ decays
have been measured and constitute a gold mine for weak and hadronic
physics, allowing in principle to extract different combinations of
the Unitarity Triangle angles.

Besides presenting an update of the standard {\utfit} analysis, we have
shown in this paper that new measurements at $B$ factories begin to have
an impact on the overall picture of the Unitarity Triangle
determination. In particular the angle $\gamma$ is today measured 
through charged $B$ decays into $DK$ final states within 20$\%$ accuracy
and only a twofold ambiguity.
In the following years the precise measurements of the UT angles 
will provide further tests of the Standard Model in the flavour 
sector to an accuracy up to the percent level.

Finally, by introducing the compatibility plots, we have studied the impact
of future measurements for testing the SM and looking for new physics.
In the near future the measurement of $\Delta m_s$ and of the UT angle 
$\gamma$ will play the leading r\^ole.

\section{Acknowledgements}

We would like to warmly thank people who provided us the experimental
and theoretical inputs which are an essential part of this work and
helped us with useful suggestions for the correct use of the
experimental information. We thank: A.~Bevan, T.~Browder,
C.~Campagnari, G.~Cavoto, M.~Danielson, R.~Faccini, F.~Ferroni, P. Gambino,
G. Isidori, M.~Legendre, O.~Long, F.~Martinez, L.~Roos, A.~Poulenkov, M.~Rama,
Y.~Sakai, M.-H.~Schune, W.~Verkerke, M.~Zito.  We also thank
A.~Soni for useful discussions.  Finally, we thank
M.~Baldessari, C.~Bulfon, and all the BaBar Rome group for help in the
realization and for hosting the web site.

\end{document}